\documentclass[aps,pre,superscriptaddress,citeautoscript, amsmath,amssymb,twocolumn]{revtex4-1}
\usepackage{graphicx}
\usepackage{amsmath}
\usepackage{hyperref}
\usepackage{color}
\usepackage{bigints}
\usepackage{mathrsfs}
\usepackage{algorithm}
\usepackage{algpseudocode}
\usepackage{comment}
\definecolor{darkblue}{RGB}{8,81,156}
\hypersetup{colorlinks,breaklinks,
            linkcolor=darkblue,urlcolor=darkblue,
           anchorcolor=darkblue,citecolor=darkblue}

\date{\today}

    \definecolor{dark-purple}{RGB}{118,42,131}
    \definecolor{dark-green}{RGB}{27,120,55}
    \definecolor{light-purple}{RGB}{231,212,232}
    \definecolor{LIGHT-PURPLE}{RGB}{194,165,207}
    \definecolor{light-green}{RGB}{168,216,183}
    \definecolor{gray}{RGB}{186,186,186}
    \definecolor{super-dark-green}{RGB}{0,69,41}
    \definecolor{super-dark-purple}{RGB}{63,0,125}
    \definecolor{super-dark-blue}{RGB}{8,48,107}
    \definecolor{super-dark-red}{RGB}{165,0,38}
    
    \definecolor{super-dark-purple}{RGB}{64,0,75}
    \definecolor{super-dark-green}{RGB}{0,68,27}

\usepackage{tabularx,booktabs} 

\usepackage{array}
\usepackage{longtable}
\newcolumntype{L}[1]{>{\raggedright\let\newline\\\arraybackslash\hspace{0pt}}p{#1}}
\newcolumntype{C}[1]{>{\centering\let\newline\\\arraybackslash\hspace{0pt}}m{#1}}
\newcolumntype{R}[1]{>{\raggedleft\let\newline\\\arraybackslash\hspace{0pt}}m{#1}}

\setcitestyle{super}

\begin{document}

\title{Estimating position-dependent and anisotropic diffusivity tensors from molecular dynamics trajectories: Existing methods and future outlook}

\author{Tiago Domingues, Ronald Coifman, Amir Haji-Akbari}
\email{amir.hajiakbaribalou@yale.edu}
\affiliation{Department of Chemical and Environmental Engineering, Yale University, New Haven, CT  06520}

\begin{abstract}
\noindent
Confinement  can substantially alter the physicochemical properties of materials by breaking translational isotropy and rendering all physical properties position-dependent. Molecular dynamics (MD) simulations have proven instrumental in characterizing such spatial heterogeneities and  probing the impact of confinement on materials' properties. For static properties, this is a straightforward task and can be achieved via simple spatial binning. Such an approach, however, cannot be readily applied to transport coefficients due to lack of natural extensions of autocorrelations used for their calculation in the bulk.   The prime example of this challenge is diffusivity, which, in the bulk, can be readily estimated from the particles' mobility statistics, which satisfy the Fokker-Planck equation. Under confinement, however, such statistics will follow the Smoluchowski equation, which lacks a closed-form analytical solution. This brief review explores the rich history of estimating profiles of the diffusivity tensor from MD simulations and discusses various approximate methods and algorithms developed for this purpose. Beside discussing heuristic extensions of bulk methods, we overview  more rigorous algorithms, including kernel-based methods, Bayesian approaches, and operator discretization techniques. Additionally, we outline methods based on applying biasing potentials or imposing constraints on tracer particles. Finally, we discuss approaches that estimate diffusivity from mean first passage time or committor probability profiles, a conceptual framework originally developed in the context of collective variable spaces describing rare events  in computational chemistry and biology. In summary, this paper offers a concise survey of diverse approaches for estimating diffusivity from MD trajectories, highlighting challenges and opportunities in this area.
\end{abstract}

\maketitle

\section{Introduction}

\noindent 
Confinement refers to situations in which a material is encapsulated by one or more physical interfaces, and can trigger substantial  changes to its physical properties. Such changes arise due to the breaking of both translational and rotational symmetry, which renders  all physical properties functions of position. Such position dependence becomes most pronounced within an interfacial region  that has a characteristic thickness of a few diameters of the material's building blocks. Whenever the confinement length scale is also comparable to the interfacial length scale, materials' properties exhibit the most pronounced deviations from the bulk.\cite{AlcoutlabiJPhys2005} Depending on the size of these building blocks and the range of their interactions with the interface, such deviations can be observed at different length scales. Confinement, therefore, is a potent means of fine-tuning the thermodynamic,\cite{JacksonJChemPhys1990, SwallenScience2007, GiovambattistaPhysRevLett2009, ZhangMacromolecules2011, ChabanACSNano2012, MooreJPhysChemC2012, RodriguezLarreaNatureNanotech2013} structural,\cite{HayesJPhysChemB2009, FumagalliScience2018,  FengNatMater2019, LeNanoscaleHoriz2020} and transport\cite{HuPRL1991, DemirelPRL1996, ZhangPhysRevB2000, RavivNature2001, KimJPCL2013, PouraliChemPhys2014, BerrodNanoscale2016, TuMacromolecules2022} properties of materials, while also influencing the kinetics and mechanisms of rare events.\cite{LucentPNAS2007, MittalPNAS2008, JiangChemSocRev2014, HajiAkbariPNAS2017, AltabetPNAS2017-p, HussainJACS2021, ShoemakerACSNano2024}
Among the spectrum of physical properties that can be substantially impacted by confinement, transport properties stand out prominently, as properties such as  diffusivity,\cite{BerrodNanoscale2016} viscosity,\cite{HuPRL1991, DemirelPRL1996, RavivNature2001} and thermal\cite{PouraliChemPhys2014} and ionic\cite{KimJPCL2013, TuMacromolecules2022} conductivity experience substantial alterations under confinement. 

Since their advent in the mid-20th century,\cite{RosenbluthJChemPhys1954, AlderJChemPhys1955, AlderJChemPhys1957, AlderJChemPhys1959, RahmanPhysRev1964} molecular simulations have emerged as indispensable tools for studying confined states of matter, \cite{GubbinsPhysChemChemPhys2011} and computational studies characterizing  the position dependence of physical properties within confined materials can be traced back to 1970's.\cite{CroxtonJPhysC1971, LeeJChemPhys1974, LiuJChemPhys1974, AbrahamJChemPhys1975, ToxvaerdJChemPhys1977, SubramanianMolPhys1979} It is fairly straightforward to determine spatial profiles of the thermodynamic and structural properties that can be unambiguously computed for a particular region within the simulation box from a single snapshot. This is simply achieved by partitioning the simulation box into suitable spatial bins and estimating the mean of the quantity of interest within each bin via a combination of time- and ensemble-averaging.  However, applying this approach to transport properties is not feasible, as those are typically computed using autocorrelations of appropriate mechanical observables, and autocorrelations of such nature cannot be unambiguously defined for open systems. Therefore generalizing the linear response formalisms, originally devised for translationally isotropic materials, to confined materials is a nontrivial undertaking.

This review is primarily dedicated to exploring methodologies for computing spatial profiles of transport properties from molecular simulations of confined materials, with a particular emphasis on diffusivity-- arguably the most widely computed transport property in molecular simulations.  Indeed, reports of self-diffusivity calculations can be traced back to the early days of molecular simulations, such as Rahman's pioneering work in 1964.\cite{RahmanPhysRev1964} We first describe the fundamental framework employed for computing transport properties in the bulk, and discuss the challenges of generalizing such frameworks to confined geometries. We then overview different classes of strategies aimed at effectively estimating position-dependent anisotropic diffusivity profiles.

This paper is organized as follows. In Section~\ref{section:transport-bulk}, we provide phenomenological definitions of transport coefficients, overview the computational methodologies employed for their estimation from MD, and discuss the specific challenges associated with estimating them in confined geometries. Section~\ref{section:adhoc-methods}  describes \emph{ad hoc} extensions of classical methods, with brief discussions of their plausible theoretical foundations. The subsequent sections navigate more rigorous approaches, including kernel-based methods (Section~\ref{section:kernel-based}), Bayesian techniques (Section~\ref{section:Bayesian}), and operation discretization approaches (Section~\ref{section:operator-discretization}). A comprehensive overview of methods based on applying biasing potentials to tracer particles is presented in Section~\ref{section:bias-based}. In Section~\ref{section:diff-CV-spaces}, we shift focus to using mean first passage time and committor probability profiles to estimate diffusivity, methodologies developed in the context of the mathematically associated problem of characterizing diffusive behavior along collective variables employed for describing rare events.  Finally, Section~\ref{section:conclusion} is dedicated to providing a broader perspective and outlining potential avenues for future exploration.

\section{Estimating transport properties from molecular dynamics trajectories}
\label{section:transport-bulk}

\subsection{Estimators in the bulk}
\label{section:transport-bulk:bulk}

\noindent 
Transport coefficients are phenomenological constants that establish a connection between macroscopic fluxes of physical properties, and external fields or thermodynamic driving forces. More precisely, suppose a system simultaneously exposed to a sequence of sufficiently small gradients, $\nabla{X}_1, \nabla{X}_2, \cdots, \nabla{X}_p$. Generally, there is always a natural (conjugate) flux associated with each such gradient. For instance, a temperature gradient will result in a heat flux if no other gradient is present. However, in the presence of multiple gradients, the flux associated with the $k$-th property, $\mathbf{J}_k$, will, in principle, be linked to all other gradients through the following relationship:
\begin{eqnarray} \label{eq:general-def-transport-coeff}
\textbf{J}_k &=& -\sum_{j=1}^p \mathbf{L}_{kj}\cdot\nabla X_j
\end{eqnarray}
Here, $\mathbf{L}_{kj}$'s, which are tensors of appropriate ranks, are referred to as \emph{transport coefficients}. Note that for every $j$ and $k$,  $\mathbf{L}_{jk} = \mathbf{L}_{kj}$ according to Onsager's reciprocity principle.\cite{onsager_reciprocal_1931} At a microscopic level, these coefficients can be viewed as rates at which a system responds to microscopic fluctuations.

While transport coefficients can, in principle, be defined for any pair of fluxes and driving forces,  certain transport coefficients hold particular significance for physicists and materials scientists, and are widely reported  in both experimental and computational studies of materials. One such transport coefficient is \emph{diffusivity}, which is historically defined for multicomponent systems. More precisely, $\mathbf{D}_i$, the diffusivity of component $i$, establishes a  connection between its diffusive flux, $\mathbf{J}_i$, and  its concentration gradient:
\begin{eqnarray}
    \textbf{J}_i &=& -\mathbf{D}_i\cdot\nabla c_i.
    \notag
\end{eqnarray}
In general, $\mathbf{D}_i$ is a second-rank symmetric positive-definite tensor,  but in the case of bulk simple liquids, it often exhibits isotropic behavior. Similarly, the \emph{self-diffusivity} of a pure material can be defined as the proportionality factor relating the flux and the gradient of the conditional probability of particle displacements. Serving as a proxy for the rate of structural relaxation in materials, self-diffusivity is well-defined but is difficult to measure experimentally. Nevertheless, it stands as one of the most widely computed transport coefficients in molecular simulation studies.

Analogous constitutive relationships govern the relationships between heat flux, $\textbf{q}$, and temperature gradient, $\nabla T$, as well as electric current, $\mathbf{i}$, and electrostatic potential gradient, $\nabla\varphi$:
\begin{eqnarray}
    \textbf{q} &=& -\mathbf{k}\cdot\nabla T\notag\\
    \textbf{i} &=& -\pmb\sigma\cdot\nabla\varphi\notag
\end{eqnarray}
Here, $\mathbf{k}$ and $\pmb\sigma$, both second-rank symmetric positive-definite tensors, denote the \emph{heat conductivity} and \emph{electrical conductivity}, respectively. Finally, a linear relationship can also be postulated between the stress tensor, $\pmb\tau$, and the symmetric part of the shear rate tensor, $\nabla\mathbf{v}+\nabla\mathbf{v}^T$:
\begin{eqnarray}
\tau_{ij} &=& \mu_{ijkl}\left[
\frac{\partial v_k}{\partial x_l} + \frac{\partial v_l}{\partial x_k}
\right]\notag
\end{eqnarray}
The proportionality constant is referred to as \emph{viscosity} and is naturally a fourth-rank tensor.\cite{avron_odd_1998}

In principle, all transport coefficients can be estimated from nonequilibrium MD simulations\cite{HooverAnnRevPhysChem1983, EvansCompPhysRep1984} in which an external driving force, such as temperature gradient, shear deformation, or electric field is applied to the system. The resulting macroscopic fluxes can then be readily computed, providing a means for estimating the relevant proportionality constants.\cite{VogelsangMolPhys1988, SvishchevPhysChemLiq1994, WangPhysRevB2012, JadhaoPNAS2017} Conversely, a constant-flux boundary condition can be imposed to the system, with the transport coefficient estimated from the computed conjugate gradient.\cite{MullerPlatheJChemPhys1997, TenneyJChemPhys2010}  The major limitation of all such nonequilibrium approaches is the substantial magnitudes of the necessary driving forces (or fluxes), which often surpass their experimental counterparts by several orders of magnitude. This raises profound questions regarding the validity of the linear flux-driving force relationships under such extreme conditions. It is therefore unclear whether the proportionality constants estimated from such simulations will be applicable to experimentally relevant conditions.

An alternative approach, conceptually akin to non-equilibrium techniques in the limit of small gradients, involves utilizing linear response theory\cite{marconi_fluctuationdissipation_2008}  to express transport coefficients in terms of autocorrelations of microscopic fluxes. Linear response theory quantifies a system's response when its Hamiltonian is perturbed from equilibrium by a small external field,~i.e.,~$\mathcal{H}(\mathbf{P},\mathbf{Q},t)=\mathcal{H}_{\text{eq}}(\mathbf{P},\mathbf{Q}) - \mathcal{F}(t)\mathcal{A}(\textbf{P},\textbf{Q})$. Here, $\mathbf{Q}\equiv(\mathbf{q}_1,\mathbf{q}_2,\cdots,\mathbf{q}_N)$ and  $\mathbf{P}\equiv(\mathbf{p}_1,\mathbf{p}_2,\cdots,\mathbf{p}_N)$ correspond to the positions and momenta of the constituent particles, respectively. The temporal evolution of the expected value of another mechanical observable, $\mathcal{B}(\mathbf{P},\mathbf{Q})$, will be given by:
\begin{eqnarray}\label{eq:lin-resp}
    \langle\Delta \mathcal{B}(t)\rangle &=& \int_{-\infty}^t \mathcal{F}(t')\phi_{\mathcal{A}\mathcal{B}}(t-t')\,dt' .
\end{eqnarray}
Here, $\Delta\mathcal{B}(t) = \mathcal{B}(t) - \langle \mathcal{B}\rangle_{\mathcal{H}_{\text{eq}}}$, and $\phi_{\mathcal{A}\mathcal{B}}(\tau)$ can be interpreted as a response function and is given by:
\begin{eqnarray}\label{eq:resp-fcn}
    \phi_{\mathcal{A}\mathcal{B}}(\tau) &=&\beta \langle
    \mathcal{B}(\tau)\dot{\mathcal{A}}(0)
    \rangle_{\mathcal{H}_{\text{eq}}} 
\end{eqnarray}
For any specific transport coefficient, an external perturbation can be applied to establish a suitably small macroscopic gradient within the system. Equations~\eqref{eq:lin-resp} and \eqref{eq:resp-fcn} can then be employed to relate the corresponding flux of interest to the gradient. This approach leads to a collection of equations known as \emph{Green-Kubo} relationships\cite{green_markoff_1954, kubo_statistical-mechanical_1957}, characterizing transport coefficients in terms of auto-correlations of mechanical observables.

As an illustration, consider self-diffusivity, where a plausible perturbation to the Hamiltonian cab be formulated as:
\begin{eqnarray}\label{eq:pert-Hamil-GK}
\mathcal{A}(\mathbf{P},\mathbf{Q}) &=& -\alpha\mathbf{w}\cdot\sum_{i=1}^N\mathbf{q}_i.
\end{eqnarray}
Here, $\alpha$ represents a fixed force pulling the particles along a unit vector $\mathbf{w}$, resulting in a net flux of particles along $\mathbf{w}$. The response function for momentum flux along another unit vector $\mathbf{u}$, $\mathcal{B}(\mathbf{P},\mathbf{Q}) = \mathbf{u}\cdot\sum_{i=1}^N\mathbf{p}_i$, can be enumerated using Eq.~\eqref{eq:resp-fcn}:
\begin{eqnarray}
\phi_{\mathcal{A}\mathcal{B}} (t) &=& -\frac{\alpha\beta}{m}\left\langle 
\sum_{i,j=1}^N \mathbf{u}^T\mathbf{p}_i(t)\mathbf{p}_j^T(t)\mathbf{w}
\right\rangle\notag\\
&\overset{\text{(a)}}{=}& -m\alpha\beta\mathbf{u}^T\left\langle 
\sum_{i=1}^N \mathbf{v}_i(t)\mathbf{v}_i^T(0)
\right\rangle\mathbf{w}\notag\\
&\overset{\text{(b)}}{=}& -mN\alpha\beta\mathbf{u}^T\left\langle
\mathbf{v}(t)\mathbf{v}^T(0)
\right\rangle\mathbf{w}\notag
\end{eqnarray}
Here, (a) follows from the fact that momentum degrees of freedom are uncorrelated, while (b) results from the indistinguishability of particles. This expression can be utilized to evaluate the mean velocity along the unit vector $\mathbf{u}$:
\begin{eqnarray}
\lim_{t\rightarrow\infty}\mathbf{u}^T\left\langle\mathbf{v}(t)\right\rangle =-\alpha\beta \int_0^{\infty}\mathbf{u}^T\left\langle
\mathbf{v}(t)\mathbf{v}^T(0)
\right\rangle\mathbf{w}\,dt.\notag
\end{eqnarray}
This observation allows us to compute $\mathbf{u}^T\mathbf{D}\mathbf{w}$, given by:\cite{marconi_fluctuationdissipation_2008}
\begin{eqnarray}
\mathbf{u}^T\mathbf{D}\mathbf{w} &=& -\frac{\mathbf{u}^T\left\langle\mathbf{v}(\infty)\right\rangle}{\alpha\beta} = \int_0^{\infty} \mathbf{u}^T\left\langle \mathbf{v}(t)\mathbf{v}^T(0)\right\rangle\mathbf{w}\,dt.\notag
\end{eqnarray}
Choosing $\mathbf{u}$ and $\mathbf{w}$ from among the basis vectors in Cartesian coordinates yields the well-known relationship:
\begin{eqnarray}
    \mathbf{D} &=& \int_0^{\infty} \bigg\langle\mathbf{v}(t)\mathbf{v}^T(0)\bigg\rangle\,dt
\end{eqnarray}
where the integrand is typically referred to as the \emph{velocity autocorrelation function (VACF)}. Similar expressions can be derived for other transport coefficients. For instance, the shear viscosity tensor can be estimated from,\cite{bradlyn_kubo_2012}
\begin{eqnarray}
    \mu_{\alpha\beta\gamma\delta} &=& \frac{V}{k_BT}\int_0^{\infty}\left\langle
    \delta \tau_{\alpha\beta}(t)\delta \tau_{\gamma\delta}(0)
    \right\rangle\,dt
\end{eqnarray}
where $\pmb\tau$ is the second-rank stress tensor computed from the virial relationship,\cite{shi_perspective_2023} and $\delta{\pmb\tau} = {\pmb\tau} - \langle{\pmb\tau}\rangle$. Likewise, thermal conductivity can be related to autocorrelations of heat flux:\cite{schelling_comparison_2002}
\begin{eqnarray}
    \mathbf{k} &=& \frac{1}{k_BT^2V}\int_0^{\infty}\left\langle
    \mathbf{J}_h(t)\mathbf{J}_h^{T}(0)
    \right\rangle\,dt
\end{eqnarray}
with the instantaneous heat flux, $\mathbf{J}_h$, defined as,
\begin{eqnarray}\label{eq:heat-flux}
    \mathbf{J}_h(t) &=& \frac{d}{dt}\,\sum_{i=1}^N \mathbf{r}_i(t)\epsilon_i(t).
\end{eqnarray}
Here, $\epsilon_i(t)$ is the sum of the kinetic and potential energy of particle $i$.  Finally, electric conductivity can similarly be obtained from  autocorrelations of the electric current:\cite{ma_directly_2018}
\begin{eqnarray}
    \pmb\sigma &=& \frac{1}{k_BTV}\int_0^\infty \mathbf{J}_e(t)\mathbf{J}_e^T(0)\,dt
\end{eqnarray}
with electric current, $\mathbf{J}_e$, given by $\mathbf{J}_e = \sum_{i=1}^Nq_i\textbf{v}_i$.

A conceptually related class of relationships, developed by Helfand \cite{helfand_transport_1960}, estimate transport coefficients through the asymptotic slopes of time- and ensemble-averaged generalized displacements. These displacements are time integrals of microscopic fluxes. The most well-known example is the \emph{Einstein relationship} \cite{einstein_uber_1905}, which links diffusivity to the asymptotic slope of \emph{mean-squared displacement (MSD)}:
\begin{eqnarray}
    \mathbf{D}_i &=& \lim_{t\rightarrow\infty} \frac{\left\langle \left[\mathbf{r}_i(t)-\mathbf{r}_i(0)\right]\left[\mathbf{r}_i(t)-\mathbf{r}_i(0)\right]^\dagger\right\rangle}{2t}
\end{eqnarray}
Helfand\cite{helfand_transport_1960} expanded upon this approach by linearizing the corresponding conservation laws and solving them over an infinite domain. When it comes to the transport of linear momentum, viscosity can be expressed as:
\begin{eqnarray}
\mu_{\alpha\beta\gamma\delta} &=&\lim_{t\rightarrow\infty}\frac1{2tk_BTV}\Bigg\langle\Big[\sum_{i=1}^N\left[p_{i,\alpha}(t)r_{i,\beta}(t)-p_{i,\alpha}(0)r_{i,\beta}(0)\right]\notag\\
&& \left[p_{i,\gamma}(t)r_{i,\delta}(t)-p_{i,\gamma}(0)r_{i,\delta}(0)\right]\Big]^2\Bigg\rangle
\end{eqnarray}
Similar expressions can be obtained for other transport coefficients. For instance, heat conductivity is given by:\cite{KinaciJChemPhys2012}
\begin{eqnarray}
\mathbf{k} &=& \lim_{t\rightarrow\infty} \frac{\langle\left[\mathbf{h}(t)-\mathbf{h}(0)\right]\left[\mathbf{h}(t)-\mathbf{h}(0)\right]^T\rangle}{2tVk_BT^2}
\end{eqnarray}
Here, $\mathbf{h}(t) = \int_0^t\mathbf{J}_h(\tau)\,d\tau$ is the integrated heat flux where $\mathbf{J}_h(\tau)$ is defined by \eqref{eq:heat-flux}. 

The Green-Kubo formalism is commonly regarded as the primary method for estimating transport coefficients from MD trajectories. In contrast, the Helfand approach is more frequently applied in the estimation of diffusivity only and is less commonly employed for other transport coefficients. Its limited usage can be partly attributed to the challenges associated with its proper implementation, particularly when dealing with periodic boundary conditions, as highlighted by Viscardy and Gaspard.\cite{viscardy_viscosity_2003}

\subsection{Challenges in confined geometries}
\label{section:transport-bulk:confined}

\noindent
Under confinement, all physical properties become functions of position. Moreover, transport coefficients also become anisotropic, making it necessary to account for their tensorial nature. The main challenge in computing them, however, lies in the absence of well-defined autocorrelation-based estimators for open systems.  Specifically, spatial profiles of mechanical observables that are unambiguously defined for arbitrary regions within the simulation box can be accurately estimated using spatial binning. For instance, the spatial profile of a mechanical observable $\mathcal{R}$ that is well defined for every particle can be expressed as:
\begin{eqnarray}\label{eq:profile-static}
\mathcal{R}(\textbf{r}) = \frac{\left\langle\sum_{i=1}^N\mathcal{R}_i\delta(\mathbf{r}_i-\mathbf{r}) \right\rangle}{\left\langle\sum_{i=1}^N\delta(\mathbf{r}_i-\mathbf{r}) \right\rangle}.
\end{eqnarray}
In practice, $\mathcal{R}(\textbf{r})$ is estimated by partitioning the simulation box into non-overlapping bins and determining the mean of $\mathcal{R}(\mathbf{r})$ within each bin. For ergodic systems, $\mathcal{R}_k$, the average of $\mathcal{R}$ over the $k$-th bin, can be estimated as:
\begin{eqnarray}\label{eq:profile-static-binned}
\mathcal{R}_k &=& \frac{\displaystyle\int_0^{t_{\text{sim}}}\mathcal{R}_i(t)\chi_k[\mathbf{r}_i(t)]\,dt}{\displaystyle\int_0^{t_{\text{sim}}}\chi_k[\mathbf{r}_i(t)]\,dt},
\end{eqnarray}
where $\chi_k(\cdot)$ is the characteristic function of the $k$-th bin. However, there exists no natural extension of \eqref{eq:profile-static-binned} for quantities such as MSD or VACF, which are autocorrelations of mechanical observables. This limitation arises due to particle exchange among bins and the ambiguity in quantifying the contribution of exchanged particles to the autocorrelation associated with a specific bin. As a consequence, methodologies discussed in Section~\ref{section:transport-bulk:bulk} are only robustly applicable in the bulk. Moreover, the statistical behavior of microscopic fluxes in confined geometries, such as particle mobilities, diverges significantly from established conservation laws valid in bulk systems. In the case of self-diffusivity, for instance, the self part of the van Hove correlation function,\cite{vanHovePhysRev1954} $G_s(\textbf{r},t|\textbf{r}_0,0)$, satisfies the \emph{Fokker-Planck equation}\cite{FokkerAnnPhys1914, PlanckSitzberPreussAkad1917} in the bulk:
\begin{eqnarray}\label{eq:FokkerPlanck}
\frac{\partial G_s}{\partial t} = \nabla\cdot\left[\mathbf{D}\cdot\nabla G_s\right].
\end{eqnarray}
However, under confinement, both diffusivity and equilibrium density become position-dependent, and $G_s(\textbf{r},t| \textbf{r}_0,0)$ will satisfy the \emph{Smoluchowski equation}:\cite{SmoluchowskiPhysZ1916}
\begin{eqnarray}
\frac{\partial G_s}{\partial t} &=& \nabla\cdot\left[
\mathbf{D}(\mathbf{r})\cdot\left(
\nabla G_s+ \beta G_s\nabla\mathcal{F}
\right)
\right] = \mathcal{L}^{\dagger}_{\mathbf{r}}G_s.\label{eq:Smoluchowski}
\end{eqnarray}
Here, $\mathcal{F}(\textbf{r}) = -\beta^{-1}\ln\rho_0(\textbf{r})$  represents the free energy profile wherein $\rho_0(\textbf{r})$ denotes the number density profile.  Unlike Eq.~\eqref{eq:FokkerPlanck}, which possesses straightforward analytical solutions in simple geometries, Eq.~\eqref{eq:Smoluchowski} lacks a closed-form solution. Consequently, it is not trivial to extract position-dependent diffusivity tensors from   MD trajectories. The remainder of this review is dedicated to a comprehensive discussion of various numerical approaches developed for tackling this nontrivial task.

\section{Ad hoc Extensions of classical methods}
\label{section:adhoc-methods}

\noindent
As discussed above, well-established methodologies based on MSD or VACF cannot be readily applied to confined geometries since the Smoluchowski equation lacks a simple analytical solution. Nevertheless, many researchers have still presumed the local validity of such formalisms, and have accordingly devised \emph{ad hoc} extensions of MSD and VACF for the purpose of estimating position-dependent diffusivities. This section is dedicated to a comprehensive discussion of such efforts, including the common practices and conventions underpinning such \emph{ad hoc} extensions. Moreover, we discuss the merits and limitations of the theoretical arguments that could be made for making such \emph{ad hoc} frameworks more rigorous.

\subsection{\emph{Ad hoc} mean squared displacements}
\label{section:ad-hoc-MSD}

\noindent
As discussed in Section~\ref{section:transport-bulk:confined}, it is not feasible to analytically establish a linear relationship between the asymptotic slope of a localized notion of MSD and local diffusivity. Nonetheless, this has been the most widely adopted approach in the molecular simulations community for estimating position-dependent diffusivity, wherein \emph{ad hoc} localized notions of MSD are constructed, and local diffusivity is extracted through a linear regression between local MSDs and the observation window (i.e.,~the time lag). As an illustration, in the case of one-dimensional confinement along the $z$ axis, a lateral MSD for bin $i$ can be formulated as:
\begin{eqnarray}
    \mathcal{M}(z_i,t) &=& \Big\langle
    \left[(x_{t+\tau}-x_\tau)^2+(y_{t+\tau}-y_\tau)^2\right]\notag\\
    && \xi_i\left[z(t')_{\tau\le t'\le t+\tau}\right]
    \Big\rangle_\tau\label{eq:adhoc-MSD}
\end{eqnarray}
Here, $\xi_i[z(t')]$ specifies the weight assigned to a particle's contribution to the \emph{ad hoc} MSD of bin $i$, based its the trajectory $z(t')$ over the time interval $\tau\le t'\le t+\tau$. The simplest choice of $\xi_i[\cdot]$ is given by,
\begin{eqnarray}\label{eq:MSD-beginning}
    \xi_i[z(t')] = \chi_i(z_\tau)
\end{eqnarray}
wherein $\chi_i(\cdot)$ is the characteristic function of the $i$-th bin. In other words,  Eq.~\eqref{eq:MSD-beginning} only allows particles that are within a particular bin at the beginning of an observation window to contribute to the local MSD of that bin.\cite{teboul_properties_2002, desai_molecular_2005} However, this straightforward approach can lead to significant errors over extended timeframes, as particles originating from bin $i$ may travel to distant bins. To address this issue, alternative definitions have been proposed. For instance, some authors use the average $z$ along $z(t')$ to allocate the particle to a specific bin.\cite{marrink_simulation_1994-1} More restrictive definitions, such as only including particles present within the bin at both the beginning and the end of the observation window,\cite{lancon_brownian_2002, haji-akbari_effect_2014, haji-akbari_thermodynamic_2015} or requiring the trajectory to remain within the bin throughout the observation window,\cite{liu_calculation_2004, shi_properties_2011}  have also been employed.

Despite inherent limitations of such \emph{ad hoc} approaches, such localized notions of MSD can be modified in creative ways to yield more realistic proxies for position-dependent dynamics. A notable example is the approach proposed by Liu and Berne,\cite{liu_calculation_2004} who approximates lateral diffusivity as,
\begin{equation}\label{eq: Lateral diffusivity Berne}
    D_{xx} = D_{yy} \approx\frac{\mathcal{M}(z_i,t)}{4t P_i(t)}
\end{equation}
Here, $\mathcal{M}(z_i,t)$ is a localized MSD as in Eq.~\eqref{eq:MSD-beginning}, with the convention that the trajectory should remain within the same bin throughout the entire time interval. Additionally, $P_i(t)$ denotes the survival probability,~i.e.,~the probability that a particle starting within the $i$-th bin will still remain in that bin after time $t$.

Another approach, akin in spirit, is proposed by Nagai and Okazaki, \cite{nagai_position-dependent_2020} wherein a biasing potential is introduced, which is flat within a designated spatial bin but becomes strongly repulsive outside the bin. The biased Hamiltonian is then employed to launch MD trajectories from which the localized MSD, $\mathcal{M}_{\text{FB}}(t)$,  is computed, with "FB" denoting the "force-biased" nature of these simulations. The authors argue that the distortion introduced within the MSD due to force biasing is independent of the magnitude of the local diffusivity, and is instead determined by a geometric factor:
$$
\mathcal{G}(\tau) = \frac{\overline{\mathcal{M}}_{\text{FB}}(\tau)}{\overline{\mathcal{M}}(\tau)}
$$
Here, $\mathcal{M}(\cdot)$ is the unbiased MSD in the bulk, $\tau=tD_0/l_0^2$ represents dimensionless time, and $\overline{\mathcal{M}} = \mathcal{M}/l_0^2$ corresponds to dimensionless MSD. After determining the geometric factor, the diffusivity within each bin can be computed by iteratively solving the following equation:
\begin{equation}\label{eq: Nagai}
\frac{\mathcal{M}_{FB,\textbf{x}_0}(t)}{\mathcal{G}\left[tD(\textbf{r}_0)l_0^{-2}\right]} = 2dD(\textbf{r}_0)t+C
\end{equation}
wherein $C$ is a constant corresponding to the nonzero intercept of MSD. 

\begin{figure}
	\includegraphics[width=.4\textwidth]{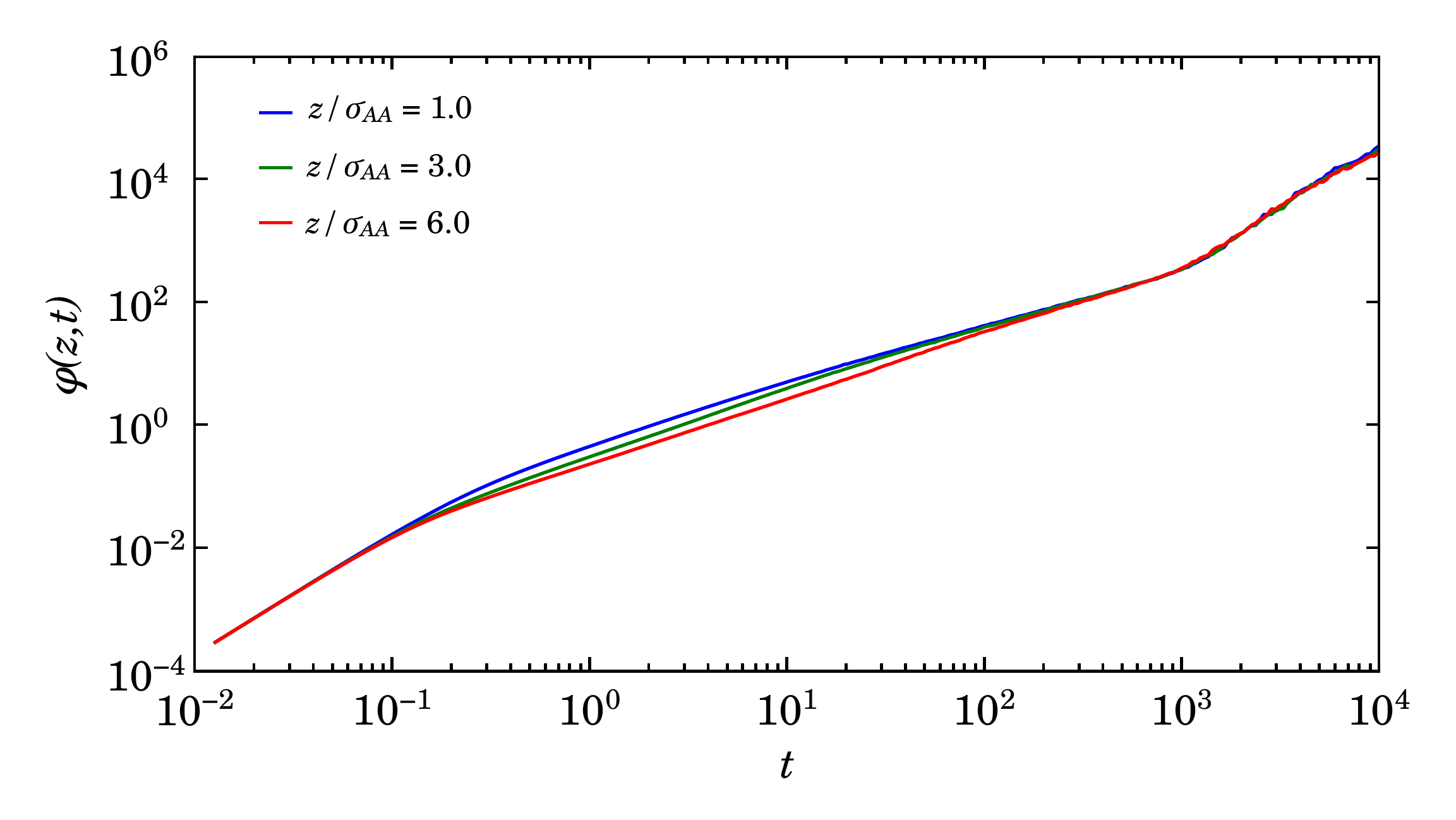}
	\caption{(Reproduced from Ref.~\citenum{haji-akbari_effect_2014}) Breakdown of linear scaling of the \emph{ad hoc} MSD with time for MSDs computed at different distances from the substrate within a Kob-Andersen\cite{KobPRL1994}  liquid.\label{fig:adhoc-MSD-scaling-breakdown}}
\end{figure}

While diffusivity profiles obtained from these \emph{ad hoc} approaches serve as semi-quantitative proxies for spatial dynamic heterogeneities in confined materials, they do not strictly align with the Smoluchowski equation in the sense that the mobility statistics implied by these approaches, in accordance with \eqref{eq:Smoluchowski}, may not necessarily match actual $G_s(\textbf{r},t)$ obtained from MD simulations. Additionally, these methods provide only lateral components of the diffusivity tensor. There have been attempts to overcome this latter limitation,~e.g.,~by employing dual simulation approaches for estimating the normal component of the diffusivity tensor, as demonstrated by Piu~\emph{et al.} \cite{liu_calculation_2004} Finally, the \emph{ad hoc} MSDs do not capture diffusive behavior over extremely long timescales due to inter-bin mixing\cite{haji-akbari_effect_2014} (Fig.~\ref{fig:adhoc-MSD-scaling-breakdown}). Consequently, fitting procedures must be constrained to intermediate timescales to yield reliable estimates.

While \emph{ad hoc} definitions of MSD may be heuristic in nature, they possess inherent fundamental merit. Specifically, the Smoluchowski equation can be readily reformulated as:
\begin{eqnarray}\label{eq:FK-general}
&& \frac{\partial\rho}{\partial t} = -\nabla\cdot(\pmb\mu\rho) + \nabla\cdot\nabla\cdot\left[\mathbf{D}\rho\right],\\
&& \rho(\textbf{r},0) = \delta(\textbf{r}-\textbf{s}).\notag
\end{eqnarray}
Here, $\pmb\mu$ and $\mathbf{D}$ represent the drift and diffusivity profiles, respectively, and are related to the formal solution of \eqref{eq:FK-general} via \emph{Kramers-Moyal}\cite{KramersPhysica1940, MoyalJRoyStatSocBMet1949} relations:
\begin{subequations}\label{eq:KM_probability}
\begin{eqnarray}
\pmb\mu(\mathbf{s}) &=& \lim_{\tau\rightarrow0^+}\frac1\tau\int(\textbf{r}-\textbf{s})\rho_\tau(\textbf{r}|\textbf{s})\,d\textbf{r}\\
\mathbf{D}(\mathbf{s}) &=& \lim_{\tau\rightarrow0^+}\frac1{2\tau}\int(\textbf{r}-\textbf{s})(\textbf{r}-\textbf{s})^T\rho_\tau(\textbf{r}|\textbf{s})\,d\textbf{r}
\end{eqnarray}
\end{subequations}
Given the interpretation of $\rho_\tau(\textbf{r}|\textbf{s})$ as a probability density, \eqref{eq:KM_probability} can be readily recast as:
\begin{subequations}\label{eq:Kramer-Moyal}
\begin{eqnarray}
\pmb\mu(\mathbf{s}) &=&   \lim_{\tau\rightarrow0^+}\left.
\frac{\left\langle\mathbf{X}_{t+\tau}-\mathbf{X}_{t} \right\rangle}{\tau}
\right|_{\mathbf{X}_t=\mathbf{s}}\\
\mathbf{D}(\mathbf{s}) &=&   \lim_{\tau\rightarrow0^+}\left.
\frac{\left\langle\left(\mathbf{X}_{t+\tau}-\mathbf{X}_{t}\right)\left(\mathbf{X}_{t+\tau}-\mathbf{X}_{t}\right)^T \right\rangle}{2\tau}
\right|_{\mathbf{X}_t=\mathbf{s}}
\end{eqnarray}
\end{subequations}
Therefore, local diffusivity can be approximated via the following finite difference expression: 
\begin{equation}\label{eq: diffusivity_from KM}
\mathbf{D}(\textbf{s}) \approx \widehat{\mathbf{D}}^\tau(\mathbf{s}) = \left. \frac{\left\langle \left(\mathbf{X}_{t+\tau}-\mathbf{X}_t\right)\left(\mathbf{X}_{t+\tau}-\mathbf{X}_t\right)^T\right\rangle}{2\tau}\right|_{\mathbf{X}_t=\mathbf{s}}
\end{equation}
Likewise, it is possible to propose a finite-$\tau$ estimator for the drift $\pmb\mu$:
\begin{equation}\label{eq: mu_from KM}
\pmb\mu(\textbf{s}) \approx \widehat{\pmb\mu}_\tau(\mathbf{s}) = \left. \frac{\left\langle \mathbf{X}_{t+\tau}-\mathbf{X}_t\right\rangle}{\tau}\right|_{\mathbf{X}_t=\mathbf{s}}
\end{equation}
Here, the timescale $\tau$ is system dependent, and should be chosen in such a manner that single-particle trajectories behave diffusively at and beyond $\tau$.  Notably, Eq.~\eqref{eq: diffusivity_from KM} bears resemblance to the earlier \emph{ad hoc} MSD definitions, therefore belonging to the broad category of Helfand approaches.  A crucial distinction lies in the Helfand approach's consideration of the limiting slope of MSD at $\tau\rightarrow \infty$ to ensure capturing long-term diffusive behavior. In confined systems, it is instead imperative to select a $\tau$ that is as small as possible in order to avoid inter-bin mixing. A large $\tau$ will introduce considerable discretization errors, compromising the spatial resolution of diffusivity profiles.  Thus, selecting an optimal $\tau$ is pivotal to preserve accurate data without loss due to discretization errors. For systems that behave diffusively across all timescales, $\tau$ can be made as small as a single time step. MD trajectories, however, are only diffusive beyond the caging regime, and as such there is a strict lower bound on the $\tau$ that can be used in \eqref{eq: diffusivity_from KM}. (An operational procedure for the selection of $\tau$ is provided in Section~\ref{section:kernel-based}.)

To comprehend the impact of time discretization introduced by a nonzero $\tau$, consider the formal solution of Eq.~\eqref{eq:Smoluchowski}, expressed as $\rho_\tau(\mathbf{y}|\mathbf{x}) = e^{t\mathcal{L}^\dagger_{\mathbf{y}}}\delta(\mathbf{y}-\mathbf{x})$. This solution allows the reformation of $\widehat{\mathbf{D}}_\tau(\mathbf{x})$ as follows:
\begin{eqnarray}
\widehat{\mathbf{D}}^\tau(\mathbf{x}) &=& \frac1\tau\int(\mathbf{y}-\mathbf{x})(\mathbf{y}-\mathbf{x})^T e^{t\mathcal{L}^\dagger_{\mathbf{y}}}\delta(\mathbf{y}-\mathbf{x})\,d\mathbf{y} \notag\\
&=& \frac1\tau\int e^{t\mathcal{L}_{\mathbf{y}}}\left[
(\mathbf{y}-\mathbf{x})(\mathbf{y}-\mathbf{x})^T
\right]\delta(\mathbf{y}-\mathbf{x})\,d\mathbf{y}\notag\\
&=& \int\left[
\sum_{l=1}^{+\infty}\frac{\tau^{l-1}}{l!}\mathcal{L}_{\mathbf{y}}^l\left[
(\mathbf{y}-\mathbf{x})(\mathbf{y}-\mathbf{x})^T
\right]\delta(\mathbf{y}-\mathbf{x})
\right]\,d\mathbf{y}\notag
\end{eqnarray} 
where $\mathcal{L}_\mathbf{y}$ is the operator adjoint  to the $\mathcal{L}^\dagger_{\mathbf{y}}$ of Eq.~\eqref{eq:Smoluchowski} under the standard $L^2$ inner product.  Notably, the $l=1$ term upon integration precisely yields $\mathbf{D}(\mathbf{x})$, resulting in:
\begin{eqnarray}\label{eq: tau effect}
&& \widehat{\mathbf{D}}^\tau(\mathbf{x}) = \mathbf{D}(\mathbf{x}) \notag\\
&& + \int\left[
\sum_{l=2}^{+\infty}\frac{\tau^{l-1}}{l!}\mathcal{L}_{\mathbf{y}}^l\left[
(\mathbf{y}-\mathbf{x})(\mathbf{y}-\mathbf{x})^T
\right]\delta(\mathbf{y}-\mathbf{x})
\right]\,d\mathbf{y}\notag
\end{eqnarray}
 The second term on the right-hand side represents an error term, which vanishes as $\tau\rightarrow 0^+$. For nonzero $\tau$, this expression serves as a foundational reference to devise correction schemes aimed at mitigating implicit mixing effects inherent in the error term above. For instance, it has been demonstrated that in one dimension, this series can be approximated to quadratic order as,\cite{friedrich_comment_2002, gottschall_definition_2008}
\begin{subequations}\label{eq: full quad correction}
\begin{eqnarray}
&& \left.\left\langle X_{t+\tau}-X_t\right\rangle\right|_{X_t=x} = \tau\mu(x)\notag\\
&&  + \frac{\tau^2}{2}\Bigg[
\mu(x)\mu'(x) + D(x)\mu''(x)
\Bigg]+ O(\tau^3)\\
&&  \left.\left\langle \left(X_{t+\tau}-X_t\right)\left(X_{t+\tau}-X_t\right)^T\right\rangle\right|_{X_t=x}  = 2\tau D(x) \notag\\
&& + \tau^2\Bigg\{
\mu^2(x)+ \left[\mu(x)D(x)\right]' + D(x)\left[D'(x)+\mu(x)\right]' \Bigg\}  \notag\\
&& + O(\tau^3)
\end{eqnarray}
\end{subequations} 
Assuming that all derivatives of $\mu(x)$ and $D(x)$ are negligible leads to the following simplified expressions:\cite{gottschall_definition_2008}
\begin{subequations}\label{eq: partial quad correction}
\begin{eqnarray}\label{eq: simplified-taucorr}
&& \left.\left\langle X_{t+\tau}-X_t\right\rangle\right|_{X_t=x} = \tau\mu(x)+ O(\tau^3)\\
&&  \left.\left\langle \left(X_{t+\tau}-X_t\right)\left(X_{t+\tau}-X_t\right)^T\right\rangle\right|_{X_t=x}  = 2\tau D(x) \notag\\
&& + \tau^2 \mu^2(x) + O(\tau^3)\label{eq: partial quad correction-b}
\end{eqnarray}
\end{subequations} 
Note that \eqref{eq: partial quad correction-b} can be readily rearranged as:
\begin{eqnarray}\label{eq: quadratic-expression}
 && \left.\left\langle \left(\Delta X_{t,\tau}-\tau\mu(x)\right)\left(\Delta X_{t,\tau}-\tau\mu(x)\right)^T\right\rangle\right|_{X_t=x}\notag\\
 && = 2\tau D(x) + O(\tau^3)
\end{eqnarray}
wherein $\Delta X_{t,\tau}=X_{t+\tau}-X_t$. It must be noted that the left hand side of \eqref{eq: quadratic-expression} constitutes a covariance of displacements. The ability to estimate diffusivity from computing a local covariance matrix suggests that probability density function of $\mathbf{X}_\tau$, a particle's position at time $\tau$, can be approximated as,
\begin{eqnarray}\label{eq: Normal approximation}
\mathbf{X}_\tau &\sim & \mathcal{N}\left(\mathbf{r}_0+\tau\pmb\mu(\mathbf{r}_0), 2\tau\mathbf{D}(\textbf{r}_0)\right).
\end{eqnarray}
wherein $\textbf{r}_0$ is the particle's position at $t=0$, and  $\mathcal{N}(\pmb\mu,\pmb\Sigma)$ is a multivariate Gaussian distribution with mean $\pmb\nu$ and covariance matrix $\pmb\Sigma$.  Indeed, Eq.~\eqref{eq: diffusivity_from KM} has been been employed for estimating lateral diffusivity of water near interfaces,\cite{sedlmeier_water_2011} as well as diffusivity along collective variables (CVs) employed in protein folding simulations.\cite{oliveira_coordinate-dependent_2022, hinczewski_how_2010} Interestingly, it was shown by Hinczewski~\emph{et al.}\cite{hinczewski_how_2010} that even for collective variable spaces, the estimated diffusivity is acutely sensitive to $\tau$. A common strategy\cite{yang_folding_2007, yang_effective_2006, freitas_drift-diffusion_2019} in protein folding simulations involves using Eq.~\eqref{eq: Normal approximation} to fit Gaussians into empirical histograms obtained around a certain point but at different times, and use the following expression to estimate diffusivity,
 \begin{equation}\label{eq: DrDiff}
     D(\lambda_0)\approx \frac{\sigma^2(\Lambda_{\tau_2})-\sigma^2(\Lambda_{\tau_1})}{2(\tau_2-\tau_1)}\equiv \widehat{D}_{\tau_1,\tau_2}(\lambda_0)
 \end{equation}
where $\Lambda_t$ is a random variable describing the state of the system (within the CV space) at time $t$. 
In the limit of $\Delta\tau=\tau_2-\tau_1\rightarrow 0^+$, it can be demonstrated that,
\begin{equation}\label{eq: 2tau lim}
    \widehat{D}_{\tau_1,\tau_2}(\lambda_0) \operatornamewithlimits{\rightarrow}_{\Delta \tau \rightarrow 0} \int  D(\lambda) \rho_{\tau_1} (\lambda|\lambda_0)\,d\lambda
\end{equation}
where $\rho_t$ denotes the solution of the Smoluchowski equation in the collective variable space.  Methods designed with this specific application in mind will be detailed in Section~\ref{section:diff-CV-spaces}.

\subsection{\emph{Ad hoc} velocity autocorrelation functions}

\noindent
In confined geometries, suitable \emph{ad hoc} definitions of VACF cam be devised in a fashion similar to MSD.\cite{MamonovBiophysChem2006, shi_properties_2011} The local diffusivity within bin $i$ can then be evaluated as,
\begin{eqnarray}
\mathbf{D}_i &=& \int_0^{+\infty} \mathbf{C}_{v,i}(t)\,dt\notag
\end{eqnarray}
with $\mathbf{C}_{v,i}(t)$, the localized VACF defined as,
\begin{eqnarray}
\mathbf{C}_{v,i}(t) &=& \left\langle
\mathbf{v}(t+\tau)\mathbf{v}^T(\tau)\xi_i\left[\textbf{r}(t')_{\tau\le t'\le \tau+t}\right]
\right\rangle_\tau.\notag
\end{eqnarray}
Here, $\xi_i[\textbf{r}(t')]$ serves a similar role as in Eq.~\eqref{eq:adhoc-MSD}. 

It is crucial to highlight two notable differences between the \emph{ad hoc} extensions of MSD and VACF. Firstly, we expect \emph{ad hoc} estimates of local diffusivity through VACF  to exhibit reduced susceptibility to inter-bin mixing. This stems from VACF's inherent decay to zero within  timescales relevant for such mixing, thereby enhancing their practical utility. Secondly, as expounded upon later in this section, specific \emph{ad hoc} extensions of VACF can be derived utilizing linear response theory. This entails applying a suitable perturbation term to the entire system while monitoring the response of a locally defined mechanical observable to such a perturbation. The arising mobility profiles can then be linked to local diffusivity in a manner similar to the bulk. As will be discussed later, these approaches are still inherently \emph{ad hoc}  in the sense that observed mobility statistics are not guaranteed to conform to predictions based on the Smoluchowski equation\cite{SmoluchowskiPhysZ1916} or the Kramers-Klein equation.\cite{KramersPhysica1940}

An an illustration, consider Hunter~\emph{et al.}\cite{hunter_new_2022} who introduce a perturbation to the Hamiltonian given by:
$$\mathcal{A}(\mathbf{P},\mathbf{Q}) = -F_c\textbf{w}\cdot\sum_{i=1}^Nc_i\textbf{q}_i.$$ 
Here, $F_c$ represents the magnitude of the force, $\mathbf{w}$ is a unit vector, and $c_i$ signifies the 'color` associated with particle $i$, allowing for the adjustment of both the direction and the strength of the biasing force applied to different particles. The authors adopt color currents as an elegant means of handling correlations between momentum degrees of freedom. Such correlations, while typically absent in a  strict statistical mechanical sense, are frequently present in MD trajectories where the system's net linear momentum is set to zero. By using $c_i=(-1)^i$, Hunter~\emph{et al.} demonstrate that such correlations readily decay in the thermodynamic limit. 

To establish a localized notion of diffusivity in confined geometries, they examine the response of a spatially localized observable given by:
$$
\mathcal{B}(\mathbf{P},\mathbf{Q}) = \frac{\mathbf{u}\cdot\sum_{i=1}^Nc_i\textbf{p}_i\chi_A(\mathbf{q}_i)}{m\sum_{i=1}^N\chi_A(\mathbf{q}_i)}.
$$
Here, $\chi_A(\cdot)$ represents the characteristic function of set $A$. Consequently, $\mathcal{B}$ signifies a colored momentum flux directed along unit vector $\mathbf{u}$ and confined to $A$. By utilizing Eq.~\eqref{eq:resp-fcn}, on can determine the response of $\mathcal{B}$ to the perturbation given by $\mathcal{A}$:
\begin{eqnarray}
-\frac{\langle\Delta\mathbf{B}(t)\rangle}{\beta F_c} &=& \int_0^{+\infty}\mathbf{u}^T\left\langle
\frac{\sum_{i=1}^N \mathbf{v}_i(t)\mathbf{v}_i^T(0)\chi_A(\mathbf{q}_i(t))}{\sum_{i=1}^N\chi_A(\mathbf{q}_i(t))}
\right\rangle\mathbf{w}\,dt\notag\\
&=& N\int_0^{\infty}\mathbf{u}^T \mathbf{C}_{v,A}(t)\mathbf{w}\,dt.\notag
\end{eqnarray}
Here, $\xi_A[\textbf{r}(t')] = \chi_A[\mathbf{r}(t+\tau)]$. Subsequently, Hunter~\emph{et al.}\cite{hunter_new_2022} propose the mean diffusivity within $A$ to be associated with $\mathbf{C}_{v,A}(t)$ as:
\begin{eqnarray}\label{eq:diff-Hunter}
\frac{1}{\text{Vol}\,(A)}\int_A\mathbf{D}(\mathbf{r})\,d\mathbf{r} &\overset{?}{=}& \int_0^{+\infty} \mathbf{C}_{v,A}(t)\,dt
\end{eqnarray}
While this quantity is a proxy for local diffusivity, it cannot be directly mapped onto the Smoluchowski formalism.  A suitable theoretical framework to assert this proposition is the Kramers-Klein equation\cite{KramersPhysica1940}, describing Langevin dynamics in the underdamped regime:
\begin{eqnarray}\label{eq: KK}
\frac{\partial p}{\partial t} &=& -\textbf{v}\cdot\nabla_{\textbf{r}}\,p + \frac{1}{m}\nabla_{\textbf{v}}\left[p\nabla_{\textbf{r}}\mathcal{F}\right] + \frac{1}{m\beta}\nabla_{\textbf{v}}\cdot\left[
\textbf{D}^{-1}(\textbf{r})\cdot\textbf{v}\, p\right]\notag\\
&& 
+\frac{1}{\left(m\beta\right)^2}\textbf{D}^{-1}(\textbf{r}):\textbf{H}_{\textbf{v}}p =  \mathcal{L}_{\textbf{r},\textbf{v}}^{\dagger}p.
\end{eqnarray}
Here, $m$ represents particle mass, $\mathcal{F}(\textbf{r})$ denotes a conservative potential of mean force, $\left[\textbf{H}_{\textbf{v}}p\right]_{i,j} = \partial^2p/\partial v_i\partial v_j$ is the Hessian tensor with respect to velocity degrees of freedom, and ':` signifies full tensorial contraction. The adjoint operator of $ \mathcal{L}_{\textbf{r},\textbf{v}}^{\dagger}$ with respect to the standard inner product is given by:
\begin{eqnarray}\label{eq: adjoint}
\mathcal{L}_{\mathbf{r},\mathbf{v}} &\equiv& \textbf{v}\cdot\nabla_{\textbf{r}} - \frac{\nabla_{\textbf{r}}\mathcal{F}}{m}\cdot\nabla_{\textbf{v}}  - \frac{\textbf{v}}{m\beta}\cdot\mathbf{D}^{-1}(\textbf{r})\cdot\nabla_{\textbf{v}}  \notag\\
&& + \frac{1}{\left(m\beta\right)^2}\mathbf{D}^{-1}(\textbf{r}):\mathbf{H}_{\textbf{v}} 
\end{eqnarray}
The \emph{ad hoc} VACF of \eqref{eq:diff-Hunter} can be formulated as,
\begin{eqnarray}
&&\mathbf{C}_{v,A} (t) = \left\langle\mathbf{v}(t)\mathbf{v}^T(0)K\left[\mathbf{r}(t)-\mathbf{r}_0\right] \right\rangle \notag\\
&&\overset{\text{(a)}}{=} \left\langle\mathbf{v}(t)\mathbf{v}^T(0)K\left[\mathbf{r}(0)-\mathbf{r}_0\right] \right\rangle \notag\\
&&=  \int \mathbf{v}\mathbf{w}^TK(\mathbf{s}-\mathbf{r}_0)p_t(\mathbf{r},\mathbf{v}|\mathbf{s},\mathbf{w})\Theta(\mathbf{s},\mathbf{w})\,d\mathbf{r}d\mathbf{s}d\mathbf{v}d\mathbf{w}\notag\\
&&
\label{eq:VACF-hunter-revised}
\end{eqnarray}
where $K[\mathbf{r}]$ is a properly normalized kernel function that, in the case of Hunter~\emph{et al.}'s work, is taken as the indicator of set $A$ (with $\mathbf{r}_0\in A$). Note that (a) follows from the linear response theory. $p_t(\cdot)$ is the solution of the Kramers-Klein equation and can be formally expressed as:
\begin{eqnarray}
p_t(\mathbf{r},\mathbf{v}|\mathbf{s},\mathbf{w}) &=& e^{t\mathcal{L}^{\dagger}_{\mathbf{r},\mathbf{v}}}\left[
\delta(\mathbf{r}-\mathbf{s})\delta(\mathbf{v}-\mathbf{w})
\right]\notag
\end{eqnarray}
By using the adjoint operator, the \emph{ad hoc} VACF can be expressed as,
\begin{eqnarray}
\mathbf{C}_{v,A} (t)  &=& \int \left[
e^{t\mathcal{L}_{\mathbf{r},\mathbf{v}}}
\right]\mathbf{v}\mathbf{v}^TK(\mathbf{r}-\mathbf{r}_0)\Theta(\mathbf{r},\mathbf{v})\,d\mathbf{r}d\mathbf{v}\notag
\end{eqnarray}
where $\Theta(\mathbf{r},\mathbf{v})$ is given by,
$$
\Theta(\mathbf{r},\mathbf{v}) = \left(
\frac{m\beta}{2\pi}
\right)^{d/2}\exp\left[-\frac{m\beta|\mathbf{v}|^2}{2}\right]\rho_0(\mathbf{r})
$$
In the case of a trivial potential of mean force and fixed diffusivity, it is easy to show that:
\begin{eqnarray}
\mathcal{L}_{\mathbf{r},\mathbf{v}}\mathbf{v}= -\frac{\mathbf{D}^{-1}\cdot\mathbf{v}}{m\beta},
\implies
\left[e^{t\mathcal{L}_{\mathbf{r},\mathbf{v}}}
\right]\mathbf{v} =\exp\left[
-\frac{t\mathbf{D}^{-1}}{m\beta}
\right]\cdot\mathbf{v},\notag
\end{eqnarray}
which, upon integrating momenta degrees of freedom, yields:
\begin{eqnarray}
\mathbf{C}_{v,A} (t) &=& \frac{1}{m\beta} \exp\left[
-\frac{t\mathbf{D}^{-1}}{m\beta}
\right]\int K(\mathbf{r}-\mathbf{r}_0)\rho_0(\mathbf{r})\,d\mathbf{r}\notag\\
\label{eq:CvA-const-D}
\end{eqnarray}
By time integrating \eqref{eq:CvA-const-D}, one can demonstrate the canonical relationship between VACF and diffusivity, namely:
\begin{eqnarray}
\int_0^{+\infty}\left\langle
\mathbf{v}(t)\mathbf{v}^T(0)K(\mathbf{r}-\mathbf{r}_0)
\right\rangle\,dt = \int \mathbf{D}K(\mathbf{r}-\mathbf{r}_0)\rho_0(\mathbf{r})\,d\mathbf{r},\notag
\end{eqnarray}
for a properly normalized kernel. However, for non-trivial potential of mean force and position-dependent diffusivity, $\mathcal{L}_{\mathbf{r},\mathbf{v}}$ will possess the following mathematical form:
\begin{eqnarray}
\mathcal{L}_{\mathbf{r},\mathbf{v}}\mathbf{v} &=& -\frac{\nabla_{\mathbf{r}}\mathcal{F}}{m} - \frac{\mathbf{D}^{-1}(\mathbf{r})\cdot\mathbf{v}}{m\beta}\notag
\end{eqnarray}
This makes constructing the $e^{t\mathcal{L}}$ operator extremely complicated since each successive application of $\mathcal{L}$ will require computing spatial derivatives of the unknown diffusivity profile as well as the potential of mean force. More precisely, if  one denotes $\mathbf{f}_\infty(\mathbf{r},\mathbf{v}) := \lim_{t\rightarrow\infty}e^{t\mathcal{L}_{\mathbf{r},\mathbf{v}}}\mathbf{v}$, then one can demonstrate that:
\begin{eqnarray}
&&\int_0^{+\infty}\left\langle
 \mathbf{v}(t)\mathbf{v}^T(0)K(\mathbf{r}-\mathbf{r}_0)
\right\rangle\,dt\notag\\
&& = \int\mathbf{g}_{\mathcal{L}}(\mathbf{r},\mathbf{v})\mathbf{v}^TK(\mathbf{r}-\mathbf{r}_0)\Theta(\mathbf{r},\mathbf{v})\,d\mathbf{r} d\mathbf{v}\notag
\end{eqnarray}
where $\mathbf{g}_{\mathcal{L}}(\mathbf{r},\mathbf{v})$ is the solution of the partial differential equation (PDE), $\mathcal{L}_{\mathbf{r},\mathbf{v}}\mathbf{g}_{\mathcal{L}}(\mathbf{r},\mathbf{v})=\mathbf{f}_{\infty}-\mathbf{v}$, which does not lend itself easily to a solution, and is not definitely consistent with the simplified postulation of \eqref{eq:diff-Hunter}. Therefore, even \emph{ad hoc} representations constructed using linear response theory fail to yield diffusivity profiles consistent with the Smoluchowski or the Kramers-Klein picture. 

Despite this fundamental limitation, linear response theory proves to be a potent framework for crafting effective-- albeit \emph{ad hoc}-- estimators for various transport coefficients, especially those characterizing the coupling among different thermodynamic driving forces. A good illustration of such capability is presented in the work of Mangaud and Rotenberg \cite{mangaud_sampling_2020}, where the authors investigate the transport properties of a solution within a slit pore under simultaneous pressure and chemical potential gradients. In such scenarios, transport coefficients can be appropriately defined utilizing Eq.~\eqref{eq:general-def-transport-coeff}.

\begin{figure}
\centering
\includegraphics[width=.48\textwidth]{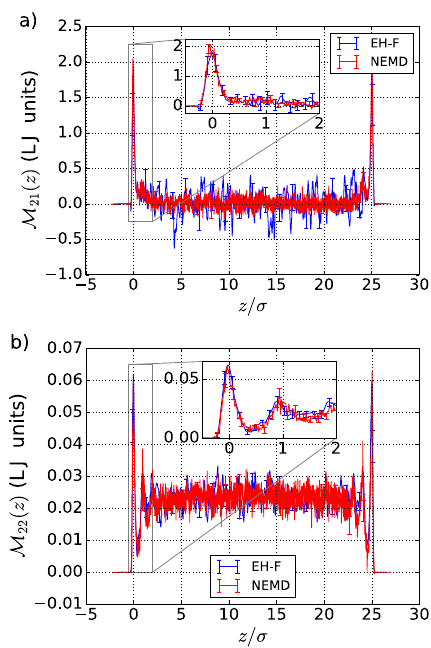}
\caption{(Reproduced from Ref.~\citenum{mangaud_sampling_2020}) (a) $\mathcal{M}_{\mathcal{P},\mu}(z)$ and (b) $\mathcal{M}_{\mu,\mu}(z)$ computed using NEMD and Eq.~\eqref{eq:GK-CrossTerms} for a binary LJ fluid within a slit pore. \label{fig:GK-Ext}}
\end{figure}

The particular geometry considered in Ref.~\citenum{mangaud_sampling_2020} is a slit pore  of thickness $h$ that is perpendicular  to the $z$ axis, while pressure and chemical potential gradients are both applied along the $x$ axis, namely parallel to the walls of the slit pore. They consider two different perturbations to the Hamiltonian, namely,
%\begin{subequations}
	\begin{eqnarray}
	\mathcal{A}_{\mathcal{P}} (\mathbf{P},\mathbf{Q}) &=& \frac{V\nabla_x \mathcal{P}}{N}\sum_{i=1}^Nx_i\notag\\
	\mathcal{A}_\mu(\mathbf{P},\mathbf{Q}) &=& -\frac{\nabla_x\mu}{1+\alpha}\left[
	\sum_{i=1}^{N_A}x_{i} - \alpha \sum_{i=N_A+1}^{N_A+N_B}x_{i}
	\right]\notag
	\end{eqnarray}
%\end{subequations}
Here, $N_A$ and $N_B$ are the number of particles of type $A$ and $B$, respectively, $x_i=\mathbf{e}_x\cdot\mathbf{q}_i$, and $\alpha=\rho_A^b/\rho_B^b$ is the ratio of bulk densities of $A$ and $B$. (Note that $A$ and $B$  particles are indexed as $1,2,\cdots,N_A$ and $N_A+1,\cdots,N_A+N_B$, respectively.)  Moreover, $\mathcal{A}_\mu$ is formulated so that no net force is exerted on the solution in the bulk. In order to use the formalism of linear response theory, they define the following localized observables that signify volume and concentration fluxes:
%\begin{subequations}
	\begin{eqnarray}
	\mathcal{B}_{\mathcal{P}} (\mathbf{P},\mathbf{Q};z) &=& \frac{h}{N}\sum_{i=1}^{N_A} v_{x,i}\delta(z_i-z)\notag\\
	\mathcal{B}_{\mu} (\mathbf{P},\mathbf{Q};z) &=& \frac{h}{V}\Bigg[\frac{1}{1+\alpha}\sum_{i=1}^{N_A} v_{x,i}\delta(z_i-z)\notag\\
	&&  - \frac{\alpha}{1+\alpha} \sum_{i=N_A+1}^{N_A+N_B} v_{x,i}\delta(z_i-z)\Bigg]\notag
	\end{eqnarray}
%\end{subequations}
One can then use Eq.~\eqref{eq:lin-resp} to enumerate the response of each localized observable to the respective global perturbation. More precisely, the coefficients of the matrix $\mathcal{M}$ given by,
\begin{eqnarray}
\lim_{t\rightarrow\infty}\left[
\begin{array}{c}
\langle\mathcal{B}_{\mathcal{P}}(z,t)\rangle\\
\langle\mathcal{B}_{\mu}(z,t)\rangle\\
\end{array}
\right] &=& -\left[
\begin{array}{cc}
\mathcal{M}_{\mathcal{P},\mathcal{P}} & \mathcal{M}_{\mathcal{P},\mu} \\
 \mathcal{M}_{\mu,\mathcal{P}} & \mathcal{M}_{\mu,\mu} 
\end{array}
\right]\left[
\begin{array}{c}
\nabla\mathcal{P}\\
\nabla\mu\\
\end{array}
\right],\notag
\end{eqnarray}
can be evaluated as,
\begin{eqnarray}\label{eq:GK-CrossTerms}
\mathcal{M}_{ij}(z) &=& \beta V\int_0^{+\infty} \left\langle
\mathcal{B}_{i}(t)\dot{\mathcal{A}_j}(0)
\right\rangle\,dt,
\end{eqnarray}
where $i,j\in\{\mathcal{P},\mu\}$. For instance, $\mathcal{M}_{\mathcal{P}\mu}(z)$ and $\mathcal{M}_{\mu\mu}(z)$ computed using Eq.~\eqref{eq:GK-CrossTerms} for a binary Lennard-Jones\cite{LJProcRSoc1924} (LJ) liquid within a slit pore is depicted in Fig.~\ref{fig:GK-Ext}, alongside their estimates obtained from NEMD.   In particular, the authors relate the diagonal components of $\mathcal{M}$ to viscosity and diffusivity. More precisely, they interpret $\mathcal{M}_{\mu\mu}$  as a Fickian binary diffusivity $D_{AB}$ along the $x$ axis. By invoking continuum approximation, they assume that $\mathcal{M}_{\mathcal{P}\mathcal{P}}(z)$ will take a form consistent with Poiseuille flow:
\begin{eqnarray}
    \mathcal{M}_{\mathcal{P}\mathcal{P}}(z) &=& \frac{\rho}{2\eta}\left[\left(\frac{h}{2}\right)^2-\left(z-\frac{h}{2}\right)^2\right]\notag
\end{eqnarray}
which enables them to fit the computed $\mathcal{M}_{\mathcal{P}\mathcal{P}}(z)$  to a quadratic profile to estimate viscosity $\eta$. Similar to the first example, these estimators are still \emph{ad hoc} in nature, as they do not yield a binary diffusivity profile that is compatible with the Smoluchowski formalism.

\section{Kernel-based approaches}
\label{section:kernel-based}

\noindent
Kernel-based methods are a class of methods in which particle positions (and displacements) are processed through the application of a \emph{kernel} function-- also referred to as a \emph{filter}. The theoretical foundation of these methods is based on It\^{o}'s Lemma,\cite{SchussTheoryStochastic2010} which specifies the temporal evolution of stochastic processes obtained by composing a standard It\^{o} process with a $C^2$ function. More precisely, suppose that $\mathbf{X}_t\in\mathbb{R}^k$ is an It\^{o} process,~i.e.,~a stochastic process whose temporal evolution is described by the following stochastic differential equation (SDE):
\begin{eqnarray}\label{eq:SDE-Ito}
d\mathbf{X}_t &=& \mathbf{p}(\mathbf{X}_t)\,dt + \pmb\pi(\mathbf{X}_t)\cdot d\mathbf{W}_t,
\end{eqnarray} 
wherein $\mathbf{W}_t$ is the $k$-dimensional Weiner process. It\^{o}'s Lemma states that the stochastic process $\mathbf{Y}_t = \mathcal{F}(\mathbf{X}_t)\equiv\left[\mathcal{F}_1(\mathbf{X}_t), \mathcal{F}_2(\mathbf{X}_t),\cdots,\mathcal{F}_n(\mathbf{X}_t)\right]\in\mathbb{R}^n$  will evolve according to the following SDE:
\begin{eqnarray}\label{eq:SDE-general-YvsX}
d\textbf{Y}_t &=& \textbf{q}\left(\textbf{X}_t\right)\,dt + \pmb\sigma(\textbf{X}_t)\cdot d\textbf{W}_t,
\end{eqnarray}
with $\mathbf{q}\in\mathbb{R}^n$ and $\pmb\sigma\in\mathbb{R}^{n\times k}$ given by:
\begin{subequations}
\begin{eqnarray}
q_i (\mathbf{X}_t) &=& \nabla\mathcal{F}_i^\dagger(\mathbf{X}_t) \mathbf{p}(\mathbf{X}_t)\notag\\&&  + \frac12\pmb\pi^{\dagger}(\mathbf{X}_t)\mathbf{H}_{i}(\mathbf{X}_t)\pmb\pi(\mathbf{X}_t)\label{eq:Ito-gen-q}\\
\pmb\sigma_i(\mathbf{X}_t) &=& \nabla\mathcal{F}_i^{\dagger}(\mathbf{X}_t)\pmb\pi(\mathbf{X}_t)\label{eq:Ito-gen-sigma}
\end{eqnarray}
\end{subequations}
Here, $\pmb\sigma_i$ is the $i$-th column of $\pmb\sigma$ and $\mathbf{H}_i$ is the Hessian of $\mathcal{F}_i$. It follows from the general theory of SDEs that the pointwise covariance matrix $\pmb\sigma\pmb\sigma^\dagger$ can be estimated from individual realization of \eqref{eq:SDE-general-YvsX} using,
\cite{SchussTheoryStochastic2010,weiss_first_1967}
\begin{eqnarray}
 \pmb\sigma\left(\mathbf{r}\right)\pmb\sigma\left(\mathbf{r}\right)^\dagger &=& \lim_{h\rightarrow0^+} \left.\frac{\left\langle (\textbf{Y}_{t+h}-\textbf{Y}_t)(\textbf{Y}_{t+h}-\textbf{Y}_t)^{\dagger}\right\rangle}{2h}\right|_{\mathbf{X}_t=\mathbf{r}}\notag\\&&\label{eq:Ito-cov}
\end{eqnarray}
 Note that the Smoluchowski equation can be viewed as a forward Kolmogorov equation associated with the overdamped Langevin SDE given by:
\begin{eqnarray}\label{eq:Langevin-overdamped}
d\mathbf{X}_t &=& -\left[
\beta\mathbf{D}(\mathbf{X}_t)\cdot\nabla\mathcal{F}(\mathbf{X}_t) + \nabla\cdot\mathbf{D}(\mathbf{X}_t)
\right]\,dt \notag\\
&& + \sqrt{2\mathbf{D}(\mathbf{X}_t)}\cdot d\mathbf{W}_t.
\end{eqnarray}
Clearly Eq.~\eqref{eq:Langevin-overdamped} falls within the broader category of SDEs described by \eqref{eq:SDE-Ito} and \eqref{eq:SDE-general-YvsX}. From It\^{o}'s lemma, an expression similar to the one yielding Kramer-Moyal coefficients can also be derived for a filtered trajectory by employing \eqref{eq:Ito-gen-q} and \eqref{eq:Ito-gen-sigma}. More specifically, suppose that $\gamma:\mathbb{R}^n\rightarrow\mathbb{C}$ is a piecewise $C^2$ function. For an $\textbf{X}_t$ satisfying \eqref{eq:Langevin-overdamped}, $Y_t=\gamma(\textbf{X}_t)$ will evolve according to the following SDE:
\begin{eqnarray}
dY_t &=& \left[\nabla\gamma(\textbf{X}_t)\cdot\pmb\mu(\textbf{X}_t) + \mathbf{D}(\textbf{X}_t):\mathbf{H}_\gamma(\mathbf{X}_t)\right]\,dt\notag\\
&& +\nabla\gamma(\mathbf{X}_t)\cdot\sqrt{2\mathbf{D}(\mathbf{X}_t)}\cdot d\mathbf{W}_t\notag
\end{eqnarray}
The associated covariance will thus be given by,
\begin{eqnarray}
&&\nabla\gamma^\dagger(\mathbf{r})\mathbf{D}(\textbf{r})\nabla\gamma(\mathbf{r}) =\notag\\
&&  \lim_{h\rightarrow0^+} \left.\frac{\left\langle\left|\gamma(\textbf{X}_{t+h})-\gamma(\textbf{X}_t)\right|^2\right\rangle}{4h}\right|_{\mathbf{X}_t=\mathbf{r}}\label{eq:Ito-filtered-cov}
\end{eqnarray}
The expectation given by \eqref{eq:Ito-filtered-cov} is proportional to the projection of the diffusivity tensor along the direction given by $\nabla\gamma$. The ability to use a filter function gives one an increased level of flexibility to design suitable estimators of diffusivity. It must be noted that Eqs.~\eqref{eq:Ito-cov} and \eqref{eq:Ito-filtered-cov} can be easily reformulated if $\mathbf{X}_t$ is drawn from a probability distribution $\nu(\cdot)$, which can be identical to the equilibrium probability distribution $\rho_0(\cdot)$,
\begin{eqnarray}
&& \int \pmb\sigma(\mathbf{r})\pmb\sigma^\dagger(\mathbf{r})\nu(\mathbf{r})\,d\mathbf{r} = \notag\\&& \lim_{h\rightarrow0^+} \left.\frac{\left\langle (\textbf{Y}_{t+h}-\textbf{Y}_t)(\textbf{Y}_{t+h}-\textbf{Y}_t)^{\dagger}\right\rangle}{2h}\right|_{\mathbf{X}_t\sim\nu(\mathbf{r})}\notag
\end{eqnarray}
One of the first attempts to use general filters (as opposed to characteristic functions that 'count' particles in a bin) to design diffusivity estimators in the context of molecular simulations was undertaken by our research group, as presented in a series of papers.\cite{domingues_robust_2023-1, domingues_robust_2023}
(Another example of using a similar expression-- albeit with a characteristic function of a set as a filter-- is the estimation of diffusivity in the collective variable space by Hegger and Stock.\cite{hegger_multidimensional_2009})

In the initial paper\cite{domingues_robust_2023-1} of the series, our focus was on the analytical derivation of the estimator and its numerical validation using synthetic data obtained from numerical integration of \eqref{eq:Langevin-overdamped} with known \emph{a priori} diffusivity profiles. All employed diffusivity profiles were functions of a single  spatial variable. The subsequent paper\cite{domingues_robust_2023} in this series delves into the application of the method to MD trajectories.

In the first paper,\cite{domingues_robust_2023-1} we propose a filter function of the form $
\gamma_{\textbf{k}}(\textbf{r}) = e^{-i\alpha\textbf{k}\cdot\textbf{r}} G(\textbf{r})
$. The complex exponential encodes information about the directionality of the diffusivity tensor (by yielding its projection along the unit vector $\mathbf{k}$) while $G(\cdot)$ is a localization function-- also known as a \emph{kernel}-- that enables estimating diffusivity around a certain point in space. In practice, $G(\cdot)$ can be defined as,
\begin{eqnarray}
G(\textbf{r}) &=& \frac{1}{\epsilon^d}K_\epsilon\left(\frac{\mathbf{r}-\mathbf{r}_0}{\epsilon}\right),\notag
\end{eqnarray}
wherein $K_\epsilon(\textbf{r})\geq 0$ is chosen in a way that approximates the delta function as $\epsilon\rightarrow0^+$. By applying Eq.~\eqref{eq:Ito-filtered-cov} to the filter functions $\gamma_{\pm}(\textbf{r}) = e^{-i\alpha\textbf{k}\cdot\textbf{r}}\left[G(\textbf{r})\pm\frac14\right]
$ and conducting some algebraic rearrangement, it can be demonstrated that:
\begin{eqnarray}
&& \int \mathbf{k}^T\mathbf{D}(\mathbf{r}) \mathbf{k} G(\mathbf{r})\nu(\mathbf{r}) \,d\mathbf{r} = \lim_{h\rightarrow0^+} \frac{1}{2\alpha^2h}\times\notag\\
&& \left\langle
\left[\gamma_{\mathbf{k}}(\mathbf{X}_{t+h})-\gamma_{\mathbf{k}}(\mathbf{X}_{t})\right]^*
\left[f_\mathbf{k}(\mathbf{X}_{t+h})-f_\mathbf{k}(\mathbf{X}_{t})\right]
\right\rangle_{\mathbf{X}_t\sim\nu(\cdot)}.\notag
\end{eqnarray} 
By letting $\epsilon\rightarrow0$, the integral on the left hand side will converge to the pointwise estimate of $D_{\mathbf{k}\mathbf{k}}$, the diffusivity projected along $\mathbf{k}$, namely,
\begin{eqnarray}
\widehat{D}_{\mathbf{k}\mathbf{k}}(\mathbf{r}_0) \approx \frac{1}{2\alpha^2 h}\frac{\Re\left[\sum_{i=1}^{N_t}\Delta_h\gamma_{\mathbf{k}}^*(\mathbf{X}_{i,t})\Delta_hf_{\mathbf{k}}^*(\mathbf{X}_{i,t})\right]}{\sum_{i=1}^{N_t}G(\textbf{X}_{i,t})}\notag
\end{eqnarray}
wherein $\Delta_hg(\mathbf{X}_t)=g(\mathbf{X}_{t+h})-g(\mathbf{X}_t)$ and the summation is conducted over $N_t$ trajectories. By letting $\alpha\rightarrow0$, one can obtain a limiting estimator given by,
\begin{eqnarray}
\widehat{D}_{\mathbf{k}\mathbf{k}}^{\alpha\rightarrow0}(\mathbf{r}_0) = \frac{1}{4h}\frac
{\sum_{i=1}^{N_t}\left[G(\mathbf{X}_{i,t+h})+G(\mathbf{X}_{i,t})\right]\left[\mathbf{k}\cdot\Delta_h\mathbf{X}_{i,t}\right]^2}
{\sum_{i=1}^{N_t}G(\mathbf{X}_{i,t})}\notag\\
&&\label{eq:estimator-filter-alpha0}
\end{eqnarray}
It is important to note that Eq.~\eqref{eq:estimator-filter-alpha0} offers a natural means of constructing an \emph{ad hoc} extension of MSD in confined geometries, by assigning equal weight to particles  that are present within a designated bin either at the beginning or at the end of an observation window. (Particles present within a bin both at the beginning and at the end of the observation window would contribute twice as much to the \emph{ad hoc} MSD.)

\begin{figure*}
\centering
\includegraphics[width=.8\textwidth]{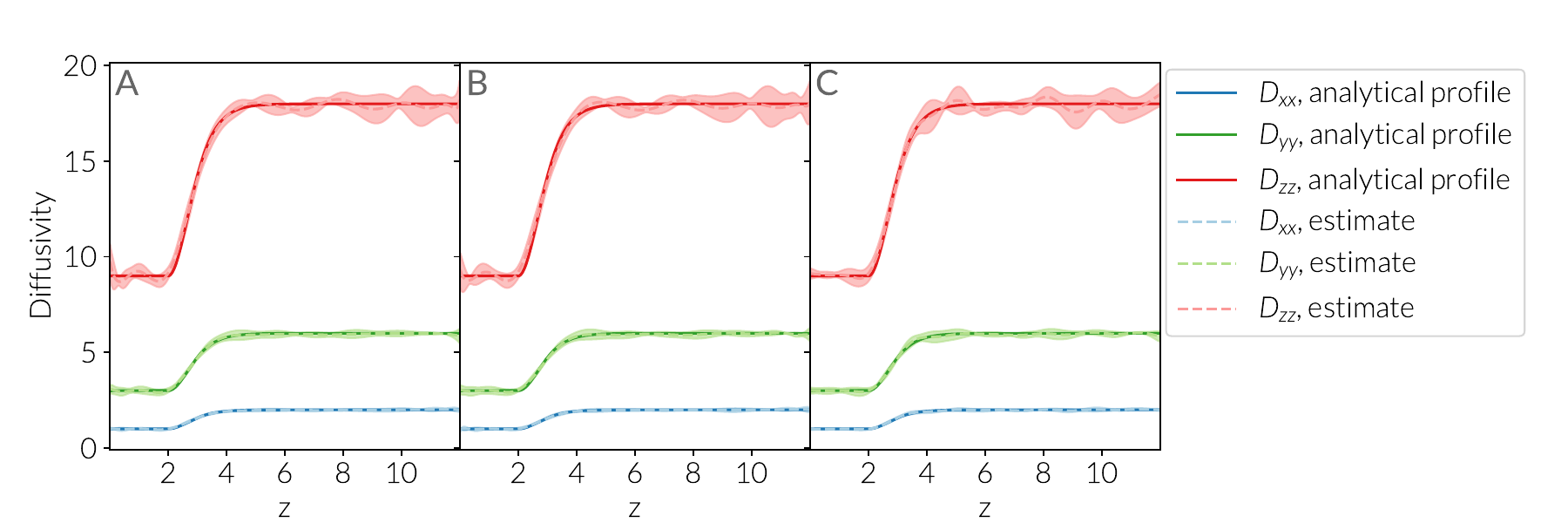}
\caption{(Reproduced from Ref.~\citenum{domingues_robust_2023-1}) Application of Eq.~\eqref{eq:projection-coeffs} to the synthetic data wherein $g_n(\cdot)$'s are chosen from (A) Legendre and (B) Chebyshev polynomials, and (C) Fourier series. Solid lines and symbols correspond to the true diffusivity and the projection-based estimates, respectively.\label{fig:projection} }
\end{figure*}

We wish to note that the filter-based estimators proposed by us in Ref.~\citenum{domingues_robust_2023-1} are closely related to Nadaraya-Watson (NW) estimators\cite{NadarayaTheorProbabAppl1964, WatsonSankhyaSerA1964} in the statistics literature. NW estimators are designed to construct a regression function $m(x)$ that relates two random variables, $Y$ and $X$, with the condition that $m(x) = \left\langle Y|X=x\right\rangle$, all without resorting to parametric expressions. To create this function, the following Taylor expansion of $m(x)$ centered at a specific point is introduced:
\begin{eqnarray}
m(x) &=& \sum_{i}\sum_{j=0}^p \beta_j(X_i-x)^jK_\epsilon(x-X_i)\notag
\end{eqnarray}
Here, $p$ is the order of the Taylor expansion, and $K_\epsilon(x-X_i)$ is a localization function that gives more weight to the $X_i$'s that are closer to $x$. 
Subsequently, a weighted least squares problem is formulated to determine coefficients of the estimator within the Taylor expansion.
\begin{equation}\label{eq: NW}
\widehat{\pmb\beta}_{\epsilon}=\arg\min_{\mathbf{\pmb\beta}\in\mathbb{R}^{p+1}}\sum_i\left[Y_i-\sum_{i=0}^p\beta_j(X_i-x)^j\right]^2K_\epsilon(x-X_i).
\end{equation}
 By setting $p=0$ in the expression provided above, we obtain the following explicit solution:
\begin{equation}\label{eq: NW_p=0}
    \widehat{m}_\epsilon(x) = \widehat{\beta}_{\epsilon,0}=\frac{\sum_i Y_i K_\epsilon(x-X_i)}{\sum_iK_\epsilon(x-X_i)}
\end{equation}
Drawing inspiration from the pioneering works of Zmirou\cite{florens-zmirou_estimating_1993} and Bandi,\cite{bandi_fully_2001} it has been proposed that a similar approach can be applied to estimate Kramers Moyal coefficient:
\begin{equation}\label{eq: LCE-NW}
 \widehat{\mathbf{D}}_{\epsilon,h}(t,\mathbf{r}) =\frac{1}{2h}\frac{\sum_{i=1}^N (\Delta _h\mathbf{X}_{i,t})(\Delta_h\mathbf{X}_{i,t})^T K_\epsilon(\mathbf{r}-\mathbf{X}_{i,t})}{\sum_iK_\epsilon(\mathbf{r}-\mathbf{X}_{i,t})}%\\
\end{equation}
where $\Delta_h\mathbf{X}^i_{i,t}$'s are displacement samples drawn from realizations of \eqref{eq:Langevin-overdamped}.
Similar to filtered estimators, Eq.~\eqref{eq: LCE-NW} can be reformulated accordingly if $\textbf{X}_{i,t}$'s (i.e.,~the starting points of stochastic trajectories) can be drawn from any probability distribution, $\nu(\mathbf{r})$:
\begin{equation}\label{limit kernel}
    \begin{aligned}
    &\operatornamewithlimits{\lim}_{h\rightarrow 0^+}\left\langle\widehat{\textbf{D}}_{\epsilon,h}(t,\mathbf{r})\right\rangle=\frac{\int \mathbf{D}(\mathbf{y}) K_\epsilon(\mathbf{y}-\mathbf{r})\nu(\mathbf{y})\,d\mathbf{y}}{\int K_\epsilon(\mathbf{y}-\mathbf{r})\nu(\mathbf{y})\,d\mathbf{y}}\\
        &\equiv \widetilde{\mathbf{D}}_\epsilon(t,\textbf{r})
    \end{aligned}
\end{equation}
Therefore, both filter- and kernel-based estimators can easily accommodate scenarios in which samples are drawn from non-equilibrium distributions. In simulations of equilibrium systems, the expression \eqref{eq: LCE-NW} can be further adapted by averaging it over $t$, with $\nu(\textbf{x})$ being substituted by $\rho_0(\mathbf{x})$ in \eqref{limit kernel}. Nevertheless, it is still imperative to observe the same considerations when selecting an appropriate timescale $\tau$ for this approach. This approach can, in principle, be used for any time series data that can be modeled by Eq.~\eqref{eq:SDE-general-YvsX}. It has been used successfully in contexts as diverse as financial data,\cite{jiang_nonparametric_1997} electroencephalographic (EEG) data,\cite{lamouroux_kernel-based_2009} and stochastic descriptions of chaotic deterministic systems.\cite{lamouroux_kernel-based_2009}

We wish to highlight a subtle-- but crucial-- distinction between the FCE estimators derived in Ref.~\citenum{domingues_robust_2023-1} and the NW-based estimators given by Eq.~\eqref{limit kernel}. Specifically, the former method applies the kernel on the particle's position at both the beginning and the end of the observation window, whereas the latter applies the kernel solely to the initial frame. This small contrast potentially affords FCEs a marginal edge in elucidating the locality of diffusivity by more effectively attenuating the impact of inter-bin mixing on displacement statistics.

In Ref.~\citenum{domingues_robust_2023-1}, the majority of numerical tests were conducted using the triangle kernel, $K_1(z)\propto (1-|z|)\chi_{|z|<1}(z)$. However, the choice of kernel did not significantly impact the reliability of diffusivity estimates. This is expected considering the observation that,
\begin{eqnarray}
\widehat{D}_{\mathbf{k}\mathbf{k}}(\mathbf{r}_0) &=& D_{\mathbf{k}\mathbf{k}}(\mathbf{r}_0)+\epsilon^2\mathbf{K}_2:\Big[\mathbf{H}_{D_{\mathbf{k}\mathbf{k}}}(\mathbf{r}_0)+\notag\\
&& 2\nabla D_{\mathbf{k}\mathbf{k}}(\mathbf{r}_0)\nabla^T\ln \rho_0(\mathbf{r}_0)\Big]
\label{eq:epsilon-scaling}
\end{eqnarray}
where $\mathbf{H}_f$ is the Hessian of the scalar function $f$ and $\mathbf{K}_2=\frac12\int\mathbf{y}\mathbf{y}^TG(\mathbf{y})\,d\mathbf{y}$. In other words, Eq.~\eqref{eq:epsilon-scaling}  illustrates that the variance of the kernel appears as a prefactor in front of the $\epsilon^2$ term, but does not alter the fundamental  scaling of systematic error with $\epsilon$. Furthermore, beyond a certain threshold, reducing the value of $\epsilon$ results in a kernel function with a significantly narrowed support, which adversely affects statistical accuracy, leading to the emergence of large error bars. This observation aligns with the theoretical expectation that, for small $\epsilon$ values, the variance of the estimator should scale proportionally to $\propto\epsilon^{-d}\int K_1^2(z)\,dz$.

One of the advantages of using kernel-based methods is that the kernel function does not have to be localized, nor does it have to be nonnegative. For instance, one can choose a collection of kernels that belong to a family of orthogonal functions, such as Fourier series, or special polynomials (Fig.~\ref{fig:projection}). One can then express,
$$
\mathbf{D}(\mathbf{r})\nu(\mathbf{r}) = \sum_{n=1}^{+\infty} \mathbf{C}_ng_n(\mathbf{r}).  
$$
The unknown coefficients within this sum can be projected onto $\mathbf{k}$ in a similar fashion, and determined using,
\begin{eqnarray}\label{eq:projection-coeffs}
C_{n,\mathbf{k}\mathbf{k}} &=& \mathbf{k}^T\mathbf{C}_n\mathbf{k} = \int D_{\mathbf{k}\mathbf{k}}(\mathbf{r})g_n(\mathbf{r})\nu(\mathbf{r})\,d\mathbf{r}\notag\\
&\approx& \frac{\left\langle\left[g_n(\mathbf{X}_{t+h})+g_n(\mathbf{X}_{t})\right]\left[\mathbf{k}\cdot\Delta_h\mathbf{X}_t\right]^2 \right\rangle}{4h}
\end{eqnarray}
Upon evaluating the computational performance of the FCE estimator, we found it to exhibit robust performance across various scenarios, provided that the drift term remains non-divergent. However, significant discretization errors were observed in cases where the drift diverges, particularly in the vicinity of hard boundaries. This phenomenon can be attributed to the increased susceptibility of the Gaussian approximation to temporal discretization in the presence of strong (diverging) drifts. For trajectories that are stochastic at all timescales, such errors can be effectively remedied by choosing a sufficiently small $h$. 

In the second paper in this series\cite{domingues_robust_2023}, we adopted the FCE estimator to trajectories generated via MD. This requires identifying a system- and position-dependent timescale $\tau$ beyond which the system exhibits diffusive behavior. This was achieved by introducing the concept of a \emph{cage escape time},~i.e.,~the characteristic timescale for a particle to escape the cage formed by its first coordination shell. More precisely, we proposed the following autocorrelation function,
\begin{eqnarray}
C(z,t) &=& \frac{\left\langle\sum_{i,j=1,i\neq j}^N\delta\left[z_i(0)-z\right]\xi(z,\textbf{r}_{ij}(0))\xi(z,\textbf{r}_{ij}(t))\right\rangle}{\left\langle\sum_{i,j=1,i\neq j}^N\delta\left[z_i(0)-z\right]\xi(z,\textbf{r}_{ij}(0))\xi(z,\textbf{r}_{ij}(0))\right\rangle}
\notag\\
&&\label{eq:CEAF}
\end{eqnarray}
Here, $r_{\text{cage}}$ the first valley of the radial distribution function at $z$, and $\xi(z,\textbf{r})=H\left[r_{\text{cage}}(z)-\|\textbf{r}\|\right]$ with $H(\cdot)$ the Heaviside function. Intuitively, $C(z,t)$ corresponds to the fraction of the neighboring particles that remain within a distance $r_{\text{cage}}(z)$  of a central particle  after time $t$ has elapsed. $C(z,t)$  can thus be computed for particles belonging to each spatial bin and be fitted to a stretched exponential\cite{PhillipsRepProgPhys1996} $C(z,t) =\exp\left[-[t/\tau_c(z)]^{\alpha(z)}\right]$ to obtain a position dependent timescale $\tau_c(z)$. We wish to note that this approach can serve as a systematic means of determining a diffusive timescale in all methods for which the specification of such a timescale is necessary.

\begin{figure}
    \centering
    \includegraphics[width=.44\textwidth]{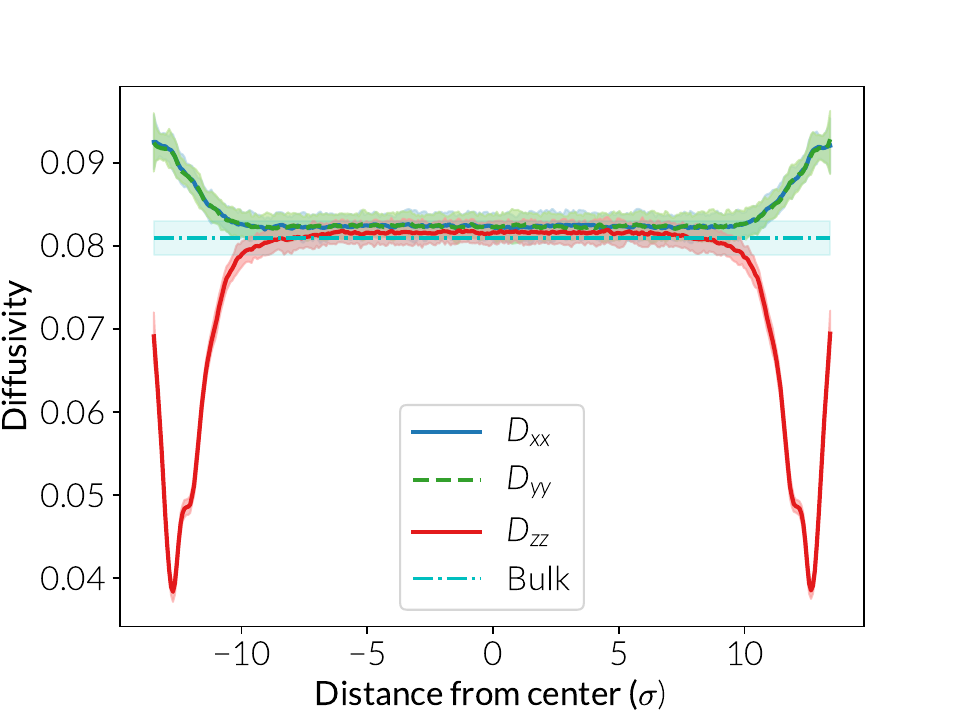}
    \caption{(Reproduced from Ref.~\citenum{domingues_robust_2023}) Diffusivity estimate from \eqref{eq:estimator-filter-alpha0} for a LJ fluid confined within a slit pore.}
    \label{fig: diff_paper2_FCE}
\end{figure}

Nevertheless, kernel-based estimators introduced in Ref.~\citenum{domingues_robust_2023-1} exhibit a small-- but systematic-- underestimation of diffusivity when applied to MD trajectories, due to the presence of the caging regime that follows the culmination of the ballistic regime. This systematic error can, however, be readily remedied using a slightly modified form of the estimator, namely,
\begin{eqnarray}\label{eq: ballistic correction}
\widehat{D}_{\mathbf{k}\mathbf{k}}^{h_1,h_2} &=& \frac{h_2\widehat{D}_{\mathbf{k}\mathbf{k}}^{h_2}-h_1\widehat{D}_{\mathbf{k}\mathbf{k}}^{h_1}}{h_2-h_1}
\end{eqnarray}
where $h_1$ and $h_2$ constitute two observation windows within the diffusive regime.
We applied the estimator given by Eq.~\eqref{eq: ballistic correction} to an LJ fluid confined within a slit-pore, with the resulting diffusivity profiles depicted in Fig.~\ref{fig: diff_paper2_FCE}. An analysis conducted for the purposes of validating the estimator demonstrated that  the computed $D_{xx},D_{yy}$ profiles are accurate, but the predicted $D_{zz}$ profile loses accuracy in the immediate vicinity of the wall due to the diverging drift. A Bayesian correction scheme based on diffusion maps\cite{CoifmanApplComputHarmonAnal2006} was then introduced and applied to rectify such inaccuracy.

Kernel-based estimators provide a robust extension of the concept of an \emph{ad hoc} mean squared displacement, due to their conceptual clarity and computational simplicity. Similar to the Kramer-Moyal estimator presented in Eq.~\eqref{eq: diffusivity_from KM}, kernel-based estimators offer the distinct advantage of accommodating a local timescale $\tau$ for each position at which diffusivity is to be estimated. This feature proves advantageous when compared to other methodologies (discussed later in this review) that require the usage of a fixed observation window for the whole system, as $\tau_c$, the timescale required to fully transition into the diffusive behavior might exhibit strong dependence on position in many circumstances.  In approaches where a single timescale must be selected for discretizing the Smoluchowski equation across the entire system, it becomes imperative to choose $\tau_{\text{max}}=\sup_z \tau_c(z)$ to ensure that ballistic effects have been overcome uniformly. It should be noted that, as the timescale $\tau$ increases, the potential influence of drift effects on the accuracy of the diffusivity estimate $\mathbf{D}(\mathbf{r})$ also becomes more pronounced.

Finally, it is pertinent to acknowledge that implementing this method requires determining certain free parameters, such as $\epsilon$ in the case of localized kernels or the total number of basis functions $N$ when employing a projection approach. The optimal selection of these parameters may necessitate a systematic exploration to attain the desired level of precision in the analysis. Schemes for choosing $\epsilon$ have been proposed in the statistics literature in the context of Nadaraya-Watson estimators.\cite{bandi_fully_2001, silverman2018density}

\section{Bayesian approaches}
\label{section:Bayesian}

\begin{figure}
\centering
\includegraphics[width=.5\textwidth]{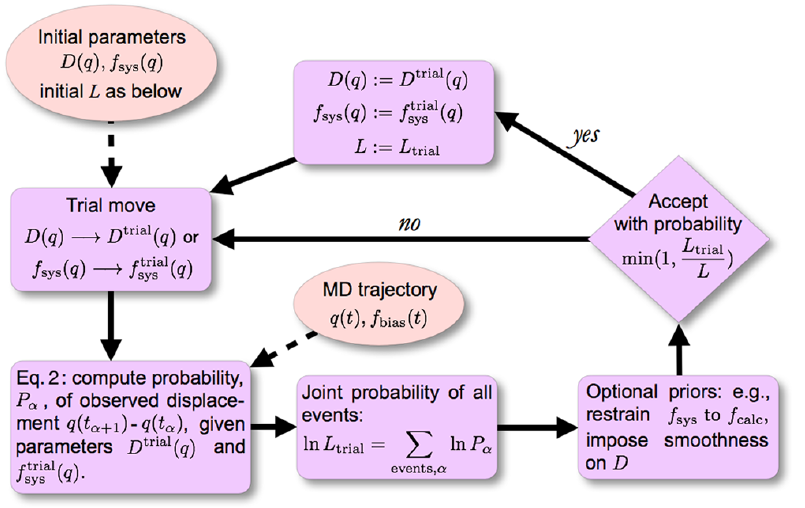}
\caption{(Reproduced from Ref.~\citenum{comer_calculating_2013}) A schematic flowchart of applying Bayesian approaches to estimate position-dependent diffusivity.\label{fig:Bayesian-flowchart}}
\end{figure}

\noindent
Bayesian approaches attempt to obtain a maximum likelihood estimate (MLE) of diffusivity from MD data, by means of constructing a suitable likelihood function,
\[\mathbb{P}(\textbf{D}|\text{data})\propto \mathbb{P}(\text{data}|\textbf{D}),\]
which is then maximized over the space of all plausible diffusivity profiles. While Bayesian approaches can be employed without any prior information about the diffusivity profile, it might, in many cases, be convenient to include preexisting information about diffusivity as a prior distribution, resulting in the following likelihood function:
\begin{eqnarray}\label{eq:MAP}
\mathbb{P}(\textbf{D}|\text{data})\propto \mathbb{P}(\text{data}|\textbf{D})\mathbb{P}(\textbf{D})
\end{eqnarray}
Maximizing \eqref{eq:MAP} results in a \emph{maximum a posteriori} (MAP) estimate of diffusivity. A schematic flowchart of Bayesian approaches is depicted in Fig.~\ref{fig:Bayesian-flowchart}.\cite{comer_calculating_2013}

Indeed, one of the most widely known and popular methods for estimating position-dependent diffusivity is a Bayesian approach proposed by  Hummer,\cite{hummer_position-dependent_2005} which is also based on spatial discretization of the Smoluchowski operator. Let $\mathbf{Q}$ be a stochastic matrix wherein $Q_{ij}$ is the expected probability of transitioning from bin $i$ to bin $j$, and let $\mathbf{P}$ the matrix that contains the actual transition probabilities obtained from MD.  The likelihood that $\mathbf{Q}$ accurately represents the data is given by
\begin{eqnarray}\label{eq: likelihood-discretized}
P(\text{data}|\mathbf{Q}) &=& \prod_{i,j=1}^{n_b} Q_{ij}^{P_{ij}n_{w,i}}
\end{eqnarray}
where $n_{w,i}$ is the total number of observed transitions starting from the $i$-th bin. 
By taking the logarithm of both sides, and adding $\pm \sum_{i=1}^{n_b}n_{w,i}\sum_{j=1}^{n_b}P_{ij}\log P_{ij}$ to the right hand side,  the log posterior probability can be expressed as:
\begin{eqnarray}\label{eq:logQcData-KL}
\log P(\mathbf{Q}|\text{data}) = -\sum_{i=1}^{n_b}n_{w,i}\sum_{j=1}^{n_b}P_{ij}\log \frac{P_{ij}}{Q_{ij}} + C_2
\end{eqnarray}
Note that maximizing \eqref{eq:logQcData-KL} over all stochastic matrices would trivially yield $\mathbf{Q}=\mathbf{P}$. One, however, needs to only conduct maximization over matrices that are consistent with diffusive behavior. In Hummer's approach, $\mathbf{Q}$ is parameterized as $\mathbf{Q}=e^{t\mathbf{R}}$ wherein $\mathbf{R}$ is a rate matrix with its entries satisfying the following properties:
\begin{equation}\label{eq:Hummer-R}
r_{ij}=
\begin{cases}
r_{ij}, & i>j\\
-\sum_{l\neq i} r_{il} & i=j\\
r_{ji}\rho_i/\rho_j & i<j
\end{cases}
\end{equation}
Here, $\rho_j$ refers to the equilibrium probability of finding a particle at bin $j$, and the condition $r_{ij}\rho_j = r_{ji}\rho_i$ is included to assure detailed balance.  Therefore, the rate matrix $\mathbf{R}$ will possess $n_f=\frac{n_b(n_b+1)}{2}-1$ free entries. The associated optimization problem can be solved using a variety of methods, such as Monte Carlo sampling from a posterior distribution of the rate matrix $\mathbf{R}$: 
\begin{eqnarray}\label{eq:logQcData-prime}
\log P(\mathbf{R}|\text{data}) &=& -\sum_{i=1}^{n_b}n_{w,i}\sum_{j=1}^{n_b}P_{ij}\log \frac{P_{ij}}{\left[e^{tR}\right]_{ij}} + C_2\notag\\
&& 
\end{eqnarray}
We wish to note that such a posterior will generally be high-dimensional considering the quadratic scaling of $n_f$ with $n_b$. However, since $Q_{ij}$ approximates $\approx\rho_t(\textbf{r}_j|\textbf{r}_i)$ as a discretized solution of the Smoluchowski equation, the ansatz $\mathbf{Q}=e^{t\mathbf{R}}$ implies that $\mathbf{R}$ can be regarded as a discrete representation of the  $\mathcal{L}^\dagger$ operator (akin to the operator discretization methods described in Section~\ref{section:operator-discretization}, such as Ref.~\citenum{palmer_correlation_2020}). 
%In contrast to these approaches, the discretized operator within Hummer's framework is obtained through sampling from the posterior distribution.  When finite difference schemes are employed to discretize $\mathcal{L}^\dagger$, the resulting matrix is typically sparse, and in the simplest one-dimensional case, assumes a tridiagonal structure. 
%In Hummer's approach, $\mathcal{L}^{\dagger}$ is discretized using a finite difference scheme. The resulting matrix is typically sparse, and in the simplest case of one-dimensional confinement, assumes a tridiagonal structure. 
%This  characteristic implies that, within the scope of this approximation, particle exchanges primarily occur between adjacent bins (i.e., the instantaneous rate of exchange between non-adjacent bins is assumed to be zero).

By drawing an analogy with finite difference discretization, constraints can be applied to the matrix $\mathbf{R}$ to ensure its sparsity, predominantly preserving non-zero elements along few off-diagonal positions.  Such constraints would effectively reduce the dimensionality of the posterior distribution, resulting in a linear scaling between $n_f$ and $n_b$. Furthermore,  this will allow the utilization of efficient diagonalization techniques for computing the matrix exponential in \eqref{eq:logQcData-prime}.  Much in the same way as the operator discretization scheme of Sicardi~\emph{et al.}\cite{sicard_position-dependent_2021} (discussed in Section~\ref{section:operator-discretization}), Hummer employs finite differences to discretize $\mathcal{L}^\dagger$ in one dimension, following Bicout and Szabo\cite{bicout_electron_1998}, yielding the following  relationship:
\begin{equation}\label{eq: R to diffusivity}
    \frac{D_i+D_{i+1}}{2}= \Delta x^2 R_{i,i+1} \left(\frac{\rho_i}{\rho_{i+1}}\right)^{1/2},
\end{equation}
where $\Delta{x}$ is the thickness of each bin. This results in a matrix $\mathbf{R}$ that is tridiagonal. In other words, within the scope of this approximation, particle exchanges primarily occur between adjacent bins (i.e., the instantaneous rate of exchange between non-adjacent bins is assumed to be zero). This expression provides a means of calculating the diffusivity profile using posterior samples of $\mathbf{R}$
or through a maximum likelihood estimate of $\mathbf{R}$, achieved by optimizing \eqref{eq:logQcData-prime}. Similarly, \eqref{eq: R to diffusivity} enables Monte Carlo sampling of $\mathbf{D}$, from which the components of the matrix $\mathbf{R}$ defining the likelihood can be specified.

In the approach formulated above, the derivative of the diffusivity profile is left unconstrained, potentially leading to the emergence of rapid oscillations in $D(x)$ due to inherent noise in the underlying MD data. To mitigate this, and consistent with the  expectation that physical properties within a single thermodynamic phase are expected to be continuous functions of position, Hummer introduced the following prior distribution within the space of diffusivity profiles:
\begin{equation}\label{eq: Hummer_prior}
    \mathbb{P}(D)\propto \prod_{i} e^{-\left[D_i-D_{i+1}\right]^2/2\gamma^2}.
\end{equation}
We wish to note that there might be a more physically motivated prior than the one described by \eqref{eq: Hummer_prior}.  Specifically, assuming the validity of the Smoluchowski equation, individual realizations of single-particle trajectories can be generated using the SDE given by Eq.~\eqref{eq:Langevin-overdamped} with a drift term given by:
\begin{eqnarray}\label{eq:Smoluchowski-drift}
\pmb\mu(\textbf{r}) = \textbf{D}(\textbf{r})\cdot\nabla\log\rho_0(\textbf{r})+\nabla\cdot \textbf{D}(\textbf{r}) 
\end{eqnarray}
Multiplying both sides of \eqref{eq:Smoluchowski-drift} by $\rho_0$ yields:
\begin{equation}
    \begin{aligned}
        &\textbf{D}(\textbf{r})\cdot\nabla\rho_0 +\rho_0\nabla\cdot\textbf{D}(\textbf{r}) =\nabla\cdot\left[\rho_0 \textbf{D}(\textbf{r})\right] =\rho_0 \pmb\mu(x)\\
        &\nabla\cdot\left[\rho_0 \textbf{D}(\textbf{r})\right] \approx  \frac{\rho_0}{\tau}\left.\left\langle \textbf{X}_{t+\tau}-\textbf{X}_t\right\rangle\right|_{\textbf{X}_t=\textbf{r}}+O(\tau^2)\\
    \end{aligned}
\end{equation}
This equation establishes a connection between diffusivity, $\mathbf{D}(\mathbf{r})$, and drift, $\pmb\mu(\mathbf{r})$, within the Smoluchowski framework.  It serves as a necessary condition as it offers $d$ equations, which are fewer than the required $d(d+1)/2$ independent components of $\mathbf{D}(\textbf{r})$. However, it provides an expression that bounds the spatial derivative of $\mathbf{D}(\textbf{r})$, offering a means to define a prior distribution for $\mathbf{D}(\textbf{r})$ as an alternative to the one in Eq.~\eqref{eq: Hummer_prior}.

\begin{figure}
\centering
\includegraphics[width=.4\textwidth]{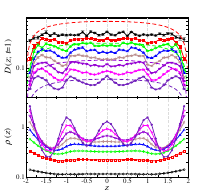}
\caption{(Reproduced from Ref.~\citenum{mittal_layering_2008}) Normal diffusivity and number density profiles for hard sphere fluids of varying packing fractions confined within a slit pore. Diffusivities are computed using Hummer's approach.\cite{hummer_position-dependent_2005}
\label{fig:hard-sphere-hummer} }
\end{figure}

As mentioned above, Hummer's method leverages a finite difference discretization of $\mathcal{L}^\dagger$ in two fundamental ways. Firstly, this discretization serves to reduce the dimensionality of the posterior, thereby enhancing the method's convergence rate. But crucially, the key contribution of the finite difference discretization method-- originally introduced by Bicout and Szabo\cite{bicout_electron_1998}-- is its ability to establish a direct connection between the matrix $\mathbf{R}$ and $\mathbf{D}(\mathbf{r})$ through Eq.~\eqref{eq: R to diffusivity}. It is important to note that this scheme is only valid in one dimension, and its extension to more intricate geometries or to alternative curvilinear coordinate systems would necessitate the development of appropriate discretization schemes. This will, in turn, result in a relationships between $\mathbf{R}$ and $\mathbf{D}$ that are more complicated that \eqref{eq: R to diffusivity}. Thus, Hummer's approach cannot be readily applied to more complex settings despite its elegance and simplicity. As explained in Appendix~\ref{Appendix2-axissim}, however, this strategy can still be applied to infer $D_{zz}(z)$ in situations wherein the diffusivity tensor is axisymmetric.

Hummer applied his Bayesian approach to a simple model system, as well as to the alanine dipeptide\cite{HermansPNAS2011} model. Since then, however, several authors have employed his approach to compute diffusion constants,~e.g.,~in confined hard sphere fluid\cite{mittal_layering_2008} (Fig.~\ref{fig:hard-sphere-hummer}), and solutes within lipid bilayers.\cite{GhorbaniJChemPhys2020, kramer_membrane_2020}

\begin{figure}
\centering
\includegraphics[width=.4\textwidth]{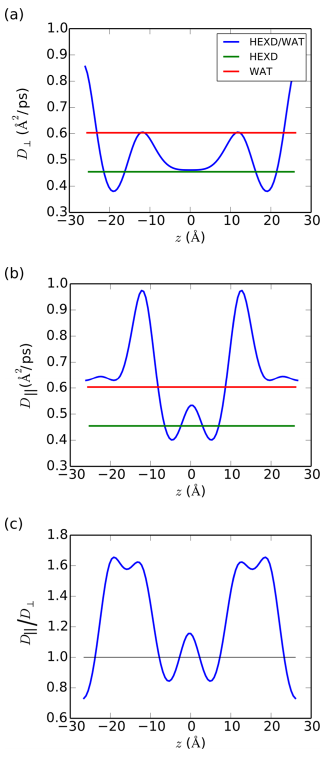}
\caption{(Reproduced from Ref.~\citenum{ghysels_position-dependent_2017}) (a) Normal and (b) lateral diffusivity profile of oxygen across a hexadecane/water film alongside (c) the anisotropy ratio computed from the Bayesian approach of Ghysels~\emph{et al}.\cite{ghysels_position-dependent_2017}
\label{fig:oxygen-Ghysels}}
\end{figure}

A conceptually similar approach was introduced by Ghysels~\emph{et al.}\cite{ghysels_position-dependent_2017} who ensured the continuity of diffusivity as a function of position via expanding it using a series of orthogonal functions $g_q(\textbf{r})$, namely $\mathbf{D}(\textbf{r}) = \sum_{q=1}^{\infty} a_qg_q(\textbf{r})$. By truncating the sum to a finite order, the authors applied Monte Carlo sampling  to infer the posterior distribution of the coefficients within this expansion.  Additionally, Ghysels~\emph{et al.} extended Hummer's approach to simultaneously infer radial and normal components of the diffusivity tensor in one-dimensional confinement in scenarios where  the diffusivity tensor could be decomposed as $\mathbf{D}(z) = D_{\|}(z)\left[\mathbf{e}_x\mathbf{e}_x^T+\mathbf{e}_y\mathbf{e}_y^T\right]+D_{\perp}(z)\mathbf{e}_z\mathbf{e}_z^T$. The differential operator $\mathcal{L}^\dagger$ of Eq.~\eqref{eq:Smoluchowski} can then be expressed in cylindrical coordinates:
\begin{equation}\label{eq: Smoluch-cylindrical}
    \mathcal{L}^\dagger f=\frac{D_\|(z)}{r}\frac{\partial}{\partial r} \left(r\frac{\partial}{\partial r}f \right)+\frac{\partial}{\partial z}\left[D_{\perp}(z) \rho_0(z)\frac{\partial }{\partial z}\left(\frac{f}{\rho_0(z)}\right)\right]
\end{equation}
The second term on the right-hand side corresponds precisely to $\mathcal{L}^\dagger$ in one dimension. Therefore, employing separation of variables, the authors demonstrated the feasibility of using Hummer's method to construct a matrix $e^{tR}$ and estimate  $D_{\perp}(z)$ by minimizing \eqref{eq:logQcData-prime} and utilizing \eqref{eq: R to diffusivity}. In order to also determine $D_{\|}(z)$, they devised a Bayesian scheme in which the  transition matrix $W_{i,mj}$ is constructed for a given $\textbf{D}(z)$  to yield the probability of transitioning from bin $i$ to bin $j$ in the $z$ direction, while exhibiting a lateral mobility corresponding to $m$-th radial bin:
\begin{equation}\label{eq: Radial propagator}
W_{i,mj}\approx\sum_{k=1}^{+\infty}\frac{J_0(\alpha_k r_m)}{\pi s^2 J_1^2(x_k)}\left[e^{[R-\alpha_k \text{diag}(D_\|)]t}\right]_{ij}
\end{equation}
Here, $\text{diag}(D_{\|})$ denotes a diagonal matrix, with its entries being $D_{\|}(z_i)$'s, where $z_i$ represents the center of the $i$-th bin along the $z$ direction. $J_0(\cdot)$ is the zeroth order Bessel function of the first kind. In order to construct a series solution (rather than an integral of Bessel functions),  the authors chose a sufficiently large distance $s$ where they imposd an artificial absorbing boundary condition,~i.e,.~$\rho(r=s,z,t|r=0,z_0,0)=0$. Therefore, $\alpha_k$'s are given by $\alpha_k=\lambda_k/s$ wherein $\lambda_k$ is the $k$-th smallest positive root of $J_0(x)$. By also defining circular bins for radial mobility, an empirical equivalent of $W$, denoted by $E_{i,mj}$, can be estimated from MD. After determining $D_{\perp}(z)$, an MLE estimate of $D_{\|}(z)$ can be obtained by minimizing:
\begin{eqnarray}
    -\log P(\mathbf{W}|\text{data}) &=& n_w\sum_{i,j,m=1}^{n_b}E_{i,mj}\log \frac{E_{i,mj}}{W_{i,mj}} + C_2\notag
\end{eqnarray}
The utilization of a finite-order expansion in terms of orthogonal functions guarantees the smoothness of both  $D_{\perp}(z)$ and $D_{\|}(z)$. They employed their methodology to characterize oxygen diffusion within organic membranes (Fig.~\ref{fig:oxygen-Ghysels}). Since its development, this approach has found widespread application in molecular simulations, particularly for probing diffusion across membranes.\cite{comer_calculating_2013, ghysels_position-dependent_2017} It has also been used to predict diffusivity of colloid suspensions \cite{beltran-villegas_self-consistent_2013} as well in  collective variable spaces for protein folding.\cite{best_coordinate-dependent_2010, hinczewski_how_2010}

Motivated by the success of these Bayesian approaches, several other Bayesian methodologies have been developed for estimating diffusivity. One example is a method proposed by Comer~\emph{et al.}\cite{comer_calculating_2013}, sometimes referred to as the adaptive biasing force (ABF) method. A crucial distinction between their approach and that of Hummer lies in the absence of spatial discretization in the former. Instead, it is based on the observation that  $\mathbf{X}_t$, the stochastic process associated with the Smoluchowski equation, exhibits a Gaussian distribution over short times. More precisely, for a sufficiently small $h$:
\begin{equation}\label{eq: asymptotic}
\mathbf{X}_t-\mathbf{r}_0\sim \mathcal{N}(\pmb\mu(\mathbf{r}_0)h,2\mathbf{D}(\mathbf{r}_0)h)
\end{equation}
which is just a restatement of \eqref{eq: Normal approximation} in terms of $\mathbf{X}_t$. Using Eq.~\eqref{eq: asymptotic}, it is possible to calculate the probability of observing a particular single-particle trajectory $\{\mathbf{x}_0,\mathbf{x}_1,\cdots,\mathbf{x}_n\}$ leading to the following expression,
\begin{eqnarray}
       & &P(\{\mathbf{x}_i\}_{1\dots n})=\prod_{i=1}^{n}p_h(\mathbf{x}_i|\mathbf{x}_{i-1})\notag\\
        & &\approx \frac{1}{\left(4\pi h\right)^{Nd/2}}
        \text{exp}\Bigg\{-\sum_{i=1}^n\Bigg[\frac{(\textbf{x}_i-\textbf{x}_{i-1}-h\pmb\mu(\mathbf{x}_{i-1}))^2}{4h|\mathbf{D}(\textbf{x}_{i-1})|^2}\notag\\
        & &+\frac{\log \left|\mathbf{D}(\mathbf{x}_{i-1})\right|}{2}\Bigg]\Bigg\},\notag
\end{eqnarray}
which can be viewed as a likelihood function for a particular diffusivity and drift profile:
\begin{equation}\label{eq: likelihood traj}
    P\left(\{\textbf{x}_i\}_{i=1}^n\right)\approx \mathbb{P}\left[\{\textbf{x}_i\}_{i=1}^n|\mathbf{D}(\mathbf{x}),\pmb\mu(\mathbf{x})\right].
\end{equation}
Here, the drift term is given by \eqref{eq:Smoluchowski-drift}.  The log-likelihood of $\mathbf{D}(\mathbf{x})$ given an observed trajectory will thus be given by:
\begin{eqnarray}\label{eq: loglikelihood traj}
        &-\log\mathbb{P}\left[\{\textbf{x}_i\}_{i=1}^{n}|\mathbf{D}(\mathbf{x}),\pmb\mu(\textbf{x})\right]\approx\dfrac{1}{2}\displaystyle\sum_{i=1}^n\log \left|\mathbf{D}(\mathbf{x}_{i-1})\right|\notag\\
        &+\displaystyle\sum_{i=1}^{n}\dfrac{\|\mathbf{x}_i-\mathbf{x}_{i-1}-h\pmb\mu(\mathbf{x}_{i-1})\|^2}{4h\left|\mathbf{D}(\mathbf{x}_{i-1})\right|}+C_1
\end{eqnarray}
While $\rho_0$ can be independently estimated from an equilibrium simulation, and the expression above could be considered as a likelihood for $\mathbf{D}(\mathbf{x})$ alone, the authors choose to treat $\mathbf{f}=\nabla \log\rho_0$ as a function to be determined through the Bayesian optimization scheme. Ultimately, the total log likelihood is derived by summing \eqref{eq: loglikelihood traj} over all observed trajectories.

Given that it is not necessary for the trajectories $\{\mathbf{x}_i\}_{i=1}^n$ entering \eqref{eq: loglikelihood traj} to be discretized, it becomes more convenient to parameterize both $\pmb\mu(\textbf{x})$ and $\mathbf{D}(\textbf{x})$ using predefined functional forms.
 The Bayesian scheme will then be used to obtain the unknown parameters. In the particular one-dimensional case considered in Ref.~\citenum{comer_calculating_2013}, the authors define a grid with regular spacing $\ell$, namely $q_i=i\ell + q_0$. Within the $i$-th cell, $D(x)$ is expressed using cubic interpolants of the following mathematical form: 
 \begin{equation}\label{eq: cubic_chipot}
 D(x) = a_i + \sum_{k=1}^3 a_i^{(k)}\left(\frac{x-q_i}{\ell}\right)^k
\end{equation}
with $a_i^{(j)}$'s given by: 
\[
\begin{cases}
    a_i^{(1)}=\dfrac{a_{i+1}-a_{i-1}}{2}\\
    a_i^{(2)}=\dfrac{-a_{i+2}+4a_{i+1}-5a_i}{2}\\
    a_i^{(3)}=\dfrac{a_{i+2}-3a_{i+1}+3a_i}{2}\\
\end{cases}
\]
A similar approach is utilized for parameterizing $\mu(x)$. Upon selecting a parameterization for $D(x)$, it will become possible to derive analytical expressions for the gradients in \eqref{eq:Smoluchowski-drift} in terms of the unknown parameters. As in the method proposed by Hummer \cite{hummer_position-dependent_2005}, the authors incorporate prior distributions, which can be applied in tandem with \eqref{eq: loglikelihood traj} to penalize rapid oscillations in $D(x)$. Specifically, they employ the prior \eqref{eq: Hummer_prior}, alongside:
\begin{equation}\label{eq: Chipot_prior}
    \log\mathbb{P}(\{a_i\})=-\sum_{i}\log a_i
\end{equation}
which ensures that $\log a_i$ are  sampled uniformly.\cite{comer_calculating_2013} It must be emphasize that the choice of  functional forms used for parameterizing $D(x)$ is by no means unique. For instance, $D(x)$ and $\mu(x)$ could be parameterized using a neural network, and \eqref{eq: loglikelihood traj} could be used as a loss function for its training. The authors present numerical comparisons between their method and Hummer's approach, indicating general qualitative agreement.% In a case where the  position-dependent diffusivity profile was \emph{a priori} known, the authors demonstrate a slightly higher level of precision in their approach.\cite{comer_calculating_2013}

We would like to comment on some numerical aspects of  this approach. While it circumvents the need for imposing a spatial discretization of the Smoluchowski operator and the associated numerical inaccuracies, the accuracy of the log-likelihood function used therein relies significantly on the Gaussian approximation of \eqref{eq: asymptotic}. This approximation is only valid for short times, and its validity diminishes as the strength of the PMF applied to the particles increases. We anticipate that this could pose challenges, particularly in situations involving hard boundaries close to which $\nabla\log\rho_0(\textbf{x})$ will diverge. Instead, methodologies that rely on spatial discretization might offer enhanced accuracy in such circumstances, as they frequently incorporate a zero-flux boundary condition to address the effects of hard boundaries

We wish to note that one can introduce another Bayesian approach by noting that the logarithm of Hummer's prior, given by Eq.~\eqref{eq: Hummer_prior} in one dimension, can be viewed as a Riemann sum. More precisely, by defining $\gamma=\widetilde{\gamma}\sqrt{\Delta{x}}$, it can be demonstrated that:
\begin{eqnarray}
 -\log\mathbb{P}(D) &=& \sum_{i}\frac{\left|D_i-D_{i+1}\right|^2}{2\gamma^2} = \frac{1}{2\widetilde{\gamma}^2}\sum_i\left|\frac{D_i-D_{i+1}}{\Delta{x}}\right|^2\Delta{x}\notag\\
 &\approx & \frac{1}{2\widetilde{\gamma}^2}\int \left|\frac{d}{dx}D(x)\right|^2\, dx .\notag
\end{eqnarray}
Therefore, in order to find a MAP estimator, one could attempt to minimize the posterior probability  by computing its functional derivative with respect to $D(x)$: 
\begin{eqnarray}\label{eq:functional derivative}
&&\frac{\delta}{\delta D(x)}\log P\big[D(x)|\text{data}\big] = \frac{\delta}{\delta D(x)}\log P\big[\text{data}|D(x)\big] \notag\\
&& +\frac{1}{2\gamma^2}\frac{\delta}{\delta D(x)}\int \left|\frac{d}{dx}D(x)\right|^2\, dx
\end{eqnarray}
Note that the second term will be proportional to $D''(x)$. Therefore, setting Eq.~\eqref{eq:functional derivative} equal to zero leads to a second-order differential equation for $D(x)$, which can be solved through a variety of means, provided that one can calculate the functional derivative of the log-likelihood. A somewhat analogous approach, while adhering to the same principles, was introduced by Chang~\emph{et al.}\cite{chang_bayesian_nodate} In line with Bayesian approaches in one dimension, the authors perform functional derivatives on $g(x)$, which is related to diffusivity through $D(x) = D_0e^{g(x)}$. This ensures that $D(x)$ remains a non-negative function. In contrast to Hummer's prior, they propose the following prior: 
\begin{eqnarray}\label{eq: chang prior}
        \log\mathbb{P}(D)&=&-\frac{1}{2}\int g(x) \left[\frac{1}{\beta\sqrt{2\pi\gamma}}e^{-\gamma\frac{d^2}{dx2}}\right]g(x)  \,dx\notag\\
        &\equiv& -\frac{1}{2}\int g(x) \mathcal{R}_{\beta,\gamma}\big[g(x)\big]  \,dx
\end{eqnarray}
The differential operator $\mathcal{R}_{\beta,\gamma}$, corresponds to the   exponential of the second derivative, with $\beta$ and $\gamma$ serving as regularization parameters-- akin to $\gamma$ in Eq.~\eqref{eq: Hummer_prior}-- that control the degree to which rapid oscillations in diffusivity are penalized.  Much like the methodology proposed by Comer~\emph{et al.}\cite{comer_calculating_2013} the log-likelihood is given by Eq.~\eqref{eq: loglikelihood traj}, and a prior with the same structure as Eq.~\eqref{eq: chang prior} is applied to the drift term $\mu(x)$. This yields a system of coupled differential equations, pertaining to $\mu(x)$ and $g(x)$, arising from the conditions $\delta\log P(g(x),\mu(x)|\text{data})/{\delta g(x)}=0$ and $\delta\log P(g(x),\mu(x)|\text{data})/{\delta \mu(x)}=0$, which are solved simultaneously. To simplify the presentation, we primarily focus on $g(x)$ given that $\mu(x)$ can be readily expressed in terms of $D(x)$ and $\rho_0(x)$. The functional derivative of Eq.~\eqref{eq: loglikelihood traj} yields:
\begin{equation}\label{eq: ODE chang}
    \mathcal{R}_{\beta,\gamma}\big[g(x)\big]=\sum_{\alpha=1}^{N_{\text{traj}}}f_\alpha(x)
\end{equation}
where summation is over the data coming from distinct trajectories. Each trajectory contributes to the corresponding ODE with a forcing term given by:
\begin{equation}\label{eq: funcder chang}
    \begin{aligned}
        & f_\alpha(x) =\\
        &-\frac{1}{2}\sum_{i=1}^{n}\frac{d}{dx}\left\{\delta\left(x-x_i^{(\alpha)}\right)\left[x_i^{(\alpha)}-x_{i-1}^{(\alpha)}-\mu\left(x_{i-1}^{(\alpha)}\right)h\right]\right\}\\
        &-\frac{1}{2}\sum_{i=1}^{n}\delta\left[x-x_i^{(\alpha)}\right]\left[1-\frac{\left|x_i^{(\alpha)}-x^{(\alpha)}_{i-1}\right|^2-\mu^2\left(x_{i-1}^{(\alpha)}\right)h^2}{2h D(x_{i-1}^{(\alpha)})}\right]
    \end{aligned}
\end{equation}
Given that the right-hand side of Eq.~\eqref{eq: ODE chang} comprises a superposition of Dirac masses, and considering the linearity of the equation, we can multiply both sides by a Green's function associated with a singular Dirac mass. Denoting this Green's function as $G_{\beta,\gamma}(x,y)$ for a source located at $y$ yields:
\begin{equation}\label{eq: integrated funcder chang}
    \begin{aligned}
        &g(x) = \\
        &\frac{1}{2}\sum_{i=1}^{n}\sum_{\alpha=1}^{N_{\text{traj}}}\Bigg\{\frac{\partial G_{\beta,\gamma}\left(x,x_i^{(\alpha)}\right)}{\partial x}\left[x_i^{(\alpha)}-x_{i-1}^{(\alpha)}-\mu(x_{i-1}^{(\alpha)})h\right]\\
        &- G_{\beta,\gamma}\big[x,x_i^{(\alpha)}\big]\left[1-\frac{|x_i^{(\alpha)}-x^{(\alpha)}_{i-1}|^2-\mu^2\left(x_{i-1}^{(\alpha)}\right)h^2}{2h D(x_{i-1}^{(\alpha)})}\right]\Bigg\}
    \end{aligned}
\end{equation}
Since this expression is valid for any point $x$ within the simulation domain, it should also hold for the $x_j^{(\alpha)}$'s, the points along all trajectories.  Evaluating $g(x)$ at each $x_j^{(\alpha)}$ results in a large system of nonlinear equations, which can be solved numerically to determine $g\big[x_j^{(\alpha)}\big]$'s. Once  known, Eq.~\eqref{eq: integrated funcder chang} can be employed to extrapolate $g(x)$ to all points. To successfully carry out this procedure, prior knowledge of $G_{\beta,\gamma}(x,y)$ is required. The precise mathematical form of $G_{\beta,\gamma}(x,y)$ is, however, only known in the absence of hard boundaries. In Ref.~\citenum{chang_bayesian_nodate}, the authors consider $x\in[0,\infty)$, which is a semi-infinite domain. As such, the Green function takes the following mathematical form:
\begin{equation}
    G_{\beta,\gamma}(x,y)=\beta\left[e^{-(x-y)^2/2\gamma}-e^{-(x+y)^2/2\gamma}\right]\notag
\end{equation}
It is worth mentioning that  as part of the procedure to determine this Green's function, the authors introduce a boundary condition at $x=0$, namely $g(0)=0$. Boundary conditions are generally necessary when formulating  MAP estimators as differential equations. In cases involving complex geometries where a closed-form expression for $G_{\beta,\gamma}(x,y)$ may not be readily available, one might start from \eqref{eq: ODE chang} and \eqref{eq: funcder chang} and explore alternative means of solving partial differential equations. 

The procedure outlined above can be applied  iteratively for different choices of regularization parameters, $\beta$ and $\gamma$. The authors provide a Bayesian framework for sampling these parameters, employing an approximate maximum marginal likelihood approach. This results in a posterior distribution for $\beta$ and $\gamma$, which can be utilized to derive error estimates for the fitting procedure. As mentioned earlier, it is crucial to emphasize that the specific form of the differential operator is contingent on the chosen regularization approach. Consequently, one can, in principle, consider suitable alternatives to  \eqref{eq: chang prior}.

Lastly, it is worth mentioning an intermediate approach proposed by T\"{u}rkcan~\emph{et al}.\cite{turkcan_bayesian_2012} bridging elements from both Hummer's\cite{hummer_position-dependent_2005} and Comer~\emph{et al.}'s\cite{comer_calculating_2013} methods. This approach is intermediate in the sense that it involves spatial discretization akin to the one used by Hummer, but the transition matrix is constructed using the Gaussian approximation rather than a matrix exponential. More precisely, the observation domain is partitioned into $n_b$ bins $\{\mathcal{B}_i\}_{i=1}^{n_b}$, and   the transition matrix is denoted with $Q$ wherein $Q^{h}_{ij}$ is the probability of transitioning for $\mathcal{B}_i$ to $\mathcal{B}_j$ over a time increment $h$,  for a given  spatial profile of $\{\textbf{D}_i,\pmb\mu_i\}_{i=1\dots N_{\text{bin}}}$. The likelihood function is thus expressed as:
$$ -\log\mathbb{P}(\text{data}|\mathbf{D}_i,\pmb\mu_i)=-\sum_{i,j}\log Q_{ij}$$
with $Q_{ij}$'s estimated from multiplying contributions from individual trajectories, namely:
\begin{equation}\label{eq: Q expansion_trajectories}
    Q_{ij}=\prod_{\alpha=1}^{n_{\text{traj}}} Q_{ij}^{(\alpha)}.
\end{equation}
Here, $Q_{ij}^{(\alpha)}$ corresponds to the likelihood associated with trajectory $\alpha$ and is given by: 
\begin{equation}\label{eq: Q likelihood per trajectory}
     Q_{ij}^{(\alpha)}=\prod_{\mathbf{x}_{k-1}^{(\alpha)}\in\mathcal{B}_i,\mathbf{x}_{k}^{(\alpha)}\in\mathcal{B}_j} p_h\left(\mathbf{x}_k^{(\alpha)}\bigg|\mathbf{x}_{k-1}^{(\alpha)}; \mathbf{D}_i,\pmb\mu_i\right).
\end{equation}
Taking logarithms from both sides of \eqref{eq: Q expansion_trajectories} yields:
\begin{eqnarray}\label{eq: loglikelihood bins}
    &&\log Q_{ij}=\sum_{\alpha=1}^{n_{\text{traj}}}\sum_{k=1}^{n}\chi_{\mathcal{B}_i}\left(\mathbf{x}_{k-1}^{(\alpha)}\right)\chi_{\mathcal{B}_j}\left(\mathbf{x}_{k}^{(\alpha)}\right)\times\notag\\ &&\log p_h\left(\mathbf{x}_k^{(\alpha)}|\mathbf{x}_{k-1}^{(\alpha)}; \mathbf{D}_i,\pmb\mu_i\right).
\end{eqnarray}
with $\log p_h(\mathbf{y}|\mathbf{x}; \mathbf{D}_i,\pmb\mu_i)$ the log of the transition probability calculated from a single trajectory using the following Gaussian approximation:
\begin{equation}
    \begin{aligned}
    &-\log p_h\left(\mathbf{x}_k^{(\alpha)}|\mathbf{x}_{k-1}^{(\alpha)}; \mathbf{D}_i,\pmb\mu_i\right) \chi_{\mathcal{B}_i}\left(\mathbf{x}_{k-1}^{(\alpha)}\right)\\
    &\approx\begin{cases}
                \dfrac{\|\mathbf{x}_k-\mathbf{x}_{k-1}-\pmb\mu_ih\|^2}{4h \left|\mathbf{D}_i\right|}+\dfrac{1}{2}\log \left|\mathbf{D}_i \right| &  \mathbf{x}_{k-1}\in \mathcal{B}_i\\
                0 & \text{ otherwise}\\
    \end{cases}
    \end{aligned}\notag
\end{equation}
Note that the indices $i$ and $j$ pertain to spatial bins, while ${\mathbf{x}_k}$'s represent points along a discrete-time trajectory, with the index $k$ tracking time increments. 

Akin to the Bayesian methods that are based on Gaussian approximations, this approach does not explicitly include any boundary information. It might therefore exhibit diminished accuracy in the presence of hard boundaries in comparison to Hummer's approach. One notable contribution of T\"{u}rkcan~\emph{et al}.\cite{turkcan_bayesian_2012} is to apply their methodology to trajectories obtained from confocal microscopy experiments. In order to account for uncertainties in position measurements, they add an isotropic noise in their definition of the transition matrix, namely:
\begin{equation}
    \begin{aligned}
    &-\log\left[p_h(\mathbf{x}_{k}|\mathbf{x}_{k-1}))\chi_{B_i}(x_{k-1})\right]\\
    &\approx\begin{cases}
                \dfrac{\|\mathbf{x}_k-\mathbf{x}_{k-1}-\pmb\mu_ih\|^2}{4h \left|\mathbf{D}_i\right|}+\dfrac{1}{2}\log \left|\mathbf{D}_i+\dfrac{\sigma^2}{h}\textbf{I} \right| &  \mathbf{x}_{k-1}\in \mathcal{B}_i\\
                0 & \text{ otherwise}\\
    \end{cases}
    \end{aligned}\notag
\end{equation}
 where $\sigma$ is the measurement noise, and $\mathbf{I}$ is the identity matrix. This approach has since been used to probe the diffusive motion of fluorescently tagged proteins at the surface of a cell.\cite{masson_mapping_2014}

\section{Operator Discretization Approaches}
\label{section:operator-discretization}

\noindent
Operator discretization approaches are developed based on the recognition that the Smoluchowski equation, Eq.~\eqref{eq:Smoluchowski}, possesses a formal solution given by $\rho_t(\textbf{y}|\textbf{x}) = e^{t\mathcal{L}^{\dagger}{\textbf{y}}}\delta(\textbf{y}-\textbf{x})$, and rely on the spatial discretization of either $\mathcal{L}^\dagger_{\textbf{y}}$ or $e^{t\mathcal{L}^\dagger_{\textbf{y}}}$ to estimate diffusivity without invoking Bayes's theorem. As an illustration, consider the approach of Sicardi~\emph{et al.}\cite{sicard_position-dependent_2021} who use Markov state models\cite{HusicJACS2018} to estimate position-dependent diffusivity. In this approach, the simulation box is discretized into $n_b$ bins, and an empirical transition matrix $M_{ij}^{(\tau)}$ is computed from MD trajectories, recording the number of times a particle moves from bin $i$ to bin $j$ over a time window $\tau$. This approach utilizes the same approximation as in \eqref{eq: diffusivity_from KM} but expresses it in terms of expectations of the solution of the Smoluchowski equation:
\begin{eqnarray}
   & &\left.\left\langle \left(\textbf{X}_{t+\tau}-\textbf{X}_t\right)\left(\textbf{X}_{t+\tau}-\textbf{X}_t\right)^T\right\rangle\right|_{\textbf{X}_t=\textbf{r}_i}\notag\\
    & &=\int(\textbf{y}-\textbf{x})(\textbf{y}-\textbf{x})^T \rho_\tau(\textbf{y}|\textbf{x})\,d\textbf{y}\label{eq:expectation_def}\\
    &&
\label{eq: MSM_diffusivity}
    \approx\sum_j (\textbf{r}_j-\textbf{r}_i)(\textbf{r}_j-\textbf{r}_i)^T M_{ij}^{(\tau)} \equiv \mathbf{C}_i^{(2),\tau}.
\end{eqnarray}
A similar expression can be obtained for the drift:
\begin{eqnarray}
         &&\left.\left\langle \textbf{X}_{t+\tau}-\textbf{X}_t\right\rangle\right|_{\textbf{X}_t=\textbf{r}_i}=\int(\textbf{y}-\textbf{x}) \rho_\tau(\textbf{y}|\textbf{x})\,d\textbf{y}\notag\\
         &&\approx\sum_j (\textbf{r}_j-\textbf{r}_i) M_{ij}^{(\tau)} \equiv \mathbf{C}_i^{(1),\tau}\label{eq:expectation_def_1}
\end{eqnarray}
Note that the integrals in \eqref{eq:expectation_def} and \eqref{eq:expectation_def_1} are expressed in terms of the nominal solution of the Smoluchowski equation, namely $\rho_t(\textbf{y}|\textbf{x})$, and are further discretized based on the employed spatial binning. One can then relate Eq.~\eqref{eq: MSM_diffusivity} to diffusivity by employing Eqs.~\eqref{eq: diffusivity_from KM} and \eqref{eq: tau effect}. More specifically, it can be demonstrated using the partial correction in \eqref{eq: tau effect} that:
\begin{eqnarray}
\mathbf{C}_i^{(2),\tau} - \mathbf{C}_i^{(1),\tau} \left[\mathbf{C}_i^{(1),\tau}\right]^T = 2\tau\mathbf{D}(\mathbf{r}) + O\left(\tau^3\right) 
\end{eqnarray}
Similar to many other techniques discussed in this review, implementing this approach requires selecting for the entire system a uniform transition timescale, $\tau$, which cannot be chosen to be arbitrarily small due to delayed  transition into the diffusive regime within MD trajectories. As discussed earlier, this might cause difficulties in probing systems with substantial dynamical heterogeneity. A possible means of resolving this issue is to choose  $\tau$ as the smallest timescale beyond which relaxation times computed from the eigenvalues of the transition matrix become insensitive to $\tau$.\cite{sicard_position-dependent_2021} It is crucial to acknowledge that employing a larger $\tau$ diminishes the accuracy of the diffusivity estimate not only due to temporal discretization errors but also because of the prevalence of drift effects. 

It is essential to highlight that determining the full diffusivity tensor using this methodology requires binning the simulation domain across all dimensions, even in cases where confinement is unidimensional. Similar to what was discussed in Hummer's approach, however, it is feasible to  apply this method only for estimating $D_{zz}$, the normal component of the diffusivity tensor, while resorting to alternative approaches, such as kernel-based methods, to estimate $D_{xx}$ and $D_{yy}$. Additionally, it is critical to take into account overarching considerations related to system discretization to ensure  accurate construction of a Markov state model.

An alternative-- but related-- approach involves discretizing $\mathcal{L}^\dagger$ instead of $e^{t\mathcal{L}}$, as proposed by Palmer,~\emph{et al.}\cite{palmer_correlation_2020} In this method, diffusivity is treated as a free parameter to be optimized by minimizing the following objective function:
\begin{equation}\label{eq: Objective_correlations}
    \chi(\textbf{D})=\sum_s \sum_{\{ij\}}\left[e^{t_s L(\textbf{D})}\pmb\Delta_{ij}-C_{ij}(t_s)\right]^2
\end{equation}
Here, the index $s$ runs over time increments, while $i$ and $j$ correspond to the bins employed for spatial discretization of the simulation box. $C_{ij}(t)$ is a time correlation function for the number of particles in $i$-th and $j$-th bins, and is defined as:
\[C_{ij}(t)=\left\langle N_i(t)N_j(0)\right\rangle,\]
while $\pmb\Delta_{ij}$'s correspond to equilibrium correlations between  bin occupancies:
\[\pmb\Delta_{ij}=\left\langle N_i(0)N_j(0)\right\rangle\]
The main intuition behind this method is the expectation that the temporal evolution of $C_{ij}(t)$ will be suitably described by $e^{tL}\pmb\Delta_{ij}$ wherein $L$ is a discretized version of the continuous operator $\mathcal{L}^{\dagger}$ using the same spatial discretization employed in computing $\pmb\Delta_{ij}$ and $C_{ij}(t)$. However, instead of using the operator that defines the spatial part of the Smoluchowski equation, the authors consider the following simplified operator:
\[\mathcal{L}^\dagger_{\text{P}} f=\nabla\cdot \left[\textbf{D}(\textbf{r}),\cdot\nabla f\right],\]
which corresponds to a diffusive process with a position-dependent diffusivity tensor but without a drift term. The authors incorporate drift indirectly by means of applying a no-flux boundary condition.  In turn, the operator is discretized as a matrix using finite differences, and the boundary conditions appear explicitly in the discretization scheme. The discretized operator $L(\textbf{D})$ is a matrix whose entries depend on the diffusivity profile, which can then be used as an independent variable to minimize the objective function \eqref{eq: Objective_correlations}.

It is worth noting that the approach, as currently described, assumes that the drift term can be satisfactorily represented by a no-flux boundary condition. We wish to note that many confined liquids exhibit strong layering at substrates, even when fluid-wall interactions are short-range in nature. Under such circumstances, the absence of an explicit drift term can cause systematic errors in the estimated diffusivity profiles. Moreover, it is crucial to emphasize that the outlined procedure involves the computation of matrix exponentials, which, in turn, requires efficient matrix diagonalization during each iteration.

Finally, we wish to discuss the work of Schulz~\emph{et al.}\cite{schulz_data-based_2017}, which is strikingly similar to Hummer's Bayesian approach\cite{hummer_position-dependent_2005}. In this approach, the simulation domain is discretized into bins, and the transition probability between bins is described using a tridiagonal rate matrix, $\mathbf{R}$:
\[Q_{ij}(t)=\left[e^{t\mathbf{R}}\right]_{ij}.\]
It's important to note that this approach is fundamentally an operator discretization approach, as the matrix $\mathbf{R}$ serves as a discretization of the differential operator $\mathcal{L}^\dagger$. However, instead of using Bayes' theorem, as in Hummer,\cite{hummer_position-dependent_2005} Schulz~\emph{et al.} minimize the mean-squared error (MSE) given by:
\begin{equation}\label{eq: quadratic objective}
    C(Q(\mathbf{R}),P)=\frac{1}{N_T}\sum_{k=1}^{N_T}\sum_{i,j} |P_{ij}(t_k)-Q_{ij}(t_k)|^2
\end{equation}
Here, $P_{ij}$ is the observed transition probabilities obtained from MD simulations or experiments. The diffusivity is then inferred from $\mathbf{R}$ using Eq.~\eqref{eq: R to diffusivity}. In Ref.\citenum{schulz_data-based_2017}, the authors directly apply their methodology to concentration profiles inferred from light absorption experiments to probe drug diffusion over the skin. One potential drawback of minimizing MSE rather than maximizing likelihood is that the latter is based on Kullback-Leibler (KL) divergence, which usually results in stronger gradients that lead to faster convergence.

\begin{figure}[h]
\centering
\includegraphics[width=.4\textwidth]{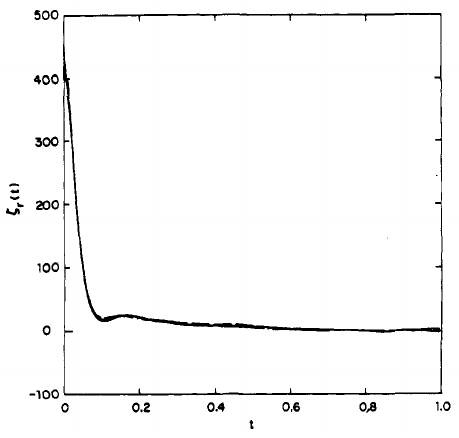}
\caption{(Reproduced from Ref.~\citenum{straub_calculation_1987}) Typical decay characteristics of the friction kernel of Eq.~\eqref{eq:GLE} computed for a model diatomic molecule within a solvent of LJ particles. \label{fig:friction-kernel}}
\end{figure}

\section{Bias-based Methods}
\label{section:bias-based}

\noindent
One intriguing approach for estimating diffusivity profiles in confined systems involves explicitly applying a suitable biasing potential to the system's Hamiltonian and utilizing the resulting mobility statistics to infer local diffusivity. At a fundamental level, these methods are exact as long as the underlying assumptions about the approximate mathematical form of the Hamiltonian are satisfied. However, introducing a bias to the Hamiltonian will inherently alter the free energetics and dynamics of the system in nontrivial ways, potentially impacting the magnitudes and functional forms of transport properties. Furthermore, it typically requires multiple MD simulations (each with biasing potentials centered at different positions) to reconstruct the complete diffusivity profile. This stands in sharp contrast to the methods discussed in previous sections, which infer the full diffusivity profile from a single unbiased MD trajectory.

The very first method of this kind, and one of the earliest methods for estimating position-dependent diffusivity profiles, was introduced by Straub~\emph{et al.}\cite{straub_calculation_1987, straub_spatial_1990} This approach is based the generalized Langevin equation (GLE) given by:
\begin{eqnarray}
d\mathbf{X}_t &=& \mathbf{V}_t\,dt \notag\\
m\,d\mathbf{V}_t &=& -\left[
\nabla\mathcal{F}(\mathbf{X}_t) + \int_0^t\pmb\zeta(t-s)\cdot\mathbf{V}_s\,ds
\right]\,dt + \mathbf{f}_t\notag\\
\label{eq:GLE}
\end{eqnarray}
Here, $\mathcal{F}$ is the system's Hamiltonian, $\pmb\zeta(\cdot)$ is a memory kernel friction tensor, and  $\mathbf{f}_t$ is a zero-mean random force. In the one-dimensional case, assuming a harmonic PMF, $\mathcal{F}(x)=\frac12m\omega^2(x-x_0)^2$, the dimensionless velocity autocorrelation function,
$$
\overline{C}_v(t) = \frac{\langle \dot{x}(t)\dot{x}(0)\rangle}{\langle \dot{x}^2\rangle}
$$
will satisfy the following differential equation:
\begin{eqnarray}
\frac{d\overline{C}_v}{dt} &=& -\int_0^tK(\tau)\overline{C}_v(t-\tau)\,d\tau\label{eq: Berne-autocorr}
\end{eqnarray}
with
\begin{eqnarray} 
K(t)&=&\omega^2+\frac{\zeta(t)}{m}.\label{eq:Berne-friction}
\end{eqnarray} 
Typical decay characteristics of a friction memory kernel computed for a simple model system is depicted in Fig.~\ref{fig:friction-kernel}. In practice, the Hamiltonian $\mathcal{F}$ is rarely  harmonic in atomic and molecular systems. Therefore, a harmonic biasing potential with a sufficiently large angular velocity is added to the system's Hamiltonian to restraint a tracer particle at $x_0$. The time-dependent friction coefficient, $\zeta(t)$ is then estimated from $\overline{C}_v(t)$. The friction coefficient at $x_0$ is obtained from the following integral:
\begin{eqnarray}
\gamma(x_0) &=& \frac1m\int_0^{\infty} \zeta(t)\,dt \label{eq:gamma-vs-zeta-Berne}
\end{eqnarray}
The local diffusivity is subsequently related to the local friction coefficient using the Stokes-Einstein relationship:\cite{MillerProcRoyal1924}
\begin{eqnarray}\label{eq:Stokes-Einstein}
m\beta\gamma(x_0)D(x_0)=1.
\end{eqnarray}
As stated above, it is  crucial for the force constant of the biasing potential to be sufficiently large in order  for the PMF to behave as a harmonic oscillator.\cite{chahine_configuration-dependent_2007}

A refined and elegant alternative to this method was introduced by Woolf and Roux.\cite{woolf_conformational_1994} By taking a Laplace transform from both sides of \eqref{eq:Berne-friction}, they demonstrate that:
\begin{eqnarray}%\label{eq: Diff WR}
\begin{array}{l}
 \mathcal{D}_\omega(s,x_0) =\\
  \dfrac{\overline{\mathcal{C}}_v(s)\left\langle (x-x_0)^2\right\rangle\left\langle \dot{x}^2\right\rangle}{\overline{\mathcal{C}}_v(s)\left[
s\left\langle (x-x_0)^2\right\rangle + \dfrac{\langle\dot{x}^2\rangle}{s}
\right] - \left\langle (x-x_0)^2\right\rangle\left\langle \dot{x}^2\right\rangle}
\end{array}
\label{eq:D-LaplaceTrans-Roux}
\end{eqnarray}
Here, $\overline{\mathcal{C}}_v(s)$ and $\mathcal{D}_\omega(s,x_0)$ corresponds to the Laplace transforms of $\overline{C}_v(t)$ and $D_\omega(t,x_0)$, respectively, and $D_\omega(t,x_0)$ denotes the time-dependent diffusivity obtained by changing the upper limit of integration in \eqref{eq:gamma-vs-zeta-Berne} from $\infty$ to $t$. (A more accessible step-by-step derivation of \eqref{eq:D-LaplaceTrans-Roux} is provided  by Gaalswyk and Rowley.\cite{gaalswyk_generalized_2016}) It is noteworthy that the infinity limit in \eqref{eq:gamma-vs-zeta-Berne} corresponds to the $s\rightarrow 0^+$ limit in \eqref{eq:D-LaplaceTrans-Roux}. In practice, $\mathcal{D}_\omega(s,x_0)$ can be computed at multiple values of $s$ and extrapolated to zero.

Finally, Hummer\cite{hummer_position-dependent_2005} demonstrated that the $s\rightarrow 0^+$ limit in \eqref{eq:D-LaplaceTrans-Roux} can be analytically estimated, resulting in the following expression:
\begin{eqnarray}\label{eq: hummer limit}
D_H &=& \lim_{s\rightarrow0^+}\mathcal{D}_\omega(s,x_0)  = \frac{\left\langle(x(t)-x_0)^2 \right\rangle^2}{\displaystyle\int_0^{+\infty}C_q(t)\,dt},
\end{eqnarray}
where $C_q(t)$ is the position autocorrelation function (PACF)\cite{hummer_position-dependent_2005, gaalswyk_generalized_2016} defined as:
\begin{equation}\label{eq: pos-autocorr}
    C_q(t)\equiv\left\langle\left[x(t)-x_0\right]\left[x(0)-x_0\right]\right\rangle
\end{equation}
Note that Eq.~\eqref{eq: hummer limit} can be re-expressed as:
\begin{eqnarray}\label{eq: hummer equivalent}
D_H &=& \frac{\langle(x-x_0)^2\rangle}{\tau}
\end{eqnarray}
with the timescale, $\tau$, given by:
$$
\tau = \frac{\displaystyle\int_0^{+\infty}C_q(t)\,dt}{\left\langle(x(t)-x_0)^2 \right\rangle}.
$$
Upon closer inspection, Eq.~\eqref{eq: hummer equivalent} is reminiscent of  the concept of mean-squared displacement, specifically the estimators based on Kramers-Moyal coefficients given by~\eqref{eq: diffusivity_from KM}. However, a notable limitation of these approaches is their assumption that diffusivity is a scalar (i.e.,~isotropic) position-dependent quantity, a condition almost never met in confined systems. Consequently, it is necessary to adapt these expressions to such circumstances.  In Appendix~\ref{Appendix2-axissim}, we demonstrate the adaptability of this methodology to estimate anisotropic diffusivities that are axisymmetric. Due to its suitability for one-dimensional collective variables, this method has been applied in studies of protein folding to compute diffusivity along a reaction coordinate.\cite{chahine_configuration-dependent_2007,best_coordinate-dependent_2010,socci_diffusive_1996}

Despite their differences, these methodologies fall under the broad category of \emph{static restraint (SR)} methods, as per Holland~\emph{et al.}\cite{holland_calculating_2012}  since they all rely on restraining the position of a tracer particle at a fixed location using a harmonic spring. An alternative approach for estimating $\pmb\zeta(t)$ in Eq.~\eqref{eq:GLE} can be devised by applying the fluctuation-dissipation theorem, demonstrating that:\cite{kubo_fluctuation-dissipation_1966,roux_ion_1991,marrink_simulation_1994-1}
\begin{eqnarray}\label{eq: fluctuation-dissipation GLE}
\pmb\zeta (t) &=& \beta\left\langle \mathbf{f}(t)\mathbf{f}^T(0) \right\rangle,
\end{eqnarray}
Here, $\mathbf{f}(t) = \textbf{F}(t) - \langle \mathbf{F}(t)\rangle_t$ is the residual force exerted on the tracer particle at time $t$. The diffusivity tensor can then be estimated as (refer to Appendix~\ref{section:force-autocorrelation}),
\begin{eqnarray}
\mathbf{D} &=& \beta^{-2} \left[
\int_0^{+\infty}\left\langle
\mathbf{f}\left(\mathbf{X}_t\right) \mathbf{f}^T\left(\mathbf{X}_0\right)
\right\rangle\, dt
\right]^{-1}\label{eq: diff from force}
\end{eqnarray}
Note that Eq.~\eqref{eq: diff from force} is valid even without a restraining force as long as the diffusivity is constant. It can also be applied, akin to SR methods, to estimate local diffusivity when the restraining force is sufficiently strong to locally restrain the tracer particle. However, since the force autocorrelation function is independent of the tracer's temporal evolution, it can be computed even for a constrained tracer particle. This corresponds to taking the limit of \eqref{eq: diff from force} as $\omega\rightarrow\infty$. In such a scenario, the diffusivity at $\mathbf{r}_0$ can be computed by pinning a particle at that position and computing the force autocorrelation function for that particle:\cite{saito_cholesterol_2011,marrink_simulation_1994-1}
\begin{eqnarray}
\mathbf{D}(\mathbf{r}_0) &=& \beta^{-2} \left[
\int_0^{+\infty}\left\langle
\mathbf{f}\left(\mathbf{r}_0,t\right) \mathbf{f}^T\left(\mathbf{r}_0,0\right)
\right\rangle\, dt
\right]^{-1}\label{eq: diff from force_MD}
\end{eqnarray}
Eq.~\eqref{eq: diff from force_MD} features a constraint rather than a restraining force. Therefore,  following Holland~\emph{et al.}'s\cite{holland_calculating_2012} terminology, the corresponding method can be labelled as a \emph{static constraint (SC)} method.

The aforementioned approaches, whether based on pinning the particle to a fixed position or employing a restraining force, prove valuable in capturing diffusivity profiles within regions of the simulation box with low probabilities of being visited by certain solutes. Indeed, these approaches were historically developed for the study of membrane permeation, and have been extensively utilized in estimating the diffusivity of permeants across lipid membranes\cite{saito_cholesterol_2011, carpenter_method_2014, sugii_molecular-dynamics_2005, gaalswyk_generalized_2016, shinoda_molecular_2004, marrink_simulation_1994-1} and ion transport through pores.\cite{roux_ion_1991} A detailed review of such approaches is given by Shinoda.\cite{shinoda_permeability_2016} 

One can devise an alternative\cite{marrink_simulation_1994-1} method for estimating the friction coefficient (and diffusivity) starting with the SDE describing underdamped Langevin dynamics:
\begin{equation}\label{eq: langevin}
    \begin{cases}
        d\mathbf{X}_t=\mathbf{V}_tdt\\
        d\mathbf{V}_t=-(\nabla \widetilde{\mathcal{F}}(\mathbf{X}_t)+\pmb\gamma(\mathbf{\mathbf{X}_t})\cdot\mathbf{V}_t)dt+\sqrt{2\widetilde{\mathbf{D}}(\mathbf{X}_t)}\cdot d\mathbf{W}_t
    \end{cases}
\end{equation}
Here, $m$ is the mass of the particle, $\widetilde{\mathcal{F}}(\mathbf{r})=m^{-1}\mathcal{F}(\mathbf{r})$ is the reduced PMF, and $\widetilde{\mathbf{D}}$ is the 'velocity diffusivity` (i.e.,~with units of velocity squared divided by time). Taking the expectation value of both sides of the velocity equation yields:
\begin{eqnarray}\label{eq:balance-underdamped}
m\langle d\mathbf{V}_t\rangle &=& \left[
\left\langle \mathbf{F}_{\text{ext}}  \right\rangle
+m\pmb\gamma(\mathbf{r})\cdot\left\langle \mathbf{V}_{t}  \right\rangle
\right]
\end{eqnarray}
where $\langle\mathbf{F}_{\text{ext}}\rangle$ is the mean external force exerted on the tracer (due to both the PMF and possibly non-equilibrium external forces). Note that $\langle\mathbf{F}_{\text{ext}}\rangle$ does not include any random forces, which average out to zero. Assuming that the external force can be kept sufficiently small for the change in velocity to be negligible, one can use the magnitude of the net force to compute the friction coefficient. In the simplest case of one-dimensional diffusivity, this will result in:
\begin{eqnarray}\label{eq: Diff from ext force}
D(z) &\approx& \frac{\langle V_t\rangle}{\beta\langle F_{\text{ext}}\rangle}
\end{eqnarray}
One way of implementing \eqref{eq: Diff from ext force}, originally proposed by Cicotti and Jacucci,\cite{ciccotti_direct_1975} is to apply vanishingly small external forces such that the friction can overcome $F_{\text{ext}}$. However, it must be emphasized that $F_{\text{ext}}$ also encompasses the effects of the PMF and is not solely comprised of the external force exerted during non-equilibrium MD. This may cause some issues in applying this methodology in the vicinity of hard boundaries.

The most interesting applications of this expression come from the works of McKinnon~\emph{et al.}\cite{mckinnon_nonequilibrium_1992} and Holland~\emph{et al.}\cite{holland_calculating_2012} Both use steered molecular dynamics in which a harmonic potential is applied to the tracer particle with a moving minimum $z(t)$. In McKinnon~\emph{et al.}'s work,\cite{mckinnon_nonequilibrium_1992} this minimum moves with a constant velocity $v_d$  and the instantaneous magnitude of the external force is calculated such that the left-hand side of \eqref{eq:balance-underdamped} remains zero at all times.  The forcing term can therefore be replaced by a term involving the work of the restraining force:
\begin{equation}\label{eq: diff from restraint force}
    D\left[z(t)\right]=\frac{v_d}{\beta\,{d\langle W_{\text{ext}}\rangle}/{dz}}\approx v_d\frac{\langle l\rangle}{\beta \langle W_{\text{ext}}\rangle}\approx \frac{\langle l\rangle^2}{t\beta \langle W_{\text{ext}}\rangle}
\end{equation}
Using Holland et al.'s terminology,\cite{holland_calculating_2012} this method can be categorized as a \emph{dynamic restraints (DR)} method. Clearly, neglecting the effect of PMF will affect the validity of equation \eqref{eq: diff from restraint force}. Indeed, a more rigorous derivation, which explicitly accounts for the effect of the PMF and uses overdamped Langevin dynamics, is given by Park and Schulten.\cite{park_calculating_2004}

\begin{figure}
    \centering
    \includegraphics[width=0.45\textwidth]{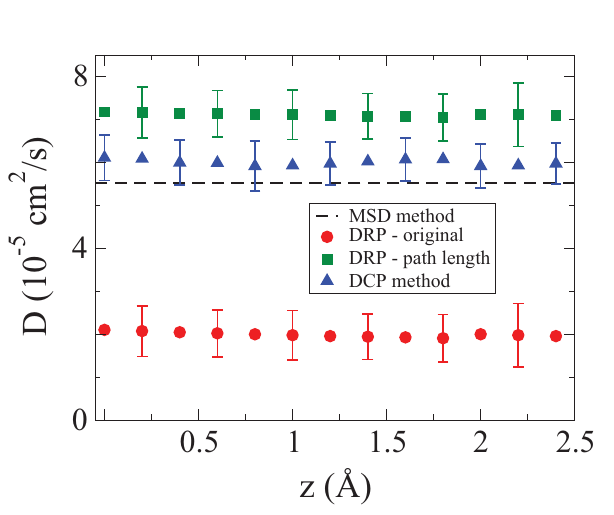}
    \caption{(Reproduced from Ref.~\citenum{holland_calculating_2012}) The effect of the particular choice of $\langle l\rangle$ on the performance of the DR method described by Eq.~\eqref{eq: diff from restraint force}. Using the path length results in better agreement with the true value of diffusivity (estimated from MSD). Overall, the DC method exhibits a superior performance in comparison with different implementations of the DR method.}
    \label{fig:holland1}
\end{figure}

Holland et al.\cite{holland_calculating_2012} highlight several practical issues in numerically implementing the method proposed by McKinnon et al.\cite{mckinnon_nonequilibrium_1992} and introduce a new method, which they call a \emph{dynamic constraint (DC)} method, to address some of those issues. They consider the limiting case in which the tracer particles are constrained to follow a prescribed path $z(t)$ rather than being subjected to a harmonic force centered at $z(t)$. Intuitively, this could be thought of as a special case of the latter approach in the limit of very stiff springs. The external force, $F_{\text{ext}}(t)$, needed to drive the particle along $z(t)$ is then back-calculated by estimating the force exerted on the tracer by the environment. The work from this force is then subtracted from the change in PMF and employed in \eqref{eq: diff from restraint force}.

They also remark on proper choices of $\langle l\rangle$ and $v_d$ in Eq.~\eqref{eq: diff from restraint force} when the DR approach is employed. They argue that the observation window $t$ needs to be partitioned into smaller windows of duration $\Delta t\ll t_s$ where $t_s$ is the characteristic oscillation period of the harmonic spring. They then argue that $\langle l\rangle$ and $v_d$ need to be chosen as:
%\begin{subequations}
\begin{eqnarray}
\langle l\rangle = \sum_{i=1}^{n_w} \langle |\Delta z_i|\rangle, &~~~&
v_d =  \frac1t\left|
\sum_{i=1}^{n_w} \langle \Delta z_i\rangle
\right|\notag
\end{eqnarray}
%\end{subequations}
In other words,  $\langle l\rangle$ needs to be chosen as the average arc length of tracer paths, while $v_d$ should be estimated from the net displacement. Failing to do so will result in a systematic underestimation of diffusivity, as shown in Fig.~\ref{fig:holland1}. In contrast, the DC approach, in which the particle follows a prescribed path, is not impacted by such uncertainties and provides more accurate estimates of diffusivity.

\begin{figure}[h]
    \centering
    \includegraphics[width=.4\textwidth]{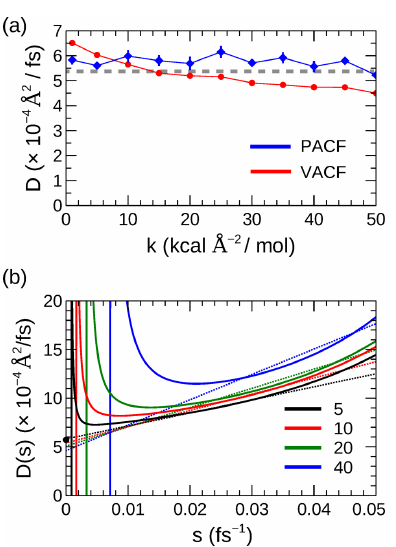}
    \caption{(Reproduced from Ref.~\citenum{gaalswyk_generalized_2016}) (a) Effect of spring constant on (a) estimates of diffusivity obtained using the Eqs.~\eqref{eq:D-LaplaceTrans-Roux} and \eqref{eq: hummer limit}, and (b) the qualitative behavior of $\mathcal{D}_{\omega}(s)$ in Eq.~\eqref{eq:D-LaplaceTrans-Roux}. }
    \label{fig:springK}
\end{figure}

Since restraining force-based methods have been in use for a long time, known numerical issues with their implementation have been reported. For instance, it is known that Langevin-based thermostats, such as the Berendsen thermostat,\cite{BerendsenJChemPhys1984} introduce extra unphysical friction despite correctly sampling the equilibrium Boltzmann distribution. This additional friction results in a systematic underestimation of diffusivity,\cite{gaalswyk_generalized_2016} which is absent from deterministic thermostats, such as the Nos\'{e}-Hoover thermostat.\cite{NoseMolPhys1984, HooverPhysRevA1985} Another obvious issue is that using stiff springs might necessitate employing a smaller MD time step.

An interesting implementation question is the sensitivity of the estimated friction coefficient to the particular choice of the spring constant. Gallswyk~\emph{et al.}\cite{gaalswyk_generalized_2016} investigated this issue, comparing the VACF-based method of Woolf and Roux,\cite{woolf_conformational_1994} with its reformulation by Hummer,\cite{hummer_position-dependent_2005} given by Eqs.~\eqref{eq:D-LaplaceTrans-Roux} and~\eqref{eq: hummer limit}, respectively. They observed that for moderate values of $k$, the VACF-based approach of Eq.~\eqref{eq:D-LaplaceTrans-Roux} is more sensitive to $k$. They conjecture that this increased sensitivity arises from the uncertainties in the numerical extrapolation of \eqref{eq:D-LaplaceTrans-Roux} to $s\rightarrow0^+$ for larger values of $k$ (Fig.~\ref{fig:springK}). Despite the reduced sensitivity of the PACF-based method of Hummer to  $k$, it can suffer from slow decay of PACF in some circumstances.\cite{gaalswyk_generalized_2016}  In bulk liquids, a few picoseconds might be sufficient for the position autocorrelation function to decay to zero at room temperature. In contrast, a solute restrained deep inside a lipid bilayer might exhibit long-lived oscillations that last up to hundreds of picoseconds, resulting in slow decay of PACF. This observation is corroborated by Daldrop and Netz,\cite{daldrop_external_2017} who demonstrate a 100-ps decay for the PACF of a methane molecule restrained within water.

\begin{figure}
    \centering
    \includegraphics[width=0.45\textwidth]{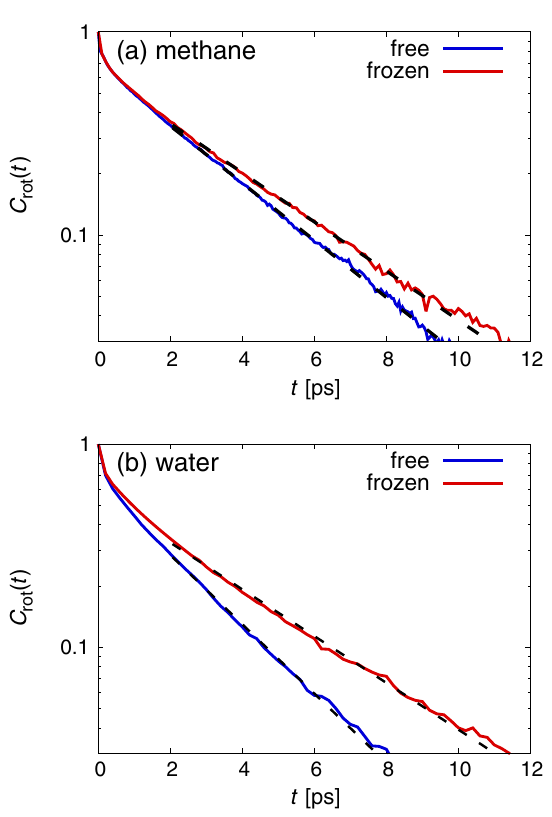}
    \caption{(Reproduced from Ref.~\citenum{daldrop_external_2017})
    Sensitivity of the orientational autocorrelation functions of water molecules within the first hydration shell of a fixed and a freely moving (a) methane and (b) water molecules. }
    \label{fig:orientational}
\end{figure}

Daldrop and Netz\cite{daldrop_external_2017} also analyze the effect of larger spring constants on the performance of Hummer's PACF-based approach, something not considered by Gaalswyk~\emph{et al.}\cite{gaalswyk_generalized_2016} They demonstrate that while larger $k$'s   lead to faster decay of PACF, they might make it highly oscillatory, which could also make its accurate numerical integration more challenging.  Moreover, they observe that larger spring constants lead to a systematic overestimation of the friction coefficient, and hence an underestimation of diffusivity. For instance, they report that the friction coefficient of a methane solute within SPC/E water\cite{BerendsenJPhysChem1987} increases by 60\% in the limit of $k\rightarrow\infty$. They attribute the source of this systematic error to the alteration of the dynamics of the first hydration shell by a frozen molecule. For instance, they compute an orientational autocorrelation function for the water molecules in the first hydration shell and find it to decay more slowly when the solute is frozen. This discrepancy becomes larger when stronger interactions are present between the solute and the solvent. Similarly, they observe an increase in the mean escape time of water molecules around the solute. In other words, solvent molecules tend to be less mobile around a pinned particle. This, once again, underscores the fundamental issue with bias-based methods in general that can alter the dynamics of the underlying system in nontrivial ways, and highlights the merits of utilizing alternative approaches,~e.g.,~based on path sampling techniques,\cite{MalmirMatter2020} to study membrane permeation.

Various other details in the implementation of bias-based methods have been  examined in the literature. For instance, Fujimoto et al.\cite{fujimoto_momentum_2021} examined the impact of finite size effects, especially in the context of long-range electrostatic interactions. Additionally, Holland et al.\cite{holland_calculating_2012} illustrated that increasing the speed of the moving spring might adversely impact the accuracy of diffusivity estimates. These findings collectively emphasize the importance of exercising careful consideration when selecting parameters for the implementation of bias-based methods.

\section{Collective variable-based approaches}
\label{section:diff-CV-spaces}

\noindent
In this section, we will discuss methodologies originally developed in the context of collective variables,~i.e.,~mechanical observables carefully chosen to accurately represent the free energy landscapes of physical and biological systems, particularly with the aim of characterizing the kinetics and mechanisms  of rare events.\cite{ValssonAnnuRevPhysChem2016, HussainJChemPhys2020} These variables can be formulated through various approaches, including physical intuition, experimental insights, or data science methodologies such as principal component analysis,\cite{SittelJChemPhys2018, KaciraniJPCB2024} diffusion maps,\cite{FergusonPNAS2010, EvansApplComputHarmonAnal2023} and machine learning.\cite{NadlerAppComputHarmonAnal2006, AppeldornJPhysChemB2022, BeyerleCurrOpinSolidSt2023, DominguesJPCL2024} From a mathematical standpoint, CVs can be seen as embeddings or projections that map the high-dimensional configuration space onto a lower-dimensional space. The projection formalism introduced by Mori and Zwanzig\cite{ZwanzigJChemPhys1960, MoriProgTheoPhys1965} demonstrates that the temporal evolution of a collective variable can be modeled as a stochastic process with memory, even in cases where the underlying equations of motion are Hamiltonian. In situations with negligible memory effects, or just as a first approximation, a Smoluchowski-type equation is postulated within the CV space:
\begin{equation}\label{Smoluch_collective}
\frac{\partial\rho}{\partial t}=\nabla\cdot\Big[\textbf{D}(\pmb\lambda)\cdot\left[\nabla_{\pmb\lambda}\rho+\beta\rho\nabla_{\pmb\lambda} F(\pmb\lambda)\right]\Big] = \mathcal{L}^{\dagger}\rho
\end{equation}
Here, $\pmb\lambda$ represents a vectorial CV that exhibits diffusive behavior, and $F(\pmb\lambda)$ denotes the Landau free energy profile\cite{MendelsJPhysChemLett2018} with respect to $\pmb\lambda$.

Assuming the validity of this description, algorithms can be developed to estimate $\mathbf{D}(\pmb\lambda)$ from the CV time series, $\pmb\lambda_t$. Indeed, several of the algorithms\cite{socci_diffusive_1996, hinczewski_how_2010, best_coordinate-dependent_2010, hummer_position-dependent_2005, sicard_position-dependent_2021, woolf_conformational_1994} discussed in prior sections have also been employed for characterizing diffusivity variations within a CV space. In this section, we want to focus on two classes of conceptually distinct methodologies, both historically rooted in computational chemistry and biology, which are based on estimating the mean first passage time (MFPT) and committor analysis, respectively.
We discuss both approaches within the framework of  transition path theory\cite{VandenEijndenAnnRevPhysChem2010} (TPT), specifically focusing on transitions between $A$ and $B$, two open sets within $\mathbb{R}^n$, which are also (meta)stable basins of attractions within the free energy landscape.

\begin{figure}
\centering
\includegraphics[width=.5\textwidth]{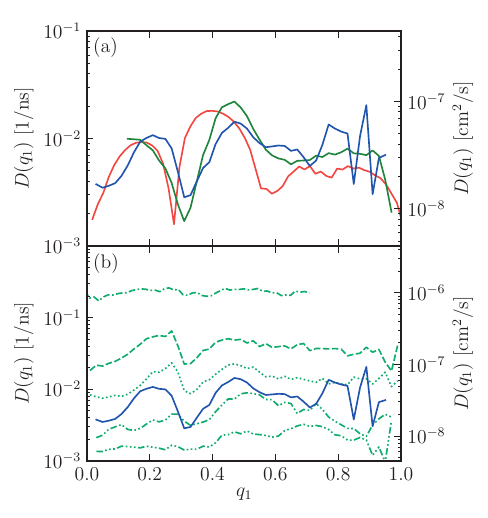}
\caption{(Reproduced from Ref.~\citenum{hinczewski_how_2010}) Computing diffusivity along $q_1$, the root mean square deviation from a perfect helix for an $\alpha$-helix forming short peptide. (a) $D(q_1)$ computed from Eq.~\eqref{eq: diff_derivatives} (green curve), Eq.~\eqref{eq: Diffusivity_RT} (blue curve) and using Hummer's Bayesian approach\cite{hummer_position-dependent_2005} (red curve). (b) $D(q_1)$ estimates from Eq.~\eqref{eq: Diffusivity_RT} (blue curve) as well as using Eq.~\eqref{eq: DrDiff} using different lag times (green curves).\label{fig:CV-peptide-helix}}
\end{figure}

 \subsection{MFPT-based approaches}

\noindent
Consider an open set $C\subset\mathbb{R}^n$. The first passage time to leave $C$ (while starting at $\pmb\lambda_0\in C$) is defined as:
$$
\tau_{\text{fp}}(C|\pmb\lambda_0) = \inf_{t} \{t>0:\pmb\lambda_t\notin C\}.
$$
First passage times can be estimated by imposing absorbing boundary conditions at $C$'s boundary, i.e.,
\begin{eqnarray}
\rho(\pmb\lambda,t) = 0,~~~~\forall \pmb\lambda^*\in C^*\subset \partial C.\label{eq:FPT-absorbing}
\end{eqnarray}
Note that the $\rho(\pmb\lambda,t)$ that satisfies \eqref{eq:FPT-absorbing} will not be properly normalized, and will instead  represent a joint probability density,~i.e.,~the probability that the particle is found at $\pmb\lambda$ at time $t$ and that it has not crossed the absorbing boundary $C^*$ up to time $t$. Within the Smoluchowski framework, $\tau_{\text{fp}}(C|\pmb\lambda_0)$ can be calculated as:\cite{weiss_first_1967}
\begin{eqnarray}\label{eq: tauFP}
        && \tau_{\text{fp}}(C|\pmb\lambda_0) =\int_0^\infty -t\frac{\partial}{\partial t}\left[\int_C \rho(\pmb\lambda,t|\pmb\lambda_0)\,d\pmb\lambda\right]\,dt\notag\\
        &&=\int_C \left[\int_0^\infty \rho(\pmb\lambda,t|\pmb\lambda_0)\,dt\right]\,d\pmb\lambda \equiv\int_C G(\pmb\lambda|\pmb\lambda_0)\,d\pmb\lambda
\end{eqnarray}
The second equality is derived through the inversion of the order of integration and the application of integration by parts. The function $G(\pmb\lambda|\pmb\lambda_0)=\int_0^\infty \rho(\pmb\lambda,t|\lambda_0)\,dt$ represents a Green's function that satisfies the following PDE:
\begin{eqnarray}\label{eq:Green-fcn-CV-nD}
&& \nabla\cdot\Big[\textbf{D}(\pmb\lambda)\cdot\left[\nabla_{\pmb\lambda}G(\pmb\lambda|\pmb\lambda_0)+\beta G(\pmb\lambda|\pmb\lambda_0)\nabla_{\pmb\lambda} U(\pmb\lambda)\right]\Big]\notag\\
&& =\rho_0(\pmb\lambda)-\delta(\pmb\lambda-\pmb\lambda_0)
\end{eqnarray}
Eqs.~\eqref{eq: tauFP} and \eqref{eq:Green-fcn-CV-nD} are valid for any open set  $C$ within a multidimensional CV space. Following Weiss,\cite{weiss_first_1967} it is possible to derive an analytical expression for the first passage time for a scalar (i.e.,~one-dimensional) CV. First, one can define:
\begin{equation}\label{eq: Udef}
    u(\pmb\lambda,t)=\int_C \rho(\mathbf{y},t|\pmb\lambda)\,d\mathbf{y}.
\end{equation}
Note that $ u(\pmb\lambda,t)$ is the probability that  $\tau_{\text{fp}}(C|\pmb\lambda)\geq t$ and satisfies the adjoint evolution equation given by:
\begin{equation}\label{eq: adjointev}
        \frac{\partial u}{\partial t} =e^{\beta F(\pmb\lambda)} \nabla_{\pmb\lambda} \cdot\left[ e^{-\beta F(\pmb\lambda)}\mathbf{D}(\pmb\lambda)\cdot \nabla_{\pmb\lambda} u\right]= \mathcal{L} u
\end{equation}
where $\mathcal{L}$ is the adjoint of the operator on the right-hand side of \eqref{Smoluch_collective}. As discussed previously, $\tau_{\text{fp}}(C|\pmb\lambda)$ can be calculated as:
\[\tau_{\text{fp}}(C|\pmb\lambda)=-\int_0^\infty  t \frac{\partial u}{\partial t}\,dt\]
It can be demonstrated that $\tau_{\text{fp}}(C|\pmb\lambda)$  will satisfy the following PDE:
\begin{equation}\label{eq: Tau green}
    e^{\beta F(\pmb\lambda)} \nabla_{\pmb\lambda} \cdot\left[ e^{-\beta F(\pmb\lambda)} \mathbf{D}(\pmb\lambda)\cdot\nabla_{\pmb\lambda} \tau_{\text{fp}}(C|\pmb\lambda)\right]=-1.
\end{equation}
All these assertions are valid for any CV space, irrespective of its dimensionality. If the CV is scalar, however, \eqref{eq: Tau green} will turn into an ordinary differential equation (ODE), which can be solved analytically for an arbitrary open interval, $C=(a,b)$ to yield:
\begin{equation}\label{eq: Gen-sol}
    \begin{aligned}
        \tau_{\text{fp}}(C|\lambda) &= C_0+C_1\int_a^\lambda \frac{1}{\rho_0(y)D(y)}\, dy\\
        &-\int_a^\lambda \frac{1}{\rho_0(y)D(y)} \left[\int_a^y \rho_0(z) \,dz\right]\,dy
    \end{aligned}
\end{equation}
Here, $\rho_0(\lambda) = {e^{-\beta F(\lambda)}}/{\int_a^b d\overline{\lambda}e^{-\beta F(\overline{\lambda)}}}$ represents the  conditional equilibrium distribution of $\lambda$ within the interval $(a,b)$. The constants $C_0$ and $C_1$ depend on the specific boundary conditions imposed at $a$ and $b$.

We now consider  the solutions of Eq.~\eqref{eq: Gen-sol} when  reflective boundary conditions are imposed at $a$ and $b$.  This scenario is widely explored in the literature, representing the case where $a$ and $b$ correspond to two (local) minima in the free energy landscape, indicative of transitions between two (meta)stable basins. Additionally, we introduce an intermediate point $\lambda_f\in (a,b)$ where an absorbing boundary condition is applied. Consequently, trajectories can originate from two distinct sets: $\lambda\in [a,\lambda_f]$ or $\lambda \in [\lambda_f,b]$, each featuring a reflective and an absorbing boundary condition. The mean first passage time of reaching $\lambda_f$ can be readily estimated as:\cite{hinczewski_how_2010}
\begin{eqnarray}\label{eq: tau reflecting}
    \tau_{\text{fp} }(\lambda_f|\lambda) = \left\{
    \begin{array}{ll}
    {\displaystyle\int}_{\lambda_f}^{\lambda}\dfrac{d\xi\,e^{\beta F(\xi)}}{D(\xi)}{\displaystyle\int}_{\xi}^b d\zeta\,e^{-\beta F(\zeta)} & \lambda \in [\lambda_f,b]\\
    {\displaystyle\int}_{\lambda}^{\lambda_f}\dfrac{d\xi\,e^{\beta F(\xi)}}{D(\xi)}{\displaystyle\int}_{a}^{\xi} d\zeta\,e^{-\beta F(\zeta)} & \lambda\in [a,\lambda_f]
    \end{array}
    \right.\notag\\
    \label{eq:MFPT-1D}
\end{eqnarray}
Note that the system can only exit the starting set at $\lambda_f$ as reflective boundary conditions are applied at the other end. This is why we denote the first passage time with $\tau_{\text{fp} }(\lambda_f|\lambda)$ instead of $\tau_{\text{fp} }(C|\lambda)$. Eq.~\eqref{eq:MFPT-1D} can be subsequently used to derive the following expressions for $D(\lambda)$ in terms of the derivative of $\tau$ with respect to $\lambda_0$:
\begin{equation}\label{eq: diff_derivatives}
	D(\lambda)=\begin{cases}
	\dfrac{e^{\beta F(\lambda)}}{{\partial \tau_{\text{fp}}}/{\partial  \lambda}}{\displaystyle\int}_\lambda^b e^{-\beta F(\xi)}\,d\xi & \lambda> \lambda_f\\
	-\dfrac{e^{\beta F(\lambda)}}{{\partial \tau_{\text{fp}}}/{\partial  \lambda}}{\displaystyle\int}_a^{\lambda} e^{-\beta F(\xi)}\,d\xi & \lambda< \lambda_f
	\end{cases}
\end{equation}
Alternative, one can define the notion of a round-trip time as,\cite{hinczewski_how_2010}
\begin{eqnarray}{\label{eq: Roundtrip}}
	\tau_{\text{rt}}(\lambda, \lambda_f)&=&\text{sign}(\lambda-\lambda_f)\left[\tau_{\text{fp}}(\lambda,\lambda_f)+\tau_{\text{fp}}(\lambda_f,\lambda)\right]\notag\\
 &=& \int_{\lambda_f}^{\lambda} d\xi\,\frac{e^{\beta F(\xi)}}{D(\xi)}\int_a^b d\zeta\,e^{-\beta F(\zeta)} 
\end{eqnarray}
By differentiating Eq.~\eqref{eq: Roundtrip} with respect to $\lambda$, it is easy to demonstrate that:
\begin{equation}\label{eq: Diffusivity_RT}
    D(\lambda) = \frac{1}{\rho_0(\lambda) \partial\tau_{\text{rt}}/\partial\lambda}
\end{equation}
Here, we adopt the notation employed by Hinczewski~\emph{et al}.\cite{hinczewski_how_2010} who use Eqs.~\eqref{eq: diff_derivatives} and  \eqref{eq: Diffusivity_RT}  to compute diffusivity along different CVs associated with folding of a helix-forming short peptide into an $\alpha$-helix (Fig.~\ref{fig:CV-peptide-helix}). However, similar expressions for the case of $\lambda\in [a,\lambda_f]$ have also been utilized by Chahine~\emph{et al.,}\cite{chahine_configuration-dependent_2007}  Sedlmeier~\emph{et al.}\cite{sedlmeier_water_2011} and  Bollinger and Truskett.\cite{bollinger_structure_2014}  It is crucial to underscore that these analytical expressions are exact when applied to one-dimensional CV spaces. Consequently, they have found widespread applications in characterizing transitions within biomolecular systems that can be adequately represented by a scalar CV. Both Hinczewski~\emph{et al.}\cite{hinczewski_how_2010} and Chahine~\emph{et al.}\cite{chahine_configuration-dependent_2007} employed these expressions for probing protein folding, while Sedlmeier~\emph{et al.}\cite{sedlmeier_water_2011} and Bollinger and Truskett\cite{bollinger_structure_2014} applied them to estimate actual diffusivity profiles within an axisymmetric geometry-- specifically for SPC/E water\cite{BerendsenJPhysChem1987} confined within a slit pore and the hard sphere fluid exposed to a one-dimensional sinusoidal potential, respectively. In both cases, diffusivity was solely a function of $z$, and the methodology was applied to estimate $D_{zz}$, which is justified due to the axisymmetric geometry of the system (see Appendix~\ref{Appendix2-axissim}).

It is important to acknowledge that the numerical implementation of this approach may pose challenges, even in cases that are truly one-dimensional. This challenge arises from the necessity to estimate the derivatives of either the mean first passage time or the round-trip time with respect to $\lambda$. Importantly, these derivatives appear in the denominators of \eqref{eq: diff_derivatives} and \eqref{eq: Diffusivity_RT}. As such, substantial instabilities might arise if such derivatives are small in magnitude.  Alternatively, one can introduce appropriate approximations to estimate these derivatives, thereby mitigating the reliance on numerical differentiation.

One such approximate approach was proposed in Belousov~\emph{et al.},\cite{belousov_first-passage_2020} who consider a sufficiently narrow interval containing $\lambda_0$, namely $C=[\lambda_0-\ell, \lambda_0+\ell]$. By applying absorbing boundary conditions at both ends of $C$, \eqref{eq: Gen-sol} will yield:
\begin{eqnarray}
\tau_{\text{fp} }(C|\lambda_0) &=& \int_{\lambda_0-\ell}^{\lambda_0}\left[\overline{\mathcal{M}} - \mathcal{M}(y) \right]\,\frac{dy}{D(y)\rho_0(y)}\notag
\end{eqnarray}
where:
\begin{eqnarray}
\mathcal{M}(y) &=& \int_{\lambda_0-\ell}^{y}\rho_0(z)\,dz\\
\overline{\mathcal{M}} &=& \frac{\displaystyle\int_{\lambda_0-\ell}^{\lambda_0+\ell}\dfrac{\mathcal{M}(y)\,dy}{D(y)\rho_0(y)}}{\displaystyle\int_{\lambda_0-\ell}^{\lambda_0+\ell}\dfrac{dy}{D(y)\rho_0(y)}}\notag
\end{eqnarray}
and $\rho_0(\lambda)\propto e^{-\beta F(\lambda)}$. The approximation works by obtaining the Taylor expansion of $F(\lambda)$ and log diffusivity around the fixed point $\lambda_0$:
\begin{eqnarray}
    &&F(\lambda) = F(\lambda_0) + F'(\lambda_0)(\lambda-\lambda_0) + O\left(|\lambda-\lambda_0|^2\right),\notag\\
    && \frac{D(\lambda)}{D(\lambda_0)} = 
    \exp\left[\frac{D'(\lambda_0)}{D(\lambda_0)}(\lambda-\lambda_0) + O\left(|\lambda-\lambda_0|^2\right)\right],
    \notag
\end{eqnarray}
which can then be used to estimate mean first passage and roundtrip times over short displacements within the CV space.  For instance, the MFPT of starting at $\lambda_0$ and reaching $\lambda_0\pm\ell$ will be given by:\cite{belousov_first-passage_2020}
\begin{eqnarray}
    &&\tau_{\text{fp}}(C|\lambda_0) \approx\notag\\
    &&\frac{1}{\beta\kappa F'(\lambda_0)D(\lambda_0)}\left[
    \frac{\cosh\left\{\left[\beta F'(\lambda_0)+\kappa\right]\ell/2\right\}}{\cosh\left\{\left[\beta F'(\lambda_0)-\kappa\right]\ell/2\right\}}-1
    \right]\notag\\
    && \label{tau_approx_taylor}
\end{eqnarray}
Here, $\kappa = D'(\lambda_0)/D(\lambda_0)$. If $\kappa\ll1$, Eq.~\eqref{tau_approx_taylor} can be further simplified to yield:
\begin{eqnarray}
\label{eq:tau_approx_2}
    \tau_{\text{fp}}(C|\lambda_0) \approx \frac{\ell\tanh\left[\beta\ell F'(\lambda_0)/2\right]}{\beta D(\lambda_0)F'(\lambda_0)}
\end{eqnarray}
Therefore, $D(\lambda)$ can be directly estimated from the first derivative of $F(\lambda)$,~i.e.,~the mean force exerted along the scalar CV, and the mean first passage time of reaching $\lambda\pm\ell$. Further approximations along the same lines can be made, particularly if the  Smoluchowski equation is replaced by a related Fokker-Planck equation \cite{belousov_statistical_2022}. 

 \subsection{Committor-based methods}

\noindent
Here, we discuss a method\cite{berezhkovskii_communication_2017} proposed by Berezhkovskii and Makarov that estimates position-dependent diffusivity from committor analysis. Assuming the existence of two basins of attractions, $A$ and $B$, within the free energy landscape, $q_B(x)$, the committor probability of reaching $B$, is defined as:
\begin{eqnarray}
    q_B(\pmb\lambda) &=& \mathbb
    {P}(T_B<T_A) \notag
\end{eqnarray}
where $T_C$ is the first passage time of reaching set $C$:
\begin{eqnarray}
    T_C &=& \inf_t \{t>0: \pmb\lambda_t\in C\}\notag
\end{eqnarray}
Assuming that the evolution of $\pmb\lambda_t$ within the CV space follows \eqref{Smoluch_collective}, it can be demonstrated that $q_B(\pmb\lambda)$ will satisfy the following elliptic PDE:\cite{e_transition-path_2010}
\begin{eqnarray}\label{Commitor_equation}
    \begin{array}{ll}
        \nabla\cdot\Big[\textbf{D}(\pmb\lambda) e^{-\beta F(\pmb\lambda)}\cdot\left[\nabla_{\pmb\lambda} q_B\right]\Big]=0 & \pmb\lambda\not\in A\cup B\\
        q_B\equiv1 & \pmb\lambda \in \partial B\\
        q_B\equiv0 & \pmb\lambda \in \partial A\\
    \end{array}
\end{eqnarray}
It is necessary to note that \eqref{Commitor_equation} is only valid when $\pmb\lambda\not\in A\cup B$, otherwise $q_B(\pmb\lambda)=0$ and $1$ for $\pmb\lambda\in\partial A$ and $\pmb\lambda\in\partial B$, respectively. 
In the case of a scalar CV, one can obtain an analytical expression for $q_B(\lambda)$. More precisely, for $A=(-\infty,a)$ and $B=(b,\infty)$, it can be demonstrated that:
\begin{equation}\label{eq: Commitor}
q_B(\lambda)=\frac{\displaystyle\int_a^\lambda\dfrac{e^{\beta F(s)}}{D(s)}\,ds}{\displaystyle\int_a^b\frac{e^{\beta F(s)}}{D(s)}\,ds}.
\end{equation}
Upon differentiation  and rearrangement,  \eqref{eq: Commitor} can be re-expressed  as:\cite{berezhkovskii_communication_2017}
\begin{equation}\label{eq:D-vs-comm}
	D(\lambda)=\frac{\displaystyle\int_a^b q_B(z)(1-q(z))\rho_0(z)dz}{\rho_0(\lambda)q_B'(\lambda)\langle\tau_{\text{tr}}(a,b)\rangle}
\end{equation}
Here, $\tau_{\text{tr}}(a,b)$, the \emph{transition time}, is the earliest time that a trajectory initiated at $a$ or $b$ leaves the interval $(a,b)$. Note that the transition time is smaller than the first passage time. For a one-dimensional CV space, the mean transition time is given by:
\begin{eqnarray}
    \langle\tau_{\text{tr}}(a,b)\rangle &=& \int_a^b\frac{e^{\beta F(s)}}{D(s)}\,ds\int_a^b q_B(u)\left[1-q_B(u)\right]e^{-\beta F(u)}\,du\notag\\
    && \label{eq:mean-tr}
\end{eqnarray}
Note that the only derivative appearing in Eq.~\eqref{eq:D-vs-comm} is $q'_B(\lambda)$, which can be evaluated using recent algorithms for the parameterization of the commitor probabilities,~e.g.,~through the application of neural networks.\cite{khoo_solving_2018} While Eq.~\eqref{eq:D-vs-comm} can, in principle, be used to estimate $D(\lambda)$, we are not aware of any instance of its application in the literature, possibly due to numerical instabilities emerging from the exceedingly small values of $q'_B(\lambda)$  within regions that are too far from the transition state.

\section{Conclusions}
\label{section:conclusion}

\noindent
In this work, we provide a brief overview of computational attempts to estimate position-dependent diffusivity tensors (and other related transport coefficients) from MD trajectories of confined systems.  At a fundamental level, these methodologies involve solving the inverse problem of inferring diffusivity profiles within the Smoluchowski framework from the observed mobility statistics of individual particles. We classify these attempts based on their underlying theoretical foundations. In addition to \emph{ad hoc} extensions (Section~\ref{section:adhoc-methods}) of rigorous algorithms developed for bulk simulations (Section~\ref{section:transport-bulk}), we discuss more rigorous methodologies, such as kernel-based methods (Section~\ref{section:kernel-based}), Bayesian approaches (Section~\ref{section:Bayesian}), operation discretization methods (Section~\ref{section:operator-discretization}), and bias-based methods (Section~\ref{section:bias-based}). We also discuss the related problem of estimating diffusivity profiles in collective variable spaces through estimating mean first passage times and committor probabilities (Section~\ref{section:diff-CV-spaces}). 

While we primarily focus on methods developed and utilized in the context of molecular simulations, we wish to note that  the Smoluchowski equation can be equivalently expressed as the forward Kolmogorov equation associated with a stochastic process. The task of deducing the PMF and position-dependent diffusivity from observed mobility statistics can thus be perceived as an inference problem in stochastic processes, which has a rich history within the statistics community and remains an active area of research.\cite{yuecai_nadaraya-watson_2022,ganguly_infinite-dimensional_nodate, RenArXiv2023} Our kernel-based method, as elaborated in our previous works,\cite{domingues_robust_2023, domingues_robust_2023-1} and detailed in Section~\ref{section:kernel-based}, represents an endeavor to adapt successful estimators from the statistics community to the problem of determining transport coefficients in molecular simulations. We contend that numerous unexplored opportunities exist in this realm. An interesting example is the study of electromagnetic wave propagation in highly scattering media, a process that is described by a PDE very similar to the Smoluchowski equation. Consequently, methodologies in the optics community have been developed to estimate diffusion tensors,\cite{markel_inverse_2001} and there exists potential in adapting such approaches to molecular dynamics trajectories. 

On a broader-- but related-- note, the question of inferring a term (or parameter) within a PDE from observations of its solutions-- generally referred to \emph{inverse problem}-- is at the heart of applied mathematics. Indeed, the applied mathematics literature is replete  with  many more strategies for discretizing the Smoluchowski operator, often validated using synthetic data, as in the work of Crommelin and Vanden-Eijnden.\cite{CrommelinMultiscaleModelSimul2011} Adapting such strategies to be applicable to MD data could provide further opportunities to develop effective operator discretization algorithms for diffusivity estimation, beyond those already considered in this review.

In addition to developing new diffusivity estimators, it is also imperative to establish standard benchmarks for validating and assessing the performance of-- new and existing-- estimators. Traditionally, the validation of new estimators has relied on their ability to accurately recover diffusivity profiles used for generating synthetic stochastic trajectories or to precisely estimate diffusivity within bulk systems. We contend that such traditional benchmarks are necessary, but not sufficient, for guaranteeing the robustness and reliability of an estimator, and more rigorous validation criteria are necessary. One such approach, as demonstrated in our previous work,\cite{domingues_robust_2023} involves feeding the predictions of the estimator into a Langevin-based SDE, and comparing the mobility statistics of the arising stochastic trajectories, with the van Hove correlation functions obtained from MD simulations. This will not only enable one to evaluate the estimator's capability to generate internally consistent diffusivity profiles, but will also make it feasible to identify and flag deviations from the Smoluchowski picture within specific systems.

It is essential to acknowledge that the Smoluchowski equation serves as an approximation for describing particle mobility over extended timescales and may not be universally applicable to all molecular systems. A key assumption within the Smoluchowski formalism pertains to the Gaussian nature and the absence of temporal correlation for the random force acting on each particle. This assumption can be relaxed by extending the GLE formalism of Eq.~\eqref{eq:GLE}, or through the incorporation of colored noise. Such extensions accommodate scenarios where the dynamics of a typical particle is non-Markovian. The GLE framework also exhibits improved agreement with VACFs computed from MD.\cite{straub_spatial_1990} Notably, there have been endeavors\cite{vroylandt_likelihood-based_2022, xie_ab_2022} within the molecular simulations community, including machine learning approaches, to fit data from MD simulations to a GLE framework without explicitly aiming to estimate diffusivity. Adapting such methodologies for the estimation of position-dependent diffusivity could be a promising avenue for future investigations.

An interesting category of systems and processes, not addressed in this discussion, include those exhibiting anomalous diffusion.\cite{JeonPRL2012, vonHansenPRL2013, KrottJCP2016} One notable example is systems comprised of interacting Brownian particles governed by the McKean-Vlasov equation.\cite{McKeanPNAS1966} Unlike the standard and generalized Langevin formalisms, which both rely on a 'mean field` treatment of a typical particle within a bath, the McKean-Vlasov equation makes it possible to account for multi-particle effects. The estimation of transport coefficients within such generalized frameworks poses an intriguing question that has not been addressed in this review. It must be noted that such complicated scenarios, such as those involving anomalous diffusion or systems described by the McKean-Vlasov equation, can still be tackled by modifying some of the methodologies discussed earlier. One viable option is to employ Bayesian or operator discretization approaches, known for their adaptability to diverse settings. In cases where the PDE governing the spatiotemporal evolution of probabilities is associated with a stochastic process, such as the McKean-Vlasov equation, kernel-based methods offer a valuable avenue. Depending on the specific characteristics of the underlying stochastic process, adapting autocorrelation-based techniques is also conceivable, although their generalization might prove more challenging.

We also do not discuss the problem of estimating transport properties other than diffusivity. Given the mathematical similarity between mass and charge transport, we expect some of the techniques described here to be applicable to estimation of position-dependent and anisotropic electrical conductivity. Indeed, methodologies such as the one proposed by Mangaud and Rotemberg\cite{mangaud_sampling_2020} have been employed by Helms~\emph{et al}.\cite{helms_intrinsic_2023} to estimate the response matrix $\mathcal{M}(z)$ in the presence of an external electrical potential.  It is, however, far more challenging to treat momentum and heat transfer in a similar fashion. Although frameworks akin to the one discussed in Ref.\citenum{mangaud_sampling_2020} can be utilized alongside closed-form solutions of macroscopic fluid mechanics problems to estimate quantities such as viscosity, their applicability to more intricate geometries remains uncertain.

We contend that more systematic approaches for estimating transport coefficients, such as viscosity, can be formulated by positing that $f(t,\mathbf{r},\mathbf{v})$, the probability density of a particle being at $\mathbf{r}$ and having a velocity $\mathbf{v}$ will satisfy the following  kinetic ansatz:
\begin{equation}\label{eq: kinetic}
    \frac{\partial}{ \partial t} f(t,\mathbf{r},\mathbf{v})+\mathbf{v}\cdot\nabla_{\mathbf{r}} f +\nabla_{\mathbf{r}} \widetilde{U}\cdot\nabla_{\mathbf{v}} f=\mathcal {O}[f]
\end{equation}
Here, the operator $\mathcal {O}[f]$ encapsulates the effective interactions among particles in the system. For instance, in the case of the Kramers-Klein equation,\cite{KramersPhysica1940} $\mathcal{O}[\cdot]$ is given by:
$$ \mathcal{O}[f] =\nabla_{\mathbf{v}}\cdot\left(\frac{1}{m\beta}\mathbf{D}^{-1}(x)\cdot\mathbf{v} f\right)+ \frac{1}{(m\beta)^2}\mathbf{D}^{-1}(x):\mathbf{H}_{\mathbf{v}}f$$
In the context of a master equation framework, $\mathcal{O}[\cdot]$ can have the following structure:
\begin{eqnarray}\label{eq: mastereq}
\mathbb{O}[f]  &=& \int\Bigg[
W(\mathbf{r}',\mathbf{r},\mathbf{v}',\mathbf{v})f(t,\mathbf{r}',\mathbf{v}')\notag\\
&& -W(\mathbf{r},\mathbf{r}',\mathbf{v},\mathbf{v}')f(t,\mathbf{r},\mathbf{v})\Big]\,d\mathbf{r}'\,d\mathbf{v}'
\end{eqnarray}
Multiplying \eqref{eq: kinetic} by $\mathbf{v}$ and integrating over velocity space allows the derivation of a momentum balance equation. A similar approach can be employed to derive an energy conservation equation. Data-driven techniques can then be applied to fit the statistics obtained from MD simulations to such a kinetic description. Transport
coefficients would be implicit to the choice of the operator, and could potentially be extracted from it. These concepts present avenues for future exploration.

\appendix

\vspace{-10pt}

\section{Axisymmetric solution to the Smoluchowski equation}\label{Appendix2-axissim}

\noindent
Here, we consider a scenario in which both the diffusivity and PMF are functions of $z$ only, and that the diffusivity tensor is axisymmetric,~i.e.,~$D_{xx}(z) \equiv D_{yy}(z) \equiv D_{rr}(z)$. This will imply translational invariance within the $xy$ plane, a condition met in systems wherein a fluid is sandwiched  between chemically uniform parallel plates. Under these conditions, the Smoluchowski equation takes the following form:
\begin{equation}\label{eq: Smoluch-symmetric}
    \frac{\partial\rho} {\partial t } = D_{rr}(z)\left[\frac{\partial^2\rho} {\partial x^2 }  +\frac{\partial^2\rho} {\partial y^2 }\right]  +\frac{\partial} {\partial z }\left[D_{zz}(z) \rho_0\frac{\partial} {\partial z } \left(\frac{\rho}{\rho_0}\right)\right] 
\end{equation}
with the operator acting on $D_{zz}(z)$ denoted as:
$$\mathcal{L}^\dagger_z f \equiv \frac{\partial}{\partial z}\left[D_{zz}(z) \rho_0(z)\frac{\partial}{\partial z}\left(\frac{f}{\rho_0}\right)\right],$$
which is the differential operator on the right-hand side of the one-dimensional Smoluchowski equation. It can be shown that:
\begin{eqnarray}
\begin{array}{l}
\dfrac{\partial\widehat{\rho}} {\partial t} = -4\pi^2D_{rr}(z)\left(k_x^2 +k_y^2 \right)\widehat{\rho} +\mathcal{L}^\dagger_z\widehat{\rho} \\
\widehat{\rho}(k_x,k_y,z,0) = \delta(z-z_0)
\end{array}\label{eq:Smol-FT}
\end{eqnarray}
with $\widehat{\rho}(k_x, k_y, z, t)$, the Fourier transform of $\rho(\cdot)$ defined as:
\begin{eqnarray}\label{eq: axiss}
&& \widehat{\rho}(k_x,k_y,z,t) = \notag\\
&& \int_{\mathbb{R}^2}e^{-2\pi i\left[k_x(x-x_0)+k_y(y-y_0)\right]} \rho(x,y,z,t|x_0,y_0,z_0)\,dx\,dy\notag\\
&&
\end{eqnarray}
In other words, $\widehat{\rho}(k_x, k_y, z, t)$ satisfies a diffusion-reaction equation in one dimension. Considering the translational invariance within the $xy$ plane for $\rho(x, y, z, t | x_0, y_0, z_0)$, which is solely a function of $\Delta{x} = x - x_0$ and $\Delta{y} = y - y_0$, we observe that the marginal probability density:
$$
p(z,t|z_0) = \int_{\mathbb{R}^2} \rho(\Delta{x},\Delta{y},z,t|z_0),d\Delta{x}\,d\Delta{y},
$$ 
is identical to $\widehat{\rho}(0, 0, z, t | z_0)$, as given in \eqref{eq: axiss}. By setting $k_x = k_y = 0$ in \eqref{eq:Smol-FT}, the reactive term vanishes, and $p(z, t | z_0) = \widehat{\rho}(0, 0, z, t | z_0)$ satisfies the one-dimensional Smoluchowski equation. This implies that for any system accurately described by \eqref{eq: Smoluch-symmetric} or \eqref{eq: Smoluch-cylindrical}, the time series of $z$ coordinates will adhere to the one-dimensional Smoluchowski picture. As such, methods developed for scalar diffusive coordinates, such as the ones discussed in Section~\ref{section:diff-CV-spaces},  can  be applied to estimate $D_{zz}(z)$.

\section{Relationship between diffusivity and force autocorrelation function}
\label{section:force-autocorrelation}

\noindent
Consider the SDE given by Eq.~\eqref{eq: langevin}, which describe underdamped Langevin dynamics. By imposing the Boltzmann distribution as the steady-state distribution of \eqref{eq: langevin}, it can be demonstrated that $\widetilde{\mathbf{D}}(\mathbf{r}) = \left(m\beta\right)^{-2}\mathbf{D}^{-1}(\mathbf{r})$ and $\pmb\gamma(\mathbf{r}) = \left(m\beta\right)^{-1}\mathbf{D}^{-1}(\mathbf{r})$. The second term on the right-hand side of the velocity equation can be interpreted as a random acceleration term resulting from interactions with the surrounding environment. Denoting this random acceleration as $d\mathbf{a}(\mathbf{X}_t)$, It\^{o} calculus can be used to demonstrate that:\cite{kubo_fluctuation-dissipation_1966}
\begin{eqnarray}\label{eq: velocity diff}
        \widetilde{\mathbf{D}}&=&\int_0^\infty\left\langle d\mathbf{a}\left(\mathbf{X}_t\right)d\mathbf{a}^T\left(\mathbf{X}_0\right)\right\rangle\,dt\notag\\
        &=& \frac1{m^2}\int_0^\infty\left\langle d\mathbf{f}\left(\mathbf{X}_t\right)d\mathbf{f}^T\left(\mathbf{X}_0\right)\right\rangle\,dt\notag
\end{eqnarray}
where $d\mathbf{f}_t = md\mathbf{a}_t$ represents the random forcing terms over the time interval $dt$. Eq.~\eqref{eq: diff from force} directly follows from the relationship between $\mathbf{D}$ and $\widetilde{\mathbf{D}}$. It is important to note that forces in molecular dynamics are continuous functions of time, so the above expression is only valid approximately, assuming the validity of underdamped Langevin dynamics.

\section*{acknowledgements}

\noindent
A.H.-A. gratefully acknowledges the support from the National Science Foundation Grants CBET-1751971 (CAREER Award) and CBET-2024473. This work was supported as part of the Center for Enhanced Nanofluidic Transport (CENT), an Energy Frontier Research Center funded by the U.S. Department of Energy, Office of Science, Basic Energy Sciences under Award \#DE-SC0019112.

\bibliography{References-LitRevDiff}

\end{document}